\documentclass[journal]{vgtc}               

\ifpdf%
  \pdfoutput=1\relax                   
  \usepackage{graphicx}               
  \DeclareGraphicsExtensions{.pdf,.png,.jpg,.jpeg} 
\else%
  \usepackage{graphicx}              
  \DeclareGraphicsExtensions{.eps}     
\fi%

\graphicspath{{figures/}{pictures/}{images/}{./images/}} 
\usepackage{microtype}                 
\PassOptionsToPackage{warn}{textcomp}  
\usepackage{textcomp}                  
\usepackage{mathptmx}                  
\usepackage{amsfonts}
\usepackage{times}                     
         
\usepackage{cite}                      
\usepackage{tabu}                      
\usepackage{booktabs}                  
\usepackage{makecell}
\usepackage{xspace}
\usepackage{enumitem}
\usepackage{amsmath}
\usepackage{subfiles}
\usepackage{overpic}
\usepackage{ccicons}
\usepackage{textcomp}
\usepackage{multirow}
\usepackage{makecell}
\newcommand{\STAB}[1]{\begin{tabular}{@{}c@{}}#1\end{tabular}}
\usepackage{cuted}

\newcommand{\etal}{et~al.\xspace}
\newcommand{\eg}{e.\,g.}
\newcommand{\ie}{i.\,e.}

\newcommand{\point}{MeTAPoint\xspace}
\newcommand{\paint}{MeTAPaint\xspace}
\newcommand{\brush}{MeTABrush\xspace}
\newcommand{\baseline}{Baseline\xspace}

\onlineid{1431}

\vgtccategory{Research}

\vgtcpapertype{Representations/Interaction}

\title{MeTACAST: Target- and Context-aware Spatial Selection in VR}

\author{%
   \authororcid{Lixiang Zhao}{0000-0001-6181-1673},
   \authororcid{Tobias Isenberg}{0000-0001-7953-8644},
   \authororcid{Fuqi Xie}{0009-0008-4728-9346},
   \authororcid{Hai-Ning Liang}{0000-0003-3600-8955}, 
   \authororcid{Lingyun Yu}{0000-0002-3152-2587}
}

\authorfooter{
\item  L. Zhao  (\raisebox{-.5pt}{\includegraphics[height=6pt]{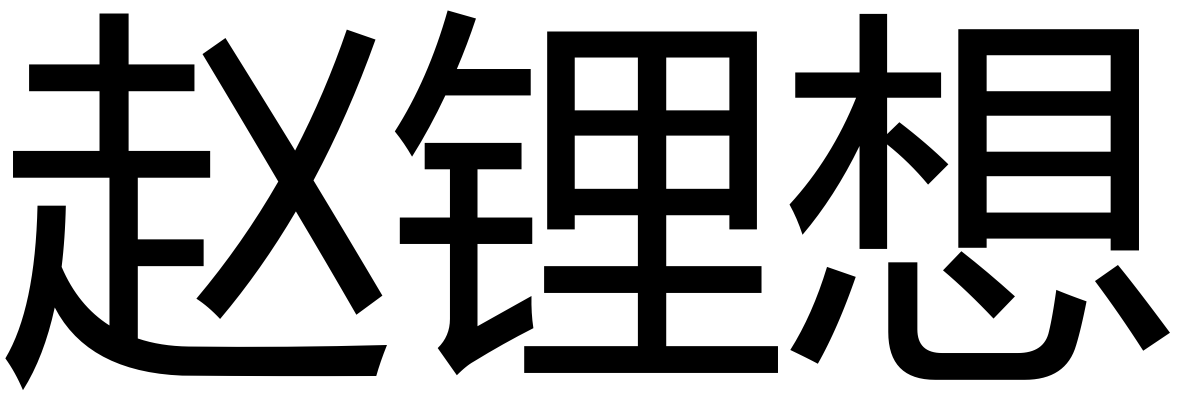}}), F. Xie  (\raisebox{-.5pt}{\includegraphics[height=6pt]{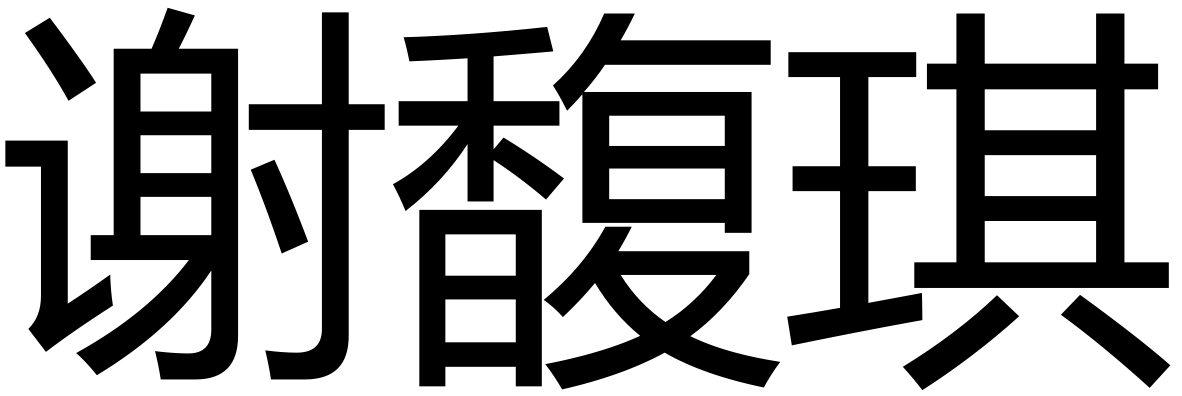}}), H.-N. Liang  (\raisebox{-.5pt}{\includegraphics[height=6pt]{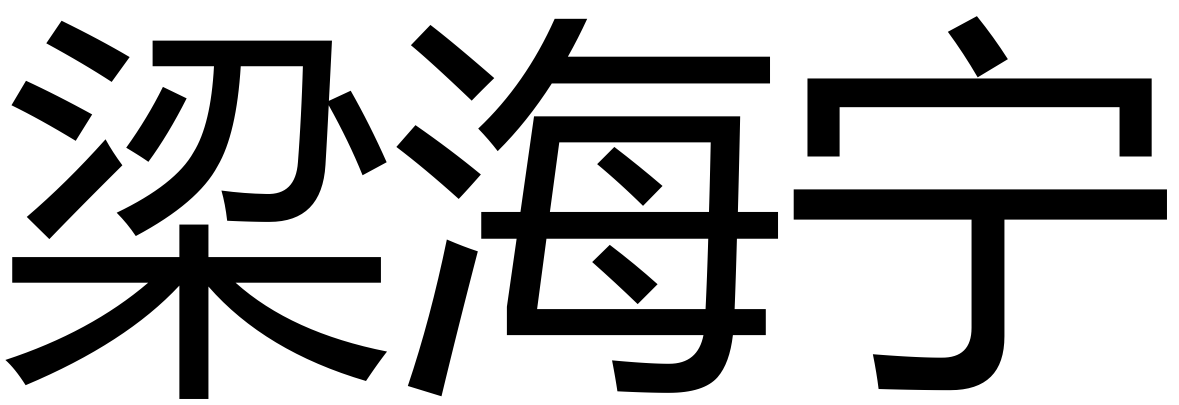}}), and L. Yu (\raisebox{-.5pt}{\includegraphics[height=6pt]{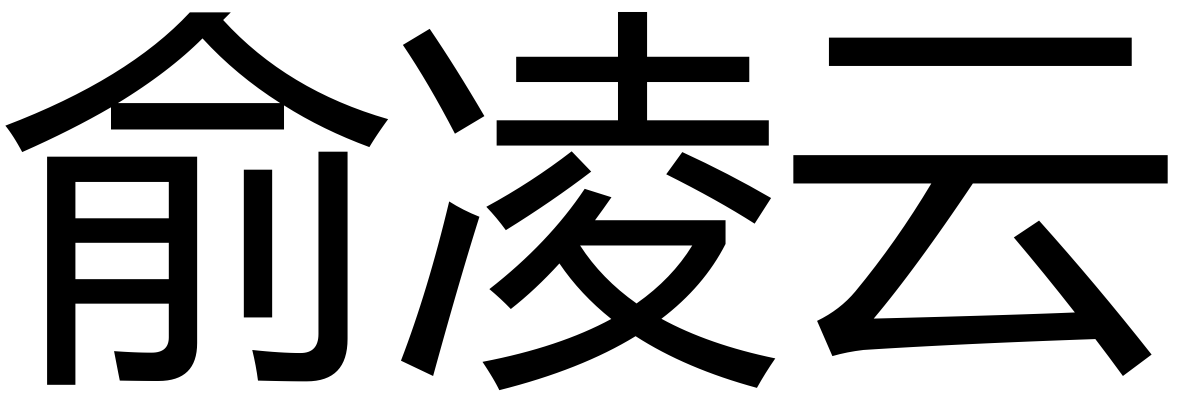}}) are with Xi'an Jiaotong-Liverpool University, China. E-mail: \{lixiang.zhao17 $|$ \textls[-10]{fuqi.xie18\} @student.xjtlu.edu.cn,  \{haining.liang $|$ lingyun.yu\} @xjtlu.edu.cn}. L. Yu is the corresponding author.
\item  T. Isenberg is with Universit{\'e} Paris-Saclay, CNRS, Inria, LISN, France. E-mail: tobias.isenberg@inria.fr.
}

\abstract{%
We propose three novel spatial data selection techniques for particle data in VR visualization environments. They are designed to be target- and context-aware and be suitable for a wide range of data features and complex scenarios. Each technique is designed to be adjusted to particular selection intents: the selection of consecutive dense regions, the selection of filament-like structures, and the selection of clusters---with all of them facilitating post-selection threshold adjustment. These techniques allow users to precisely select those regions of space for further exploration---with simple and approximate 3D pointing, brushing, or drawing input---using flexible point- or path-based input and without being limited by 3D occlusions, non-homogeneous feature density, or complex data shapes. These new techniques are evaluated in a controlled experiment and compared with the Baseline method, a region-based 3D painting selection. Our results indicate that our techniques are effective in handling a wide range of scenarios and allow users to select data based on their comprehension of crucial features. Furthermore, we analyze the attributes, requirements, and strategies of our spatial selection methods and compare them with existing state-of-the-art selection methods to handle diverse data features and situations. Based on this analysis we provide guidelines for choosing the most suitable 3D spatial selection techniques based on the interaction environment, the given data characteristics, or the need for interactive post-selection threshold adjustment.}


\CCScatlist{ 
 \CCScat{K.6.1}{Management of Computing and Information Systems}%
{Project and People Management}{Life Cycle};
 \CCScat{K.7.m}{The Computing Profession}{Miscellaneous}{Ethics}
}

\teaser{
    \centering
      \includegraphics[height=.212\textwidth]{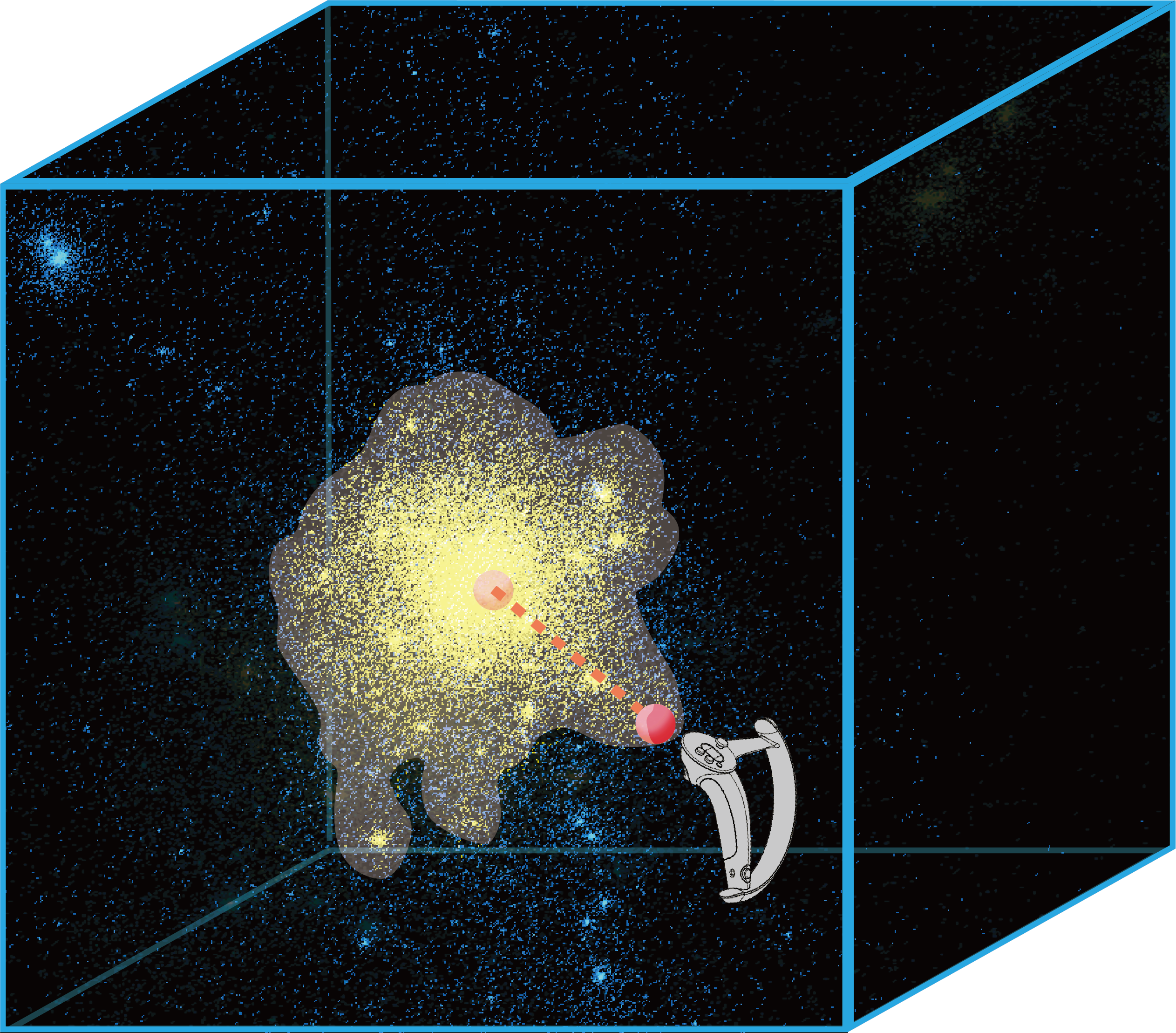}\hspace{5mm}
    \includegraphics[height=.212\textwidth]{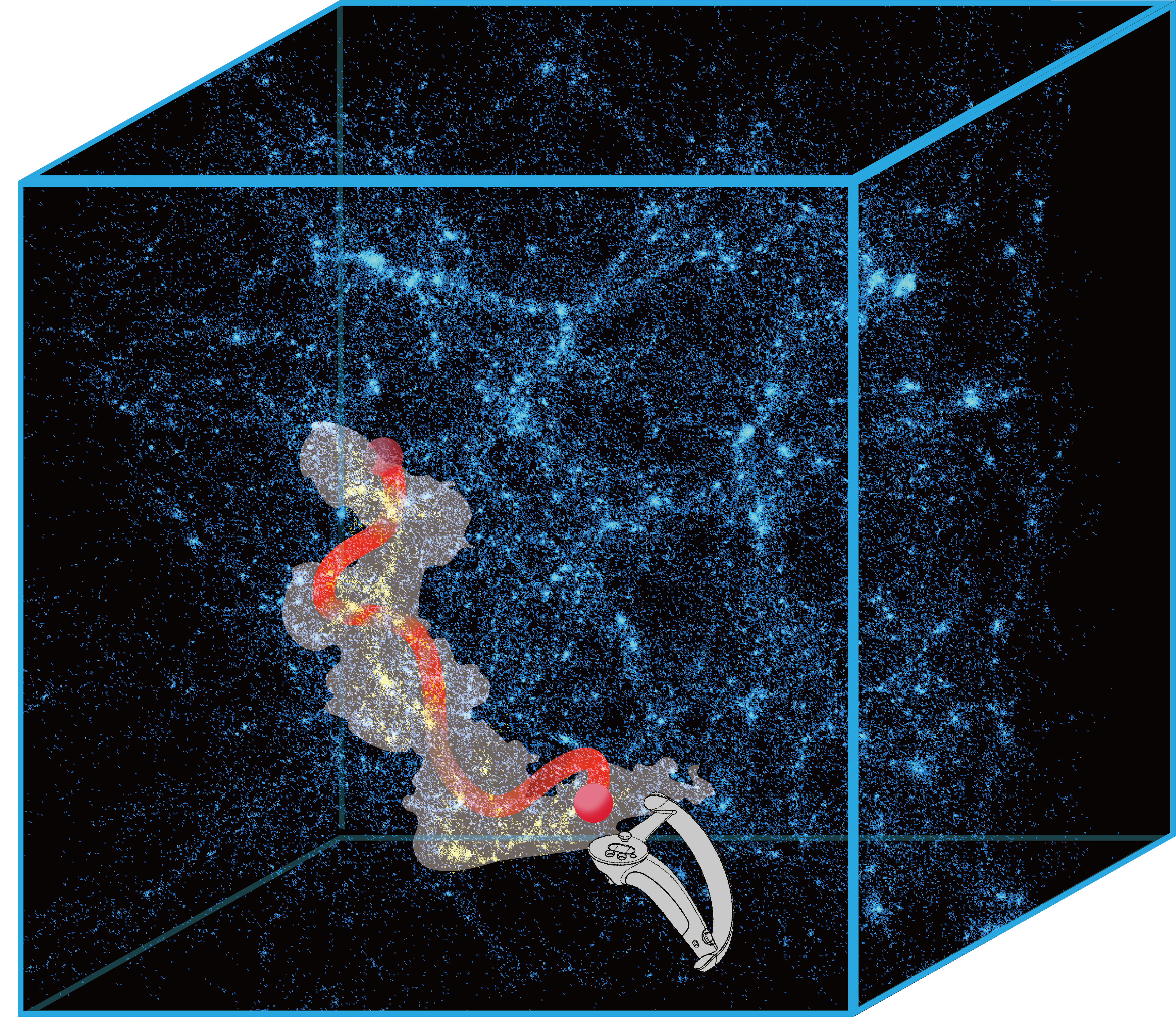}\hspace{5mm}
     \includegraphics[height=.212\textwidth]{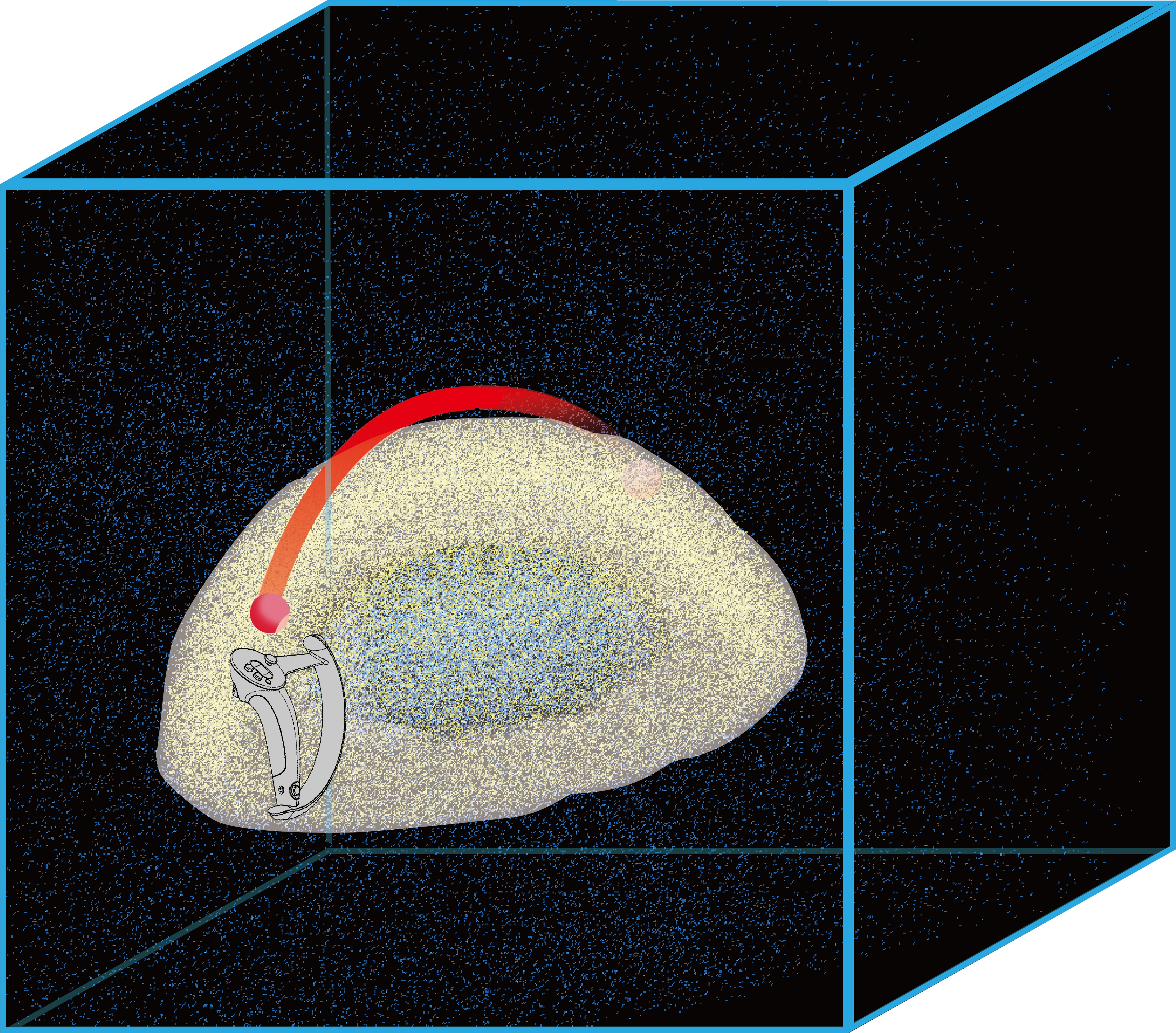}
 \caption{MeTACAST: a family of Target- and Context-aware spatial selection techniques for 3D point cloud data in VR.}
 \label{fig:teaser}
}

\begin{document}

\firstsection{Introduction}

\maketitle

\label{sec:Introduction}

Immersive environments experienced via virtual, augmented, and mixed reality (\ie, extended reality, XR) head-mounted displays (HMDs) are increasingly receiving attention from visualization researchers who aim to improve ways of understanding and exploring complex data. 
HMD-based immersive visualization is now an important research field with applications in natural sciences, in contexts that require users’ comprehension and exploration of three-dimensional spatial data.

Likewise and due to the rapid development of scientific simulation and computing technologies, the size, scale, and complexity of datasets in the natural sciences have grown exceptionally. This development substantially increases the challenges of spatial data exploration in immersive environments.
Astronomical datasets today, \eg, usually consist of multiple billions of spatial points.  
According to the cold dark matter paradigm \cite{Bond:1996:HFG,Jill:1994:TLAF}, cosmological simulations predict that a cosmic web is formed of the matter in the universe and that filaments transport matter to dense centers of clusters (formed by galaxies).
To comprehend such vast amounts of spatial data, researchers need to \emph{observe} the 3D space to gain an overview of the data, \emph{zoom} in and out of the data space to find a clear view of the structures, and \emph{select} and \emph{explore} important subsets of clusters such as those linked by cosmic filaments (Shnei\-der\-man's mantra \cite{Shneiderman:1996:ETD}). 
Dealing with multiple billions of data points, however, is a significant challenge in exploratory visualization on a 2D interface, for two reasons. First, a 2D planar surface cannot portray well the 3D shape or spatial positions of the points without constant animation. Second, it may be difficult for scientists to identify a specific target, define a precise location, and select particular parts with 2D input. 
These challenges motivate us to enhance the users' ability to explore massive 3D point cloud datasets in immersive contexts.

Data selection is a fundamental and prerequisite step of data exploration and has been explored for many years \cite{Wills:1996:S5W,besanccon:2021:state-of-art}. 
Spatial selection, by selecting sub-regions in point cloud data in 3D space, is often based on the locations, structures, and density distribution of the 3D points. 
The goal of our paper is to \emph{investigate effective techniques for selecting large-scale point cloud data in a VR environment}. 
In contrast to interactions on 2D monitor screens, immersive 3D spatial interactions are not limited by a static and planar screen for selecting objects---so interactive selection has the potential to be more expressive. In an immersive 3D space, however, users have not only the ability but also the need for six degrees of freedom for a single point of input, leading to substantial differences in their spatial interaction and selection strategies compared to those on 2D monitor screens. Nonetheless, we can take inspiration from 2D selection strategies, in particular those that use a structure- or context-aware approach \cite{Yu:2012:ESA,Yu:2016:CEE} as this technique allows us to limit the input complexity and adjust it to people's mental abilities and expectations. These strategies then have to be adapted, in particular, to unstructured point cloud data that contain vast amounts of detail distributed across multiple levels with unknown spatial distributions.

To understand these constraints and ultimately develop a suitable approach, we conducted a think-aloud user study to observe users' perception, interpretation, and actions when they were asked to select point cloud data in VR. 
The study was conducted with local university students, as we were specifically interested in how selection strategies might differ when users were asked to select parts of unknown and unstructured point cloud data.
Based on the findings, we developed \point, \brush, \paint, three context- and target-aware selection techniques for point cloud data rendered in an immersive VR environment. All three methods analyze the density distribution of the local area where users interact and as such, users can select the crucial features without precise input. 
\point selects the cluster that is closest to the location where the user is pointing, and then changes its selections as the user continues to provide input. 
\brush examines the user's input stroke and uses this information to infer their intention by identifying the primary filament that is closest to the path. 
Finally, \paint selects targets based on the user's perception of the structure and how well they can distinguish it from the surrounding context.
To achieve real-time feedback, we compute the density distribution and generate the selection volume on the GPU. Thus, immediate results are shown, and based on these, users can further adjust the density threshold. 
We then conducted a controlled experiment to test the efficiency, accuracy, and user preferences for our three selection techniques in an immersive environment, which demonstrated their effectiveness and suitability for different data scenarios. 

In summary, we make the following four main contributions:
\begin{itemize}[nosep]
\item a study on users' behavior in selecting unknown point cloud data in an immersive environment; 
\item a toolbox of three context-aware and target-aware selection techniques that allow users to precisely select regions of space they intend to further explore with simple and approximate actions;
\item algorithms to compute density fields and generate selection volumes on the GPU to facilitate fast exploratory visualization; and 
\item guidelines for designing and selecting appropriate 3D spatial selection techniques in diverse environments and contexts.
\end{itemize}
\vspace{-.5em}

 \section{Related Work}
 \label{sec:Related Work}
 
Several methods exist for selecting data points in visualizations, which differ based on the data type, user requirements, and the environment and platform used for data visualization. Our primary focus in this work is spatial selection, which entails defining a Region of Interest (ROI) in 3D space to specify a 3D subspace. We are specifically interested in selections in unstructured datasets that do not feature explicit objects (\eg, particle or volumetric data) based on their unique properties (\eg, density, spatial distribution, or given scalar fields). Below we review relevant work on spatial selection in general as well as on techniques specifically designed for immersive environments.
\vspace{-.3em}

\subsection{Spatial selection techniques}
Raycasting techniques are commonly used for selecting single objects in 3D space (in both immersive and projected settings) and are found in many applications \cite{Ferran:2009:EPS, Pierce:1997:IPI, Lee:2003:TICV, Wingrave:2005:ERS,Jonathan:2001:VOS}. The concept relies on 3D pointing: the user selects a specific pre-defined shape or object in 3D space by pointing at it, and a 3D ray is cast from the pointer to the object. The target object is identified by finding the first shape or object that intersects the ray. Adjusted techniques for selection in dense environments exist as well \cite{Hong:2021:DET}. 
Raycasting techniques can operate at a distance, are fast, and are easy to understand. Precise input, however, is critical for selecting the correct object, in particular when dealing with small objects that are possibly occluded.
To address this occlusion issue, Kopper \etal \cite{Kopper:2011:RapidAA} designed a progressive refinement technique to reduce the required precision of the selection task. Many 3D selection approaches have been proposed to provide precise input for selection tasks, such as 3D picking \cite{Tietjen:2008:METKT} and 3D lasso \cite{Keefe:2008:DDIT}. The use of raycasting can also facilitate the identification of a point of interest (POI) in spatial data, such as in scalar fields. Wiebel \etal \cite{Wiebel:2012:WYSIWYP}, \eg, devised WYSIWYP to analyze the scalar values along the selection ray to assist users in pinpointing a POI in medical volume data.

Raycasting approaches, however, are often not ideal for handling complex data distributions such as a 3D space with millions of particles. In such scenarios, selecting a single particle is often not meaningful because users are more interested in particle groups than in single data items. Selection of multiple targets or regions of interest (ROI) is often accomplished using various volume-based selection techniques that involve the use of pre-defined basic geometric shapes. These shapes can easily be manipulated in three-dimensional space, such as the use of cones to select multiple data points \cite{Grossman:2006:DES} or the use of cubes to select brain fibers \cite{Chen:2009:ANI}.
Nonetheless, these techniques' ability to select target objects far from a pre-defined shape is restricted. Therefore, freeform lasso methods are proposed to offer more flexible input to select objects or ROI in 3D space based on the 2D lasso or stroke. Lazy Selection \cite{xu:2012:lazy}, \eg, allows users to quickly select one or more desired shape elements with a novel scribble-based tool by roughly sketching through them. The Tablet Freehand Lasso (Cylinder Selection) \cite{Lucas:2005:DE3,Lucas:2005:DEM} and Volume Catcher \cite{Owada:2005:VC} allow users to flexibly select an ROI in 3D space based on a 2D lasso or a stroke. Besançon \etal \cite{Besancon:2019:TangibleBrush} proposed a hybrid approach that emphasizes the level of user control in defining the selection volume in 3D space. They introduced Tangible Brush, a technique that provides manual control over the final selection volume and which combines 2D touch lasso input with 6-DOF 3D tangible brushing to allow users to perform 3D selections in volumetric data. 

Compared to the use of pre-defined shapes, freeform lasso selection offers greater adaptability and flexibility, in particular when it comes to manual control over the final selection volume in 3D space. This approach allows users to customize their ROIs based on the local data characteristics---which is why we also target a freeform technique. These fully manual techniques, however, may not be efficient for complex selection tasks such as identifying small, dense, or complex clusters in noisy backgrounds that may be difficult to visually recognize. 
Several approaches have been proposed to address this limitation. Chen \etal proposed LassoNet \cite{Chen:2020:LassoNet}, a deep learning-based approach to lasso selection for 3D point clouds. It aims to learn a latent mapping from viewpoint and lasso to point cloud regions, enabling users to select particles that best match their selection intention. Structure- and context-aware selection techniques \cite{Yu:2012:ESA, Yu:2016:CEE} have also been designed to take data features such as the local density into account. With these techniques, users draw a 2D lasso around the target, and only clusters with densities above a certain threshold are selected. While context-aware approaches have been effective, they are not designed for immersive environments---which we target with our paper---, where users do not have a dedicated 2D surface for drawing the lasso.
\vspace{-.5em}

\subsection{Spatial selection in immersive environments}
In immersive visualization settings, researchers can gain a clearer understanding of the data, not only through interaction and exploration but also simply through the stereoscopic projection.
Studies have also demonstrated \cite{Kraus:2020:TII} that users can perform better in cluster identification tasks when using immersive scatterplot visualizations compared to traditional 3D-to-2D projected representations.

Also in immersive VR, raycasting is a common technique to help determine where a user is looking or pointing at. It is frequently used for various interactions such as object selection, manipulation, and action triggering. Hurter \etal \cite{Christophe:2018:FCS} introduced FiberClay, a raycasting-based approach to brush paths using two VR controllers to select paths that connect both targeted points.  
Selecting multiple objects in VR, however, can be challenging and imprecise, especially when the objects are small, moving, occluded, or complex.
To improve interaction precision, researchers have proposed various approaches. De Haan \etal \cite{DeHaan:2005:IntenSelect} developed IntenSelect to aid users in selecting objects in motion within cluttered and occluded VR environments with a scoring function. Baloup \etal \cite{baloup:2019:raycursor} introduced a controllable cursor positioned on top of a ray, allowing users to select nearby or occluded objects.
Maslych \etal \cite{Maslych:2023:TIA} showed how to select fully occluded objects in dense environments. Their two approaches maintain the spatial relations between minimized versions of the original objects by displaying the objects that collide with a cone-shaped selection volume in a flat or cylindrical mini-map.
Wei \etal \cite{Wei:2023:PGT} recently proposed a user intention model for gaze-based selection techniques based on eye and head endpoint distributions to estimate the intended targets. Adaptive pointing \cite{konig:2009:APD} also enhances the accuracy of pointing devices. Finally, Stenholt \cite{Stenholt:2012:SMOS} designed a magic wand, a novel multiple-object selection technique based on local proximity, enabling users to select hundred or thousands of objects. We also aim for precise selection, yet one in which the input imprecision is mediated by context-aware selection processing as well as one in which the user can easily adjust the selection after the initial specification---like in the projected 3D settings we reviewed above.

Compared to interaction on a 2D surface, selecting within immersive VR is conceptually more straightforward and is more preferred by users than using traditional 2D monitors\cite{Franzluebbers:2022:VRP}, as users can view the data in stereoscopic 3D and have manual control over the final selection.
To facilitate 3D input, various techniques have been developed, such as user-defined box virtual tool~\cite{Teylingen:1997:VDV, Bernd:2009:MVT} and cone-based selection volumes~\cite{Nicholas:2016:SPP}. Spatio-Data Coordination \cite{Cordeil:2017:DSS} and the Embodied Axes \cite{Cordeil:2020:EmbodiedAxes} proposed by Cordeil \etal enable users to accurately determine the selection region of the spatial data using sliders on the tangible axes.
These techniques rely on predefined 3D selection volumes such as boxes, cones, and cylinders to select multiple objects within the space.
Several approaches have also been proposed to select within complex datasets using freeform methods.
Lubos \etal \cite{Lubos:2014:TBP} introduced a bi-manual user interface for selecting point cloud data. Their approach involves tracking the users' hands and enabling them to touch and select a 3D point cloud in an immersive environment.
Similarly, Gemoz \etal \cite{gomez:2010:FBT} proposed a contacting method with a spherical input to select neural fiber tracts.
In addition, various hybrid approaches have been proposed to improve selection accuracy by using hybrid approaches. They leverage different technologies and techniques to make the selection process more efficient and intuitive for users.
Sereno \etal \cite{sereno:2022:HTTinAR}, \eg, integrated an AR device with a tablet that acts as a tangible device for 3D selection.
Montano-Murillo \etal \cite{Roberto:2020:HMS} proposed a hybrid VR selection technique that allows users to place a selection volume at a specified position in the VR environment. The volume is initially attached to a virtual tablet and can be adjusted for size and thickness to better fit the target object.
Researchers have also focused on incorporating data features or geometric features into the 3D selection process. 
For example, Malmberg \etal \cite{malmberg:2006:3DLW} used surface generation based on a 3D live-Wire drawn by the user to segment and select within volumetric images. This approach enables users to select objects based on their volumetric value as well as geometric properties.
Torin \etal \cite{mcdonald:2020:IUV} introduced a semi-automatic neuron tracing method based on the Morse-Smale complex in an immersive environment. This approach enables users to select the desired region even with inaccurate trace input, reducing fatigue during the exploratory task. 
Jackson \etal \cite{jackson:2013:LTI}, finally, proposed a tangible 3D interface that is capable to select the fibers that own the same orientation as the tangible prop.

All these spatial selection techniques for immersive environments can select single objects or regions based on the data's location and/or distribution. They employ tangible and hybrid interaction techniques to improve the selection accuracy for complex data. In our work, we aim for selection directly within the HMD-based VR environment, without any additional required hardware, and for an interaction design that is as flexible as the traditional context-aware techniques, but without any need for a planar interaction surface. We work toward this goal by studying users' strategies in various situations to be able to incorporate data features and user intention into the selection process.
 \vspace{-0.5em}

\section{Think-aloud Elicitation Study}
\label{sec:Think-aloud Study}

To gain a deeper understanding of how users adopt selection strategies within VR in various contexts and for various targets, we began by conducting a think-aloud study with six participants from our local \textls[-10]{university. Here we present our findings and discuss the resulting design} considerations related to selection context, target, action, and strategy.

\subsection{Study design}
Various factors such as the size of the visualization, the interaction environment, the characteristics of the data (\eg, context, data distribution, density), and input precision requirements can influence the selection strategies employed by users. To design effective selection techniques we need to understand the users' selection strategies in VR environments as they deal with these situations. In a think-aloud elicitation study we thus, instead of teaching them any specific selection methods, showed participants datasets and selection targets and first asked them to freely explore the 3D point cloud data. We then let them demonstrate their selection strategies to us with a 3D drawing tool and asked them to explain why they thought it was the most appropriate and direct way to select crucial features or regions. We recorded all user actions and input trajectories for further analysis. Both studies (think-aloud, user study) received approval from the University Ethics Committee.

The \textbf{size of the visualization}
is a crucial factor for the users' selection strategies, especially in VR. In the physical world, perspective distortion causes objects located closer to the human eye to appear larger than their actual size. This effect is even more obvious when viewing point clouds in VR due to the narrower field of view. As a result, gaps between 3D points that are closer to the user appear larger than those between points further away. This effect can make it challenging for users to accurately estimate the density distribution and, thus, to identify a dense cluster when they are ``within'' it. Kraus \etal~\cite{Kraus:2020:TII} found that cluster identification task performance differs between different visualizations (monitor screen for 2D/3D scatter plot, 
table-size VR, and room-size VR) in terms of accuracy, efficiency, memorability, sense of orientation, and user preference. To answer the question about whether users could find the entire cluster and further investigate how visualization size would impact the users' selection strategies, following the approach by Danyluk \etal \cite{danyluk:2020:TBC} we investigated three different sizes (\autoref{fig:visualizationsize} in \autoref{appendix-think-aloud}): hand-size, table-size, and room-size:
\begin{description}[nosep,leftmargin=1.5em,labelindent=0em,leftmargin=!,labelindent=!,itemindent=!,font=\normalfont\itshape]
\item[Hand-size:] We adopted a hand-size scale of 30\,cm for the data vi\-sua\-li\-za\-tion---a suitable size to observe the detailed structures clearly in most cases. The data is controlled by the left and, its center is attached to the left controller with a 35\,cm offset to prevent the left hand from obstructing interactions with the right hand in the data.
\item[Table-size:] We placed the data statically 1\,m above the floor with a base length of 64\,cm; users can walk around and reach inside.
\item[Room-size:] In this setting we also do not allow users to manipulate the data placement and for the selection, they have to rely on the picking/raycasting metaphor to reach far-away data locations. 
\end{description}

We used four distinct \textbf{datasets} (\autoref{fig:ThinkAloudData} in \autoref{appendix-think-aloud}). We rendered the target particles in \textcolor{Goldenrod}{yellow}, while we presented the interfering particles in \textcolor{DodgerBlue}{blue}. We selected each dataset based on its unique features: 
\begin{description}[nosep,leftmargin=1.5em,labelindent=0em,leftmargin=!,labelindent=!,itemindent=!,font=\normalfont\itshape]
\item[Clusters:] five uniformly dense ellipsoidal clusters in a noisy setting. 
\item[N-body simulation:] a cosmological N-body simulation that comprises a huge, extremely dense cluster at the center, which is surrounded by multiple smaller clusters (from \cite{springel:2008:aquarius}).
\item[Filament:] a cosmic web simulation with thin filaments (from \cite{Springel:2005:SimulationsOT}).
\item[Complex geometries:] an empty half-box outside, a half-ball, and a cuboid of particles inside.
\end{description}

\subsection{Findings}
\label{sec:think-aloud:findings}
\textbf{Preferred visualization size.} All participants preferred hand-size and table-size visualization settings that allowed them to select data directly. Moreover, these settings also facilitated easier recognition of point cloud clusters and their boundaries by the participants. Thus, in the following, we focus on analyzing users' actions and selection strategies with hand- and table-size environments.

\textbf{Target: Dedicated cluster.} When we asked participants to select clusters that are visually clearly separable from the background
(\eg, a separate ball-shape cluster or small, isolated clusters),
most pointed at the target clusters directly and hoped that the target clusters could be selected with minimal actions. This observation aligns well with traditional 3D selection, which relies on picking and raycasting. 

\textbf{Target: Regular, simple shape.} When the shape of the target particles is simple,
many participants attempted to draw lassos in the air that followed the geometric features of the target, while a minority also mentioned that pointing was also possibly applicable if the selection range could be determined in some way.

\textbf{Target: Filament-like structure.} In this case
some participants brushed the points following the dominant branch, while others tried to separate the target points from the noisy context through a helix in 3D space. Both strategies seem to be valid approaches.

\textbf{Context: Unclear boundary.} When participants were unable to clearly distinguish the boundary of clusters (\eg, many small clusters surrounded by particle noise), they attempted to enclose the target cluster(s) by drawing a lasso around them as they would on a monitor screen. Most of them realized, however, that drawing a 1D linear lasso to enclose points in 3D space is insufficient. So they expressed the desire for a more intuitive way to define a 3D region around the input location that would allow them to include the target points. 

\textbf{Context: Occlusion.} We found that, for easy shapes,
participants were able to estimate the 3D location and shape of target clusters even when the particles were fully or partially occluded. In the \emph{complex geometries} dataset,
\eg, participants placed a contact point (the interacting point in 3D) on the opposite side of the target without the need to rotate the data or selected occluded geometries by placing the contact point inside directly. This ability may allow us to design an effective 3D interaction for clear target clusters.

\textbf{Accuracy.} In comparison to screen-based selection, in VR participants were generally able to \emph{mark} the targets \emph{closer} to their actual locations. We observed, however, that participants often had difficulties accurately clicking on, for instance, the exact position of an intended cluster. When attempting to select data by drawing lines directly in VR using controllers (6DOF input), participants also often failed to align their input precisely with target filaments. Even the input lasso looked as if it would precisely follow the filament. After rotating the data to observe it from another viewpoint, participants saw that the input lasso was at a distance from the target due to VR's imprecise depth perception---an issue that was also noted in related work \cite{mcdonald:2020:IUV}. 

\subsection{Design considerations}
\label{sec:consideration}
Based on these findings, we propose the following design considerations and requirements for the design of selection techniques in VR. 
\begin{itemize}[nosep,leftmargin=10pt]
    \item \textit{Target-aware selection.} The techniques should be capable of handling various target shapes, such as clusters, filaments, partially occluded structures, and structures intertwined with others.
    \item \textit{Context-aware selection.} The techniques should be capable of handling various challenging situations, including non-uniform feature density, unclear boundaries, and occlusion. Users should not be limited or hindered by these interfering factors but, instead, should be able to focus on their selection tasks, including identifying the shape, location, and critical features of the data. 
    \item \textit{Accurate and intuitive selection.} The techniques should minimize a user's need to move and, instead, infer their selection intention from the approximate location and path of the input, precisely adjusting the selection to crucial data features such as geometric shape and region density. Users should be able to use simple, natural, and approximate input to select intended regions precisely in VR.
    \item \textit{Exploratory selection and immediate results.} When dealing with a complex dataset that contains a large number of noisy points and unclear boundaries, it is crucial for users to visually examine the selection results and receive them without delay. This is important, \eg, in astronomical data exploration where numerous unknown features are concealed within the data and noise is prevalent.
    \item \textit{Partial selection and threshold adjustment.} Enabling post-selection threshold adjustment is important for our selection tool to be flexible in handling various situations such as selecting complex data shapes, partial regions, and non-homogeneous density. 
\end{itemize}

\section{Selection Techniques}
 \label{sec:Selection Techniques}
Based on these design considerations we developed the MeTACAST family for selection in HMD-based VR environments, which consists of three spatial techniques, each tailored to a specific selection intent. 

\subsection{Interaction metaphor}
Our think-aloud study showed that participants preferred visualization tools that rely on direct interaction with data. We thus adopted the Worlds in Miniature (WIM) metaphor \cite{Stoakley:1995:WIM, Pivovar:2022:VRS, Zhao:2022:LWIM} and developed a hand-held miniature of the 3D point cloud data to facilitate data selection in a virtual environment. We attached the WIM to the VR controller of the non-dominant hand, providing users with an exocentric view of the 3D data and facilitating correct density comprehension of the point clouds. To select points, the user can use their dominant hand to control a \textcolor{FireBrick}{red} sphere. We positioned the sphere $1$ cm above the top of the VR controller to prevent collisions between two VR controllers. This \textcolor{FireBrick}{red} sphere follows the movement of the controller, allowing users to have full control in 3D space to point, brush, or paint within the data.

\newlength{\lineskiplimitbackup}%
\setlength{\lineskiplimitbackup}{\lineskiplimit}%
\setlength{\lineskiplimit}{-\maxdimen}%
Similar to CloudLasso~\cite{Yu:2012:ESA} and CAST~\cite{Yu:2016:CEE}, instead of relying on particle positions we use a continuous density field $\rho(\mathbf{r})$ that represents the particle density at position $\mathbf{r}$ in space. We compute the field at all nodes $i$ of a regular 3D grid (box $B$) that covers the dataset as $\rho(\mathbf{r}^{(i)})$. We then derive the value $\rho(\mathbf{r})$ of the field at any other point $\mathbf{r}$ in space through linear interpolation from the values $\rho(\mathbf{r}^{(i)})$ at the grid-nodes that are closest to $\mathbf{r}$. This approach enables us to apply our method not only to particles but also to volumetric data that samples a scalar field representing any visually salient data aspect, rather than just density.

\setlength{\lineskiplimit}{\lineskiplimitbackup}

Based on this field concept we can now design interaction techniques that are adjusted to our VR environments. In contrast to projection-based approaches such as CloudLasso and CAST, we need to allow users to directly point, brush, or draw on the dataset in the 3D virtual space according to their preferred strategies. We aim to meet the design goals we formulated in \autoref{sec:consideration} and developed three techniques---MeTAPoint, MeTABrush, and MeTAPaint---that we describe next.

 \subsection{\point}
 \label{subsec:MetaPoint}

\begin{figure}[t]
    \centering
    \captionsetup[subfigure]{labelformat=empty} 
    \subfloat
        {\begin{overpic}[width=\columnwidth]{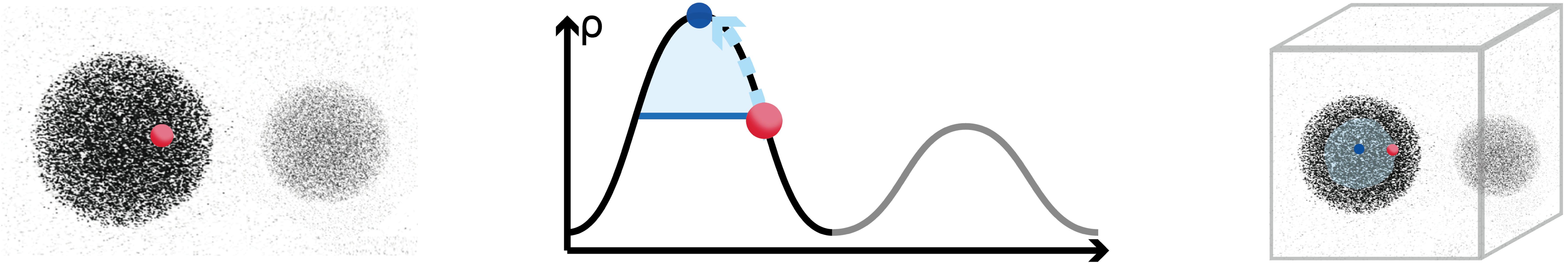}
            \put(1,2.5){\textcolor{black}{(a)}}
             \put(31,2.5){\textcolor{black}{(b)}}
              \put(76,2.5){\textcolor{black}{(c)}}
        \end{overpic}} \hspace{1 pt}
    \subfloat
        {\begin{overpic}[width=\columnwidth]{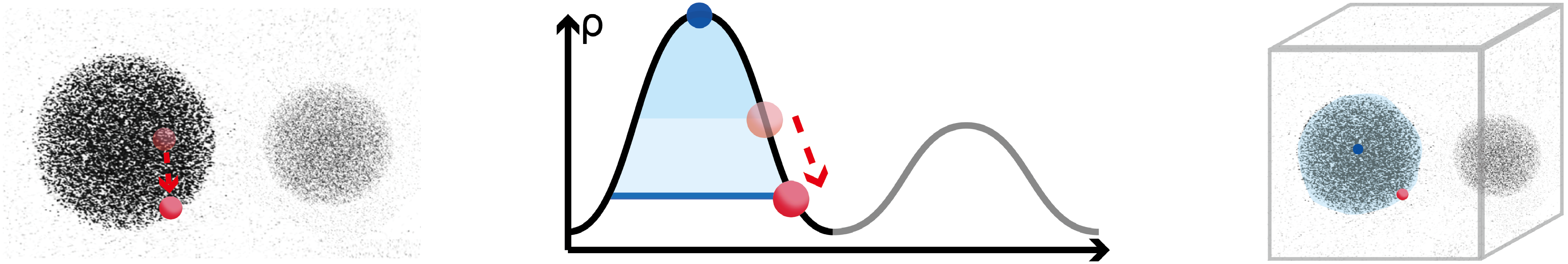}
             \put(1,0.7){\textcolor{black}{(d)}}
             \put(31,0.7){\textcolor{black}{(e)}}
              \put(76,0.7){\textcolor{black}{(f)}}
        \end{overpic}}
 \caption{MeTAPoint: (a) the user points at the target cluster (\textcolor{FireBrick}{red}); (b) we derive the closest maximum point (\textcolor{DarkSlateBlue}{blue}) and density threshold (schematic representation); (c) we compute the selection volume; (d) the user drags the controller to adjust the density threshold; (e, f) we recompute the density threshold and selection volume.}
 \label{fig:Metapoint}
\end{figure}

\setlength{\lineskiplimitbackup}{\lineskiplimit}%
\setlength{\lineskiplimit}{-\maxdimen}%
The problems in VR with general free-hand input imprecision and placement with respect to a 3D structure (\autoref{sec:think-aloud:findings}) lead to users wanting to test their initial selection results through interactive exploration to better understand the data distribution. Users 
attempted to draw a stroke on the target to explore the boundary of the target cluster. To address this need, we propose our first selection technique \point.
With it, the user begins by pointing at the target cluster, or near it, and the initial selection is based on this input position. To account for imprecision, we attempt to determine the intended target by analyzing the input position. Specifically, we follow the direction of the gradient from the input position $\mathbf{r}^{(s)}$ (\textcolor{FireBrick}{red} point in \autoref{fig:Metapoint}(a)) to find the local maximum $\mathbf{r}^{(m)}$ of the scalar field (\textcolor{DarkSlateBlue}{blue} point in \autoref{fig:Metapoint}(b)). We use this input adjustment in all MeTACAST techniques and discuss technical details in \autoref{sec:MaxPointExtraction}.

Next, we determine the selection volume $V$ that surrounds the local maximum $\mathbf{r}^{(m)}$, based on the density field. To enable users to explore the edges of the target cluster, we derive the density threshold based on the density at the initial input position $\rho(\mathbf{r}^{(s)})$. Using the Marching Cubes algorithm \cite{Wyvill:1986:DSS,Lorensen:1987:MCH} we identify all volumes where the density $\rho$ is above $\rho(\mathbf{r}^{(s)})$ in the whole space, and we pick the volume that encloses $\mathbf{r}^{(m)}$ as the initial selection volume $V$ (blue area in \autoref{fig:Metapoint}(b, c)).

The user can continue to explore the target boundary by dragging the controller with the trigger pressed. As they drag the \textcolor{FireBrick}{red} point, we adjust the density threshold accordingly (\autoref{fig:Metapoint}(d, e)) and compute a new selection volume (\autoref{fig:Metapoint}(f)). It is important to note that we designed \point to select the single volume that is closest to $\mathbf{r}^{(s)}$. Therefore, if the user drags the \textcolor{FireBrick}{red} point close to another cluster, the selected target may switch. In this way, we provide the user with an immersive experience of testing the target volume and selection boundary through continuous interaction in the 3D environment.

\setlength{\lineskiplimit}{\lineskiplimitbackup}

To enhance exploratory visualization and accelerate iso-surface computation we leveraged a GPU-based, parallel implementation of Marching Cubes. We assign a dedicated thread to each voxel, resulting in smooth threshold modification and efficient triangle generation.

\subsection{\brush}
\label{subsec:MetaBrush}
A potential issue for \point is that it requires users to carefully adjust the contact point---the input position has a direct impact on the selection result, including the target and threshold. Consequently, it may be challenging for \point to select a part of a filament in a complex dataset, which is a common task, \eg, in astronomical exploration. In our elicitation study, we observed that users usually want to brush the points following the dominant branch or create a helix in 3D space to separate the target points from the noisy context (\autoref{fig:ThinkAloudData}(c)). While their input may not be precise, their selection intent is clear from the input. To address this need, we propose our second selection technique, \brush, which infers users' intention using the entire input path, adjusts the position of the input stroke, and then provides a precise selection---without being limited by imprecise input, non-homogeneous features, or complex structures.

\begin{figure}[t]
\centering
    \captionsetup[subfigure]{labelformat=empty} 
    \subfloat
        {\begin{overpic}[width=\columnwidth]{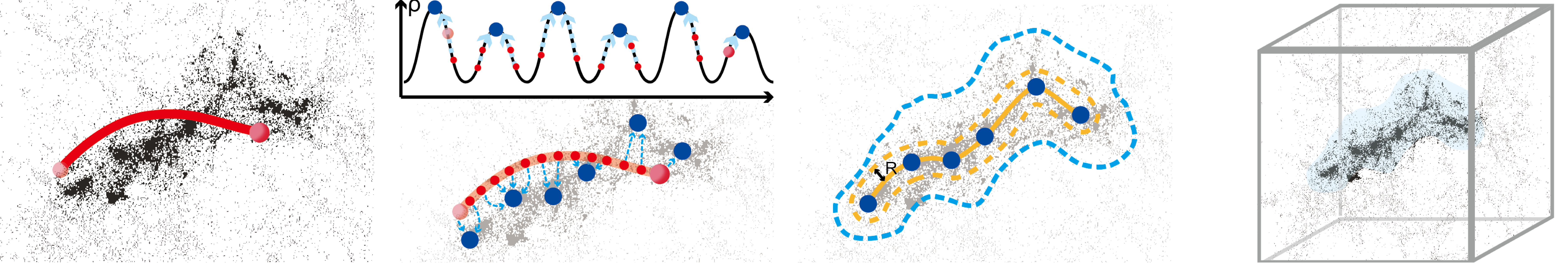}
               \put(1,0.7){\textcolor{black}{(a)}}
              \put(25,0.7){\textcolor{black}{(b)}}
             \put(50,0.7){\textcolor{black}{(c)}}
              \put(73,0.7){\textcolor{black}{(d)}}
        \end{overpic}} 
    \caption{\brush: (a) the user draws a 3D stroke (\textcolor{FireBrick}{red}); (b) the stroke points (\textcolor{FireBrick}{red}) are extracted; by following the direction of gradient (\textcolor{DodgerBlue}{blue} arrow), we identify the local maximum points (\textcolor{DarkSlateBlue}{blue}); (c) we construct a tunnel-like volume (\textcolor{Goldenrod}{yellow} dotted region) based on the MaxLine with the radius $R$; $V_{\mathrm{init}}$ is derived (\textcolor{DodgerBlue}{blue} dotted line); (d) the final selection.}   
    \vspace{-0.5ex}
    \label{fig:MeTABrush}
\end{figure}

\setlength{\lineskiplimitbackup}{\lineskiplimit}%
\setlength{\lineskiplimit}{-\maxdimen}%
With \brush, the user can brush a stroke along the target filament particles in 3D space (\autoref{fig:MeTABrush}(a)). We then take a sample of points $\mathbf{r}^{(s)}=\{\mathbf{r}^{(s_0)}, \mathbf{r}^{(s_1)}, ... \mathbf{r}^{(s_n)}\}$ on the input stroke and follow the direction of gradient from each point (\autoref{fig:MeTABrush}(b)) like before. This process yields a set of destinations, that is, local maximum points $\mathbf{r}^{(m)}=\{\mathbf{r}^{(m_0)},\mathbf{r}^{(m_1)},...,\mathbf{r}^{(m_t)}\}$ of the density field; note that $n$ may not equal $t$. Next, we connect successive pairs of destinations $\mathbf{r}^{(m_i)}$ and $\mathbf{r}^{(m_{i+1})}$ to obtain the path $P$ (indicated as MaxLine below) as follows:\vspace{-1ex}

\setlength{\lineskiplimit}{\lineskiplimitbackup}
\begin{equation}
\begin{aligned}
   & P:x \in \mathbb{R}^{+} \mapsto \mathbb{R}^{3}, P(0)=\mathbf{r}^{(m_i)},P(end)=\mathbf{r}^{(m_{i+1})},\\
   & P^\prime(x)=\frac{\nabla f\left(P(x)\right)}{\|\nabla f\left(P(x)\right)\|}+2\frac{(P(m_{i+1})-P(x))}{\|(P(m_{i+1})-P(x))\|},
\end{aligned}
\end{equation}\vspace{.001pt}

\setlength{\lineskiplimitbackup}{\lineskiplimit}%
\setlength{\lineskiplimit}{-\maxdimen}%
\noindent With this process we connect all local maximum points following roughly the direction of the gradient, while also smoothing the MaxLine $P$.
Based on $P$, we construct a tunnel-like shape $T$ (region circled by the yellow dotted line in \autoref{fig:MeTABrush}(c)) that extends along the MaxLine with a pre-defined radius of $R$ (represented by the size of the input marker, which can be adjusted by the user later). We then identify all particles whose destinations, following the direction of the gradient, fall within $T$, using the algorithm we detail in \autoref{sec:MaxPointExtraction}.
We thus identify the set of the potential target particles $J$, which we indicate by the \textcolor{DodgerBlue}{blue} dotted line in \autoref{fig:MeTABrush}(c).
Next, for each potential target particle $j\in J$, we consider the ellipsoid with semi-axes $\ell_{x}^{(j)}$\hspace{-.75ex}, $\ell_{y}^{(j)}$\hspace{-.75ex}, and $\ell_{z}^{(j)}$, which are the smoothing lengths along $x,y,z$ of the $j$\textsuperscript{th} particle (\autoref{appendix-densityestimation} has more details on smoothing length and density estimation).
We further combine these ellipsoids into an initial volume of interest $V_{\mathrm{init}}$ in the data box $B$ that covers the whole data:\vspace{-1ex}

\setlength{\lineskiplimit}{\lineskiplimitbackup}

\begin{equation}
    V_{\mathrm{init}}=\{\mathbf{r} \,\vert\, \mathbf{r} \in B,\,  \exists j \in J,\,  \|\mathbf{r}^{(r;j)}\|\le 1\},
\end{equation}\vspace{.001pt}
where $\mathbf{r}_{k}^{(r;j)}=(\mathbf{r}_{k}-\mathbf{r}_{k}^{(j)}) / \ell_{k}^{(j)}$ along the $k$\textsuperscript{th} direction ($k=x,y,z$), and $\|\mathbf{r}\|$ denotes the Euclidean norm of the vector $\mathbf{r}$. 

\setlength{\lineskiplimitbackup}{\lineskiplimit}%
\setlength{\lineskiplimit}{-\maxdimen}%
Next, we calculate the initial density threshold $\rho_0$ by as the arithmetic mean of the density of all the grid-nodes inside $V_{\mathrm{init}}$. 
Through Marching Cubes we obtain the iso-surface inside $V_{\mathrm{init}}$. 
We then obtain the volumes inside of the iso-surface and select the volumes that contain a segment of the MaxLine, which means that the selected volumes should contain at least one local maximum point $\mathbf{r}^{(m_i)}$ on the MaxLine. 
We regard these volumes as the initial selection volumes $V$\hspace{-2pt}. 
In addition, users are able to adjust the density threshold post-selection via the VR controller. 
They can modify the threshold in a range of $[\rho_0/16,16\rho_0]$, mapping with the function $\rho_s=2^s\rho_0$ with $s\in[-4,4]$. 
When $s$ is adjusted, we thus recompute the scalar quantity for all grid-nodes inside $V_{\mathrm{init}}$ with $\rho_0$ replaced by $\rho_s$ and obtain the iso-surface using Marching Cubes.

\setlength{\lineskiplimit}{\lineskiplimitbackup}

\subsection{\paint}
 \label{subsec:MeTaPaint}

In our elicitation study, we observed users trying to select regular points shapes. 
These strategies allowed users to depict the geometric features of their target accurately. The provided input, however, was inherently imprecise.
 To address this issue, we thus introduce \paint, a technique that interprets the 3D path of a drawn selection stroke and selects the candidate cluster that best fits the drawn stroke.
 
\begin{figure}[t]
\centering
  \captionsetup[subfigure]{labelformat=empty} 
    \subfloat
        {\begin{overpic}[width=\columnwidth]{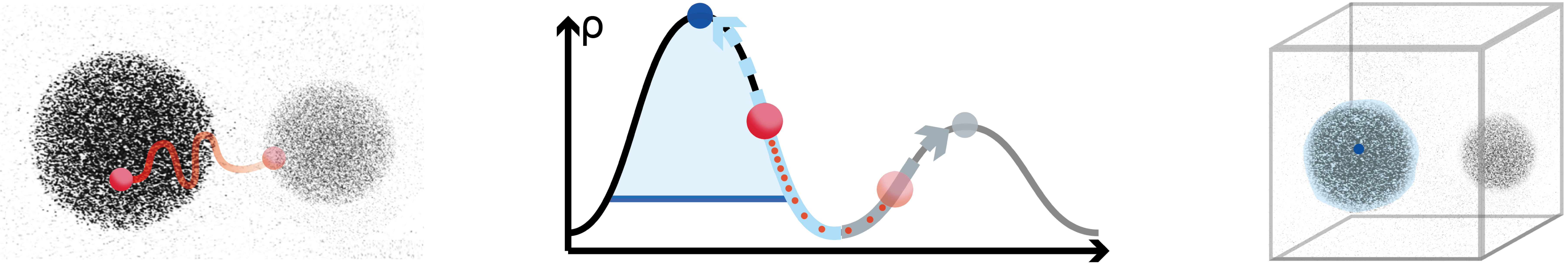}
            \put(1,0.7){\textcolor{black}{(a)}}
             \put(31,0.7){\textcolor{black}{(b)}}
              \put(76,0.7){\textcolor{black}{(c)}}
        \end{overpic}}
				
   \caption{\paint: (a) the user draws a 3D stroke (\textcolor{FireBrick}{\textcolor{FireBrick}{red}}) near the black cluster, some parts of the input are located near the gray one; (b) the stroke is split into multiple points (\textcolor{FireBrick}{red}), which flow towards local density maxima (\textcolor{DarkSlateBlue}{blue} and \textcolor{gray}{gray} points) along the direction of the gradient; (c) the black cluster is selected since it receives most seeds.}
    \vspace{-0.5ex} 
   \label{fig:MeTAPaint}
\end{figure}

\setlength{\lineskiplimitbackup}{\lineskiplimit}%
\setlength{\lineskiplimit}{-\maxdimen}%
 With \paint, the user brushes a target cluster with a 3D stroke (\autoref{fig:MeTAPaint}(a)). We first identify the initial density threshold $\rho_0$ by sampling the density of the grid-nodes surrounding the input stroke. We define a tunnel-like region $T$ that extends along the input stroke with a radius of $(\ell_{x}+\ell_{y}+\ell_{z})/3$, where $\ell_x,\ell_y,\ell_z$ are the smoothing lengths along $x,y,z$, respectively (\cite{Yu:2012:ESA,Yu:2016:CEE}). We then calculate the initial threshold $\rho_0$ as the arithmetic mean of the grid-nodes' density $\mathbf{r}^{(n)}$ within $T$, given by\vspace{-1ex}

\setlength{\lineskiplimit}{\lineskiplimitbackup}
\begin{equation}
    \rho_{0}=\frac{1}{N_{T}} \sum_{n=1}^{N_{T}} \rho(\mathbf{r}^{(n)}),
\end{equation}\vspace{.001pt}

\setlength{\lineskiplimitbackup}{\lineskiplimit}%
\setlength{\lineskiplimit}{-\maxdimen}%
\noindent where ${N_{T}}$ is the number of the nodes in $T$. 
Then we take a sample of points $\{\mathbf{r}^{(s_0)}, \mathbf{r}^{(s_1)}, ... \mathbf{r}^{(s_n)}\}$ on the input stroke and follow the direction of gradient from each point to obtain the destinations $\{\mathbf{r}^{(m_0)},\mathbf{r}^{(m_1)},...,\mathbf{r}^{(m_t)}\}$ (\autoref{fig:MeTAPaint}(b)), as we did in \brush.
We then count the number of $\mathbf{r}^{(s_i)}$ that contribute to each $\mathbf{r}^{(m_j)}$, and find $\mathbf{r}^{(m_{\mathrm{max}})}$, that is, the destination for the most stroke points $\mathbf{r}^{(s_i)}$. 
Using Marching Cubes, we obtain the iso-surface with density $\rho_0$, and the corresponding enclosed volumes.
Finally, we regard the volume containing the destination $\mathbf{r}^{(m_{max})}$ as the initial selection volume $V$ (\autoref{fig:MeTAPaint}(c)).
The post-selection threshold adjustment follows the same procedures as we described for \brush. 
Note that \paint selects the cluster that is most heavily brushed by the input stroke, meaning that users only need to brush on or around the geometric features of their target, and the intended single cluster gets selected.

\setlength{\lineskiplimit}{\lineskiplimitbackup}

\subsection{Local maximum point extraction}
 \label{sec:MaxPointExtraction}

To increase input precision, in all MeTACAST methods, we adjust the positions of the input point/stroke by extracting local maxima.
For a smooth density field $\rho:\mathbb{R}^3\to\mathbb{R}$,
a point $\mathbf{r}$ is a local maximum of $\rho$, if 
\begin{equation}
    \nabla \rho(\mathbf{r})=0 \text{ and } \lambda_{1} < 0,
\end{equation} 

\setlength{\lineskiplimitbackup}{\lineskiplimit}%
\setlength{\lineskiplimit}{-\maxdimen}%
\noindent where $\lambda_{1}$ is the largest among all the eigenvalues $\lambda_{1} \ge \lambda_{2} \ge \dots \ge \lambda_{n}$ of the Hessian of $\rho$ at $\mathbf{r}$. 
To follow the user input point $\mathbf{r}^{(s)}$ along the direction of the gradient of $\rho$ and find the local maximum point $\mathbf{r}^{(m)}$ for the density field, we define the path $P$ as

\setlength{\lineskiplimit}{\lineskiplimitbackup}
\begin{equation}
    P: t \in \mathbb{R}^{+} \mapsto \mathbb{R}^{3}, \quad P(0)=\mathbf{r}^{(s)}, \quad P^{\prime}(t)=\nabla \rho(P(t)) .
\end{equation}

\setlength{\lineskiplimitbackup}{\lineskiplimit}%
\setlength{\lineskiplimit}{-\maxdimen}%
\noindent We take $\mathbf{r}^{(m)}$ as the destination of the path $P$, that is, $\mathbf{r}^{(m)} = \mathrm{dest}(P) = \lim _{t \to \infty} P(t)$. 
According to Morse theory \cite{Martin:2001:morse}, for a smooth function with a non-degenerate Hessian matrix, the path $P$ converges to a local maximum except in the case when $\mathbf{r}^{(s)}$ is located on the stable manifold of a saddle or is a minimum of the density field. 

Given a position $P(t)$ along the path $P$, we compute $P(t+\delta t)$ along the direction of $\nabla \rho(P(t))$, which is calculated by linear interpolation from the values of $\rho$ at the closest grid-nodes to $P(t)$, obtained in \autoref{appendix-densityestimation}. Finally, we obtain the local maximum point $\mathbf{r}^{m}=\mathrm{dest}(P)$ at the end of the path $P$. 
In the case where the user's input is a stroke, we compute the destination points for all the sample points on the stroke with a parallel algorithm to increase computation speed. We report on details on the system's performance in \autoref{tab:system performance} in \autoref{appendix-system performance}.

\setlength{\lineskiplimit}{\lineskiplimitbackup}

\section{User study}
\label{sec:User study}
We conducted a controlled experiment to evaluate the performance of the MeTACAST methods in identifying various target structures with uniform or non-uniform density and shape. 
We compared our three MeTACAST methods to a conventional, region-based method, to which we refer as \baseline.
We directly derived the \baseline method based on Touching the Cloud \cite{Lubos:2014:TBP}, a technique designed for selecting particles within the region where the user brushes using 3D input in a VR environment. We compared all methods based on the accuracy and time taken for particle selection. We pre-registered the user study plan, our analysis code, and our result prediction at \href{https://osf.io/dvj9n/?view_only=07cae591f59944499691047101aecfef}{\texttt{osf.io/dvj9n}}.

\subsection{Study Design}

\noindent\textbf{\textit{Participants.}}
We recruited 32 unpaid participants (17 male, 15 female) from the local university, 18--33 years old (M=$23.1$, SD=$2.8$), all of whom reported to be right-handed.
Among them, 15 use VR at least once per week, 14 at least once per year, and 3 had never used VR devices before. Furthermore, 18 participants had obtained a Bachelor's degree or higher. They all had normal or corrected-to-normal vision and were able to distinguish clearly the colors used in our application.

\textbf{\textit{Apparatus.}}
We used Valve Index~\cite{Valve}, a PC-based VR head-moun\-ted display (HMD; 1440\,\texttimes\,1600 resolution per eye, 108\textdegree{} field of view, 90\,Hz refresh rate). We carried out the study on a PC (Intel Core\texttrademark{} i9, 3.5GHz, 64GB RAM, GeForce RTX3090, 24 GB video memory). 

\begin{figure}[t]
    \centering
    \captionsetup[subfigure]{labelformat=empty}
     
    \subfloat[\label{fig:userstudyData(a)} ]
        {\begin{overpic}[width=0.24\columnwidth]{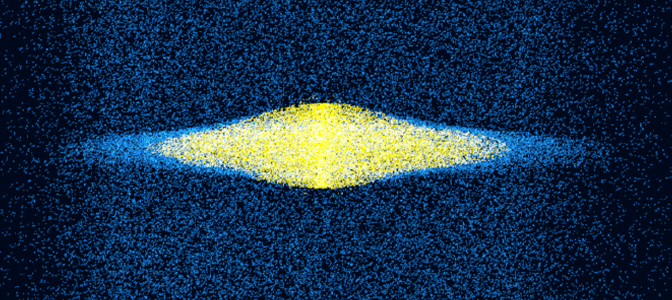}
            \put(5,5){\textcolor{white}{(a)}}
            
        \end{overpic}}\hfill
    \subfloat[\label{fig:userstudyData(b)} ]
        {\begin{overpic}[width=0.24\columnwidth]{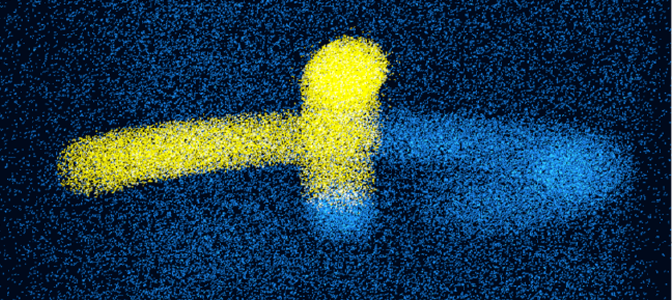}
              \put(5,5){\textcolor{white}{(b)}}
            
          \end{overpic}}\hfill
    \subfloat[\label{fig:userstudyData(c)} ]
        {\begin{overpic}[width=0.24\columnwidth]{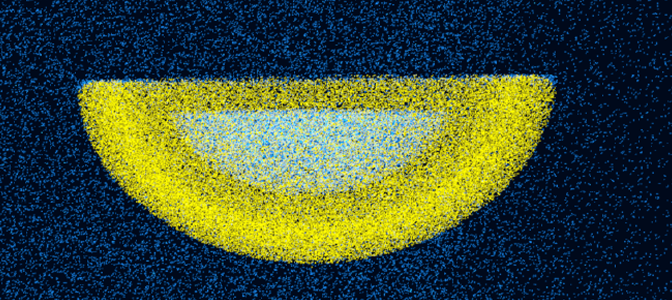}
              \put(5,5){\textcolor{white}{(c)}}
            
        \end{overpic}}\hfill
    \subfloat[\label{fig:userstudyData(d)} ]
        {\begin{overpic}[width=0.24\columnwidth]{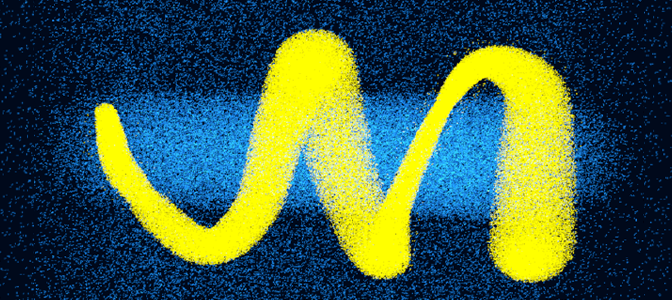}
              \put(5,5){\textcolor{white}{(d)}}
            
         \end{overpic}}\\[-3.5ex]
    \subfloat[\label{fig:userstudyData(e)} ]
        {\begin{overpic}[width=0.323\columnwidth]{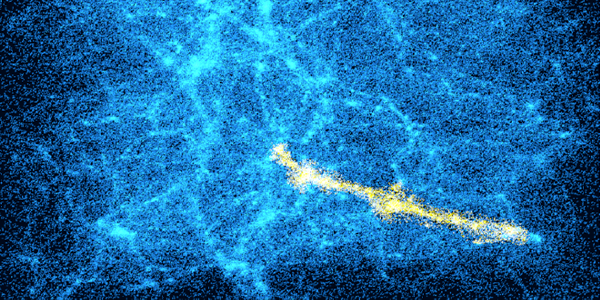}
             \put(5,5){\textcolor{white}{(e)}}
            
         \end{overpic}}\hfill
    \subfloat[\label{fig:userstudyData(f)} ]
        {\begin{overpic}[width=0.323\columnwidth]{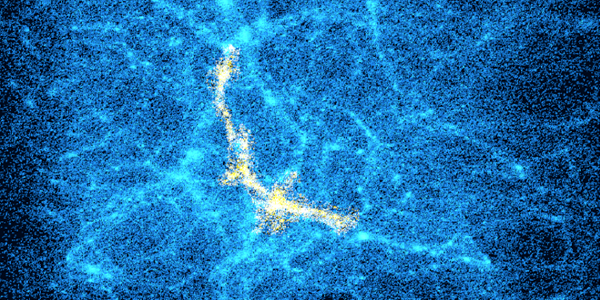}
             \put(5,5){\textcolor{white}{(f)}}
            
          \end{overpic}}\hfill
    \subfloat[\label{fig:userstudyData(g)} ]
        {\begin{overpic}[width=0.323\columnwidth]{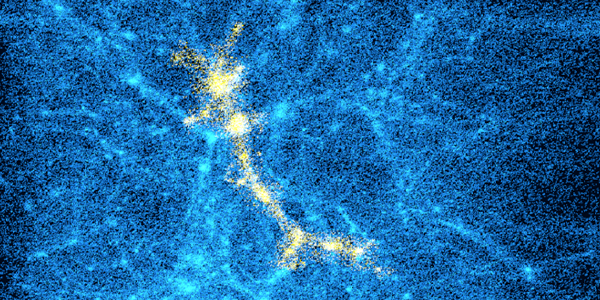}
             \put(5,5){\textcolor{white}{(g)}}
            
         \end{overpic}}\vspace{-2em}

 \caption{Datasets we used in our study: (a) \textit{Disk}, (b) \textit{Rings}, (c) \textit{Shell}, (d) \textit{Strings}, and (e--g) \textit{Filaments}.}
 \label{fig:userstudyData}
\end{figure}

\textbf{\textit{Datasets.}}
All datasets in our study (\autoref{fig:userstudyData}) have target (\textcolor{Goldenrod}{yellow}) and interfering particles (\textcolor{DodgerBlue}{blue}).
We designed or selected these datasets to have different features that made the selection of targets challenging:
\begin{description}[nosep,leftmargin=1.5em,labelindent=0em,leftmargin=!,labelindent=!,itemindent=!,font=\normalfont\itshape]
\item[Disk:] This dataset (\autoref{fig:userstudyData(a)}) exhibits a gradual decrease in density from its center towards the periphery. The high-density area around the center is the target region. 
\item[Rings:] This dataset (\autoref{fig:userstudyData(b)}) consists of two half rings with uniform density, positioned perpendicular to each other with a slight gap between them. The target area comprises a portion of each ring.
\item[Shell:] This dataset (\autoref{fig:userstudyData(c)}) comprises a half-ball of interfering particles that are partially surrounded by a semispherical shell of particles. Both structures have uniform densities and are in close proximity. The semispherical shell is the target structure.
\item[Strings:] This dataset (\autoref{fig:userstudyData(d)}) comprises two string-like structures with non-uniform density. The outer string wraps around the central string and exhibits variations in both density and perimeter along itself. We asked participants to select the outer string. 
\item[Filaments:] The Millennium-II data subset \cite{Springel:2005:SimulationsOT} (\autoref{fig:userstudyData(e)}) is a complex network of filaments that connect high-density clusters. 
We chose this real-world data to represent a realistic scenario in astronomy. Unlike the other datasets, we showed the target filaments only for two seconds, to give the participants a general sense of their 3D location and structure. This way we wanted to see what crucial features participants found important and how our methods supported their selections. To avoid a learning effect we thus asked participants to select three distinct filaments in each condition (\autoref{fig:userstudyData}(e--g)). 
\end{description}
 
\textbf{\textit{Task and Procedure.}}
We asked participants to select the \textcolor{Goldenrod}{yellow} and to avoid the \textcolor{DodgerBlue}{blue} particles, and to complete the tasks as fast and as precise as possible. We split the study into four tasks (T1--T4) with explicit goals and one task (T5) with an implicit goal. During the explicit goal tasks, which used the datasets \textit{Disk} (\autoref{fig:userstudyData(a)}), \textit{Rings} (\autoref{fig:userstudyData(b)}), \textit{Shell} (\autoref{fig:userstudyData(c)}), and \textit{Strings} (\autoref{fig:userstudyData(d)}), the target particles remained highlighted in \textcolor{Goldenrod}{yellow} until they were selected, after which all selected particles turned \textcolor{FireBrick}{red}. For each combination of trial and dataset, we chose a unique starting orientation, which changed between trials but which we used for all participants. 
In the implicit goal task (with dataset \textit{Filaments}, \autoref{fig:userstudyData(e)}) we showed the \textcolor{Goldenrod}{yellow} target particles to participants for only 2 seconds, giving them a limited period to perceive their location and structure. We allowed them, however, to inspect the target particles as many times as needed for an additional 2 seconds each time but did not allow them to make any selections during this time. Once participants had gained a general understanding of the target particles (\eg, location, structure, context, etc.), we asked them to select the ones that possessed ``important features'' based on their judgment. With this task we aimed to explore the possibility that, by employing  context- and target-aware techniques, users could rely on a general understanding of these ``important features'' when identifying target points, without them continuously being visible.

To prepare participants for the actual study, we first familiarized them with the selection techniques by practicing with additional training data. During the training trials, we instructed the participants to perform the selection using VR controllers and allowed them to take as much time as needed. In the actual study, however, we instructed them to complete their selection tasks with both speed and accuracy but did not inform them if or when they had achieved the selection goal.
We provided two possible selection modes: union and subtraction. As Yu \etal~\cite{Yu:2012:ESA, Yu:2016:CEE} had previously discussed, subtraction is best done using region-based techniques, not with structure-aware ones. We thus implemented subtraction with the \baseline technique in all trials. 
Participants could use the VR joystick to adjust the density threshold and use the VR button to adjust the size of the input marker. In addition, we provided undo and redo functions as well as a reset to the initial unselected state. The whole study lasted $\approx$\,90 minutes. Following each technique, we requested that the participants evaluate their workload and fatigue levels using NASA's Task Load Index (TLX) \cite{TLXScale}. After completing all trials, we asked them to indicate which technique they preferred for each dataset and to provide their reasoning. 

\textbf{\textit{Design.}} For the explicit goal tasks with the first four datasets, we counter-balanced the order for the methods and datasets. For a given participant with a specific $P_{ID}$, where ID is unique and $\in[0, 15]$ (the first batch of 16 participants), we presented the datasets in the same order for each method. We counter-balanced the dataset order using a Latin square as ``($P_{ID}$ / 4) mod 4'' and the method order as ``$P_{ID}$ mod 4''. We repeated the Latin square order for the second batch of 16 participants. In summary, we had 32 participant \texttimes{} 4 methods \texttimes{} 4 datasets \texttimes{} 3 repetitions $=$ 1536 trials.  
For the implicit goal tasks with the last dataset (\textit{Filaments}), we used the same order for the three ROIs across all methods, but counter-balanced the method order. In the end, we had 32 participants \texttimes{} 4 methods \texttimes{} 3 cases $=$ 384 trials.

\textbf{\textit{Measures and Analysis.}} To reduce the impact of the learning effect, we excluded the first repetition of each dataset \texttimes{} method pairs. This left us with 1280 trials. Since NHST has been criticized for analyzing experimental data \cite{Baguley:2009:ES, Cumming:2013:NS, Dragicevic:2016:FSC, Dragicevic:2014:RAH}, and APA recommends alternative approaches \cite{VandenBos:2009:PMAPA}, we report results using estimation techniques with effect sizes and confidence intervals rather than $p$-value statistics.

\textbf{\textit{Accuracy.}} Similar to Yu \etal \cite{Yu:2012:ESA,Yu:2016:CEE}, we calculated two distinct accuracy scores, F1 and MCC (Matthews correlation coefficient), to compare the accuracy of our techniques.
The accuracy scores for both cases are based on the identification of true positives (TP, \# of correctly selected particles, false positives (FP, \# of incorrectly selected particles), and false negatives (FN, \# of target particles that were not selected). F1 is calculated as ${\mathrm{F1}=2\cdot(P\cdot R)/(P+R)}$, where ${P=TP/(TP+FP)}$ and ${R=TP/(TP+FN)}$. While the F1 score provides a measure of the harmonic mean of the precision, it does not consider the true negatives (TN,\# of correctly unselected particles) rate. Thus, we also used MCC as our second error metric which is calculated as:\vspace{-5pt}%

\begin{equation}
    \mathrm{MCC}=\frac{T P \cdot T N-F P \cdot F N}{\sqrt{(T P+F P)(T P+F N)(T N+F P)(T N+F N)}}.
    \notag
\end{equation}\vspace{-10pt}

The accuracy scores for both cases are normalized. We computed means and 95\% bootstrap confidence intervals (CIs; all $n$ $=$ 32).

\textbf{\textit{Completion Time.}} We analyzed the completion time data using exact CIs on log-transformed data (all $n$ $=$ 32). We report means thus as geometric means and express comparisons between means as ratios.

\subsection{Hypotheses}
\label{subsec:Hypotheses}
\begin{figure}[tb]
\includegraphics[width=\linewidth]{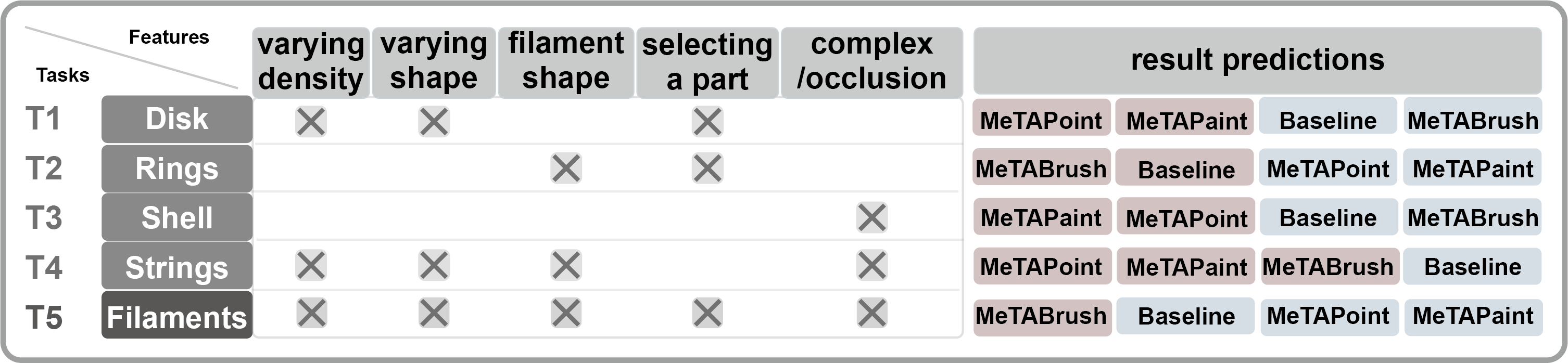}
\caption{The predicted results are determined by analyzing the data features and the principle of the selection methods: red (relatively good performance), blue (relatively poor performance).}
\label{fig:datafeature_selectiontechnique}
\end{figure}

As we designed the three MeTACAST techniques to be adaptable to specific selection intents it is reasonable to expect that their performance varies depending on the given scenario. To anticipate the results of our study, we analyzed the data features and made predictions (\autoref{fig:datafeature_selectiontechnique}) based on each technique's principles. Based on these pre-registered (\href{https://osf.io/dvj9n}{\texttt{osf.io/dvj9n}}) predictions we formulated the following hypotheses:
\newlength{\hypothesislabelwidth}%
\settowidth{\hypothesislabelwidth}{\textbf{H1}}%
\addtolength{\hypothesislabelwidth}{1.5ex}
\begin{description}[nosep,leftmargin=\hypothesislabelwidth,labelindent=0em,labelsep=1.5ex,leftmargin=!,labelindent=!,itemindent=!]
\item[H1] MeTACAST adjusts the input toward the local density maximum. All MeTACAST techniques would thus be more accurate than the non-adjusting, region-based \baseline.
\item[H2] Since we explicitly designed \brush for filament-like structures, its completion time and accuracy would be better than that of other techniques for the \textit{Filaments} datasets, especially when the target is part of a filament with varying density and shape.
\item[H3] We designed \point and \paint for selecting complete sub-structures with simple, minimum input, so they would be generally faster than other methods for selecting whole clusters.
\item[H4] Technique preference would depend on the dataset characteristics. In general, for filament-like datasets, participants would prefer \brush and \baseline, while for individual shapes they would prefer \point and \paint.
\end{description}

\label{sec:Result}
\subsection{Results---Explicit goal tasks: T1 to T4}
\label{subsec:Analysis of T1 to T4}

As we designed MeTACAST based on different selection strategies and to be able to deal with various datasets, the analysis of the selection effectiveness based on overall completion time and accuracy may not offer a complete picture of their capabilities.
We thus discuss the results for the four synthetic datasets and one real-world dataset separately and provide the overall completion time/accuracy results in \autoref{appendix-overallresults}.

\autoref{tab:perdata} in \autoref{appendix-overallresults} displays the average task completion times and two accuracy scores for each dataset and technique. For F1, 1 indicates perfect performance, while 0 represents the worst possible result. For MCC, a score of 1 indicates perfect performance, while $-1$ represents the worst performance. Mean completion times and 95\% confidence intervals for each technique are shown in \autoref{fig:time_per_dataset}, while we provide the figure for the accuracy scores for each dataset in \autoref{appendix-overallresults}. We now focus on the statistical results for each dataset based on our hypotheses. 

\noindent\textit{\textit{Disk:}} Given that the target area is part of the high-density region, which has varying density and shape, we predicted that the \point and \paint methods would outperform the \brush (designed for filament-like structures) and \baseline (region-based) methods. This is because both \point and \paint facilitate an easy initial selection and refinement of results through density threshold adjustments.
From the results, we can see \point and \paint were much faster than \baseline and \brush, and \baseline was less accurate than the other methods, which aligns with our prediction.

\textit{\textit{Rings:}} As the target area was part of two rings with uniform density and shape, we predicted that both \brush and \baseline would outperform \point and \paint, which were designed to select the entire component and change the overall selection based on the density threshold.
We saw that \brush and \baseline were indeed faster. The accuracy scores for all methods are equivalent as we predicted.

\textit{\textit{Shell:}} For the \textit{Shell} dataset, the target cluster is a complete component. We thus predicted that \paint and \point would be more effective and that \point may be slightly slower than \paint since its density threshold is determined by the input location. Due to occlusion, however, users may not have precise control over the density adjustment.
Our results show that both \point and \paint were indeed faster than the other two methods, with a substantial difference in average completion time. \point was slightly slower than \paint. These results align with our prediction. In addition, our findings indicate that \baseline is the least accurate method.

\textit{\textit{Strings:}} Based on the non-uniform density and shape of the target cluster, we predicted that all MeTACAST methods would perform better than the \baseline. We also expected that \paint and \point, requiring minimal input, would be faster than \brush. Although the target was filament-like, \brush may not be as effective as the other two as it would require participants to brush the whole ROI.
Our results show that the MeTACAST methods were both faster and more accurate than \baseline. In addition, \brush was much slower than \paint and \point, which we expected. Moreover, our results show that \baseline was again the least accurate method.

\begin{figure}[t]
    \centering%
    \subfloat{%
        \label{fig:time_for_disk}
				\begin{overpic}[width=\columnwidth,clip]{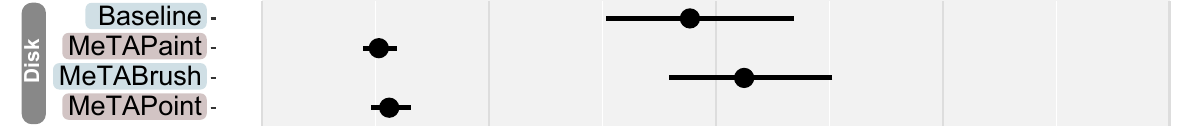}
            \put(-2.5,1){\footnotesize (a)}
        \end{overpic}
    }\\[.5ex]
    \subfloat{%
        \label{fig:time_for_rings}
				\begin{overpic}[width=\columnwidth,clip]{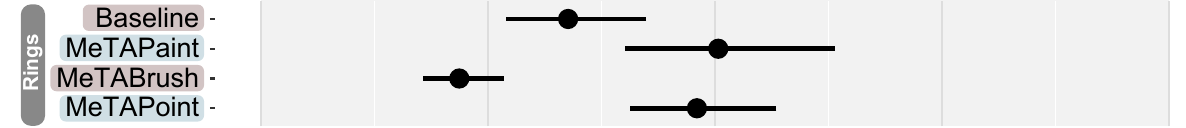}
            \put(-2.5,1){\footnotesize (b)}
        \end{overpic}
    }\\[.5ex]
    \subfloat{%
        \label{fig:time_for_shell}
				\begin{overpic}[width=\columnwidth,clip]{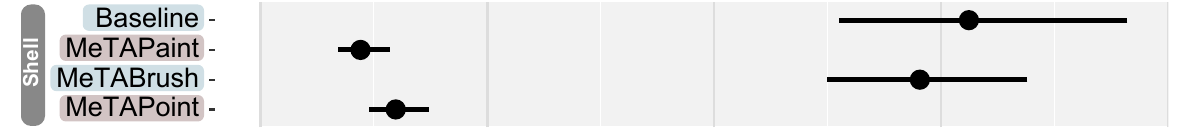}
            \put(-2.5,1){\footnotesize (c)}
        \end{overpic}
    }\\[.5ex]
    \subfloat{%
        \label{fig:time_for_strings}
				\begin{overpic}[width=\columnwidth,clip]{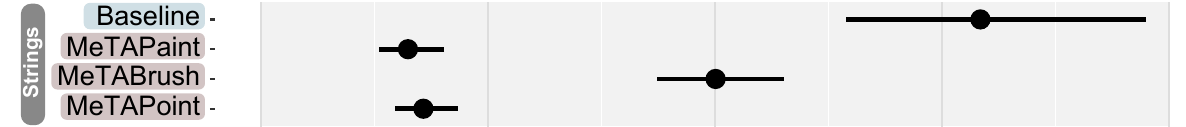}
            \put(-2.5,1){\footnotesize (d)}
        \end{overpic}
    }\\[.5ex]
    \subfloat{%
        \label{fig:time_for_filaments}
				\begin{overpic}[width=\columnwidth,clip]{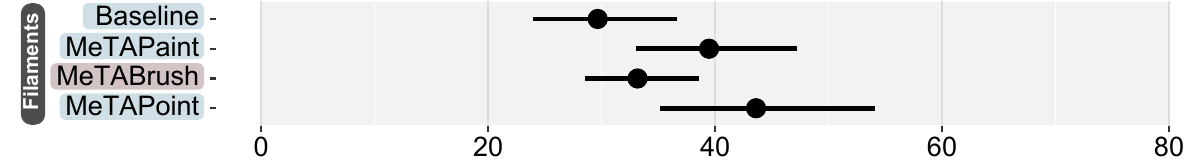}
            \put(-2.5,2){\footnotesize (e)}
        \end{overpic}
    }
    \caption{The geometric mean completion times in seconds for each selection technique in T1 to T5; (a): \textit{Disk}, (b): \textit{Rings}, (c): \textit{Shell}, (d): \textit{Strings}, (e): \textit{Filaments}. Error bars show 95\% confidence intervals (CIs).}
    \vspace{-1ex}
    \label{fig:time_per_dataset}
\end{figure}

\begin{figure}[t]
    \centering
        \includegraphics[width=1\columnwidth,trim={0 23pt 0 2pt},clip]{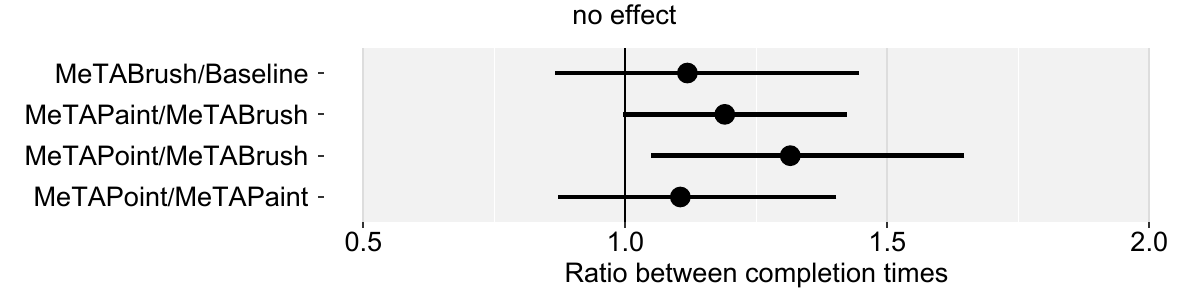}
    \caption{Pairwise completion time ratio in T5. Error bars: 95\%\,CIs.}
     \vspace{-1ex}   
   \label{fig:pairwise_ratio_time_T5}
\end{figure}

\subsection{Results---Implicit goal task: T5}
\label{subsec:Analysis of T5}
\begin{figure}[t]
    \centering%

    \includegraphics[width=1\columnwidth]{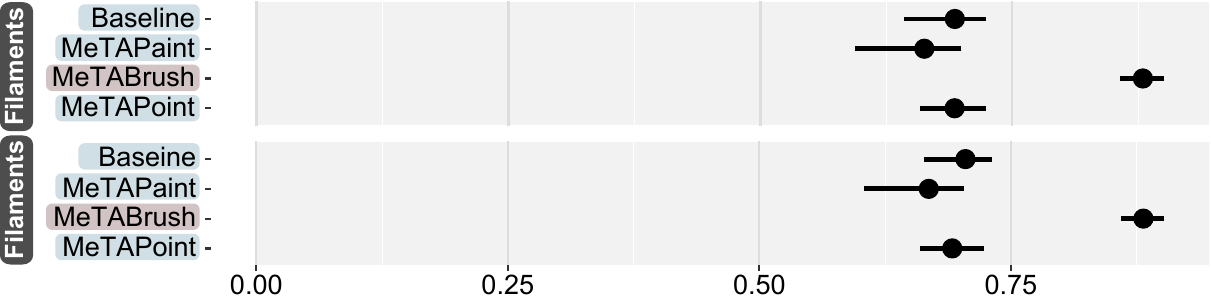}
    \vskip -2pt
    \caption{The F1 (top) and MCC (bottom) for T5. Error bars: 95\%\,CIs.}
    \label{fig:F1_and_MCC_Filaments}
    \vskip -5pt
\end{figure}

T5 was a more high-level selection task where participants could only view the features of the target cluster within a short period and make selections based on their understanding of the key data features. In addition to analyzing accuracy and completion time, we were interested in how often participants required to re-check the targets and how our selection strategies supported them in making selections based on data features. We predicted that \brush would outperform other methods since it was specifically designed for filament-like structures in a noisy environment.
The results showed that T5 had a lower accuracy level compared to T1 to T4, while \brush showed considerably higher accuracy (\autoref{fig:F1_and_MCC_Filaments}) compared to the other methods. Both \baseline and \brush techniques had a shorter mean completion time than \point and \paint (\autoref{fig:time_per_dataset}(e)), with a small difference. We further provide pairwise ratios of completion times of the MeTACAST techniques in \autoref{fig:pairwise_ratio_time_T5}. We observed that \brush took only about 0.86--1.44\texttimes{} longer than \baseline. \point took about 1.04--1.64\texttimes{} longer, and \paint took about 1.00--1.42\texttimes{} longer than \brush. Although the filament-based and region-based methods were slightly faster in selecting filament-like structures in the noisy background, the difference was not as substantial as predicted. 
Overall, we have strong evidence that \brush outperforms other methods in this task.

In conclusion, as noticed in most tasks except T5 (\textit{Filaments}, which also requires user understanding about the data feature), \baseline was the least accurate technique. Therefore, we can \textbf{partially support H1}.
For the filament datasets (\textit{Rings}, \textit{Strings}, and \textit{Filaments}), the results indicate that \brush was faster than other techniques only in the \textit{Rings} dataset. However, it is important to note that, for the \textit{Strings} dataset where the whole string needs to be selected, \point and \paint were faster. In the case of the \textit{Filaments} task, \brush took about 0.86--1.44\texttimes{} longer than \baseline, but it was substantially faster than other techniques. Yet, \brush showed considerably higher accuracy than other techniques. While we can \textbf{partially support H2}, the results thus highlight the importance of \brush in selecting filament-like structures.
For the tasks in which the whole dataset needs to be selected (\textit{Shell}, \textit{Strings}), \paint and \point outperform other methods. Thus, \textbf{we support H3}.
\vspace{-1ex}

\begin{figure}[t]
    \centering
         \includegraphics[width=1\columnwidth]{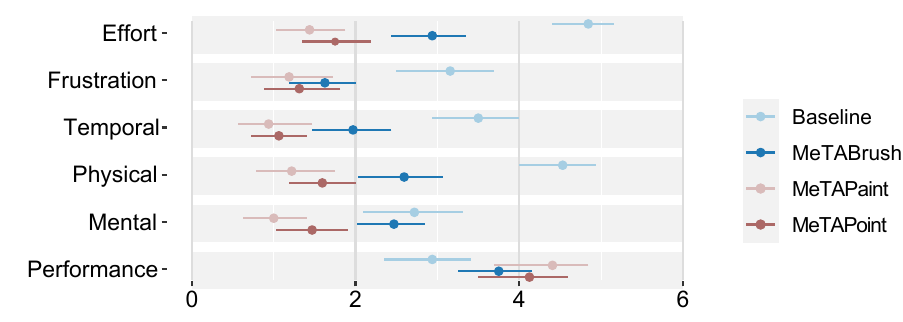}
         \vskip -6pt
    \caption{User workload and performance, T1--T4. Error bars: 95\%\,CIs.}
  \label{fig:tlx for task1-4}
\end{figure}

\begin{figure}[t]
    \centering
         \includegraphics[width=1\columnwidth]{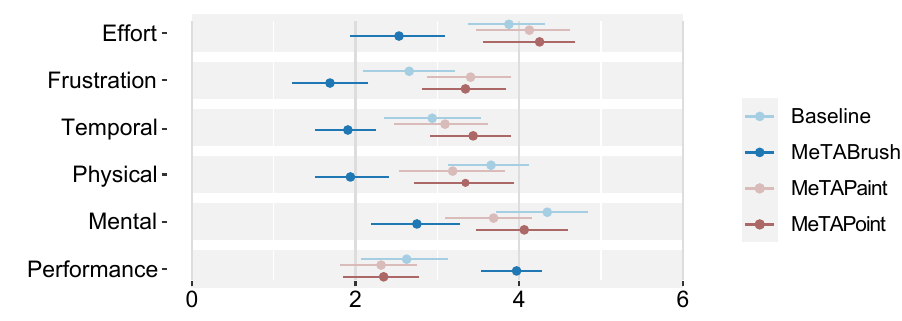}
         \vskip -4pt
    \caption{User workload and performance, T5. Error bars: 95\%\,CIs.}
    \vspace{-1ex}
    \label{fig:tlx for task5}
\end{figure}

\subsection{Results---User workload and preference}
\textbf{Workload.}
After analyzing the data, we found strong evidence that participants felt more efficient, less frustrated, less time-pressured, and needed less mental effort when using \paint and \point in T1--T4, as shown in \autoref{fig:tlx for task1-4}. These findings align with the participants' overall completion times, which indicates that their subjective experiences match their actual performance for these methods.
However, for the more complex T5 we saw that \brush was the most effective method, as it led to higher performance, less frustration, less time pressure, and less physical and mental effort. In addition, participants experienced higher mental and physical effort when using \baseline and felt more frustrated when using \point and \paint in T5.

\noindent\textbf{Preference.}
After analyzing the techniques' rank across all datasets (T1 to T5) shown in \autoref{appendix-userpreference}, we observe a correlation between the participant preference and their actual performance. Specifically, for the \textit{Disk} and \textit{Shell} datasets, the participants demonstrated a similar pattern in the ranking of the selection techniques (\paint $>$ \point $>$ \brush $>$ \baseline). Interestingly, for the two filament-like datasets, \textit{Rings} and \textit{Filaments}, most participants preferred \baseline and \brush, while \paint and \point were ranked higher for the \textit{Strings} dataset. We believe that this observation may be because participants were required to select the entire string, which can be achieved more easily with \paint that allows for a large selection with minimal input. Based on these results, \textbf{we support H4}.

During the study, we closely observed how participants used each technique to select particles. Our observations revealed that participants demonstrated a clear understanding of the methods' designs and effectively employed strategies to complete the tasks. For example, some participants intentionally brushed a stroke outside of the semispherical shell to ensure that the input stroke remained far away from the half-ball of interfering particles. They made this adjustment with the understanding that the position of the input stroke would be adjusted to align with the shell. In addition, the Baseline technique was also appropriately employed in the study, aligning with its intended design.

To further explore the impact of participants' level of VR fluency, we conducted a separate analysis of task performance on two groups. We provide and discuss the comparison results in~\autoref{appendix-comparison-seperate}.

\section{Discussion}
\label{sec:Discussion}

There are several insightful discussion points that we can derive from our experimental results, as we describe next. We also examine MeTACAST within the broader context of existing state-of-the-art selection methods, to derive general design guidelines for spatial selection.
\vspace{-1ex}

\subsection{Selection scenarios}
Different selection strategies showed clear advantages in dealing with different data characteristics and situations. When choosing a data selection technique, therefore, we need to consider various factors, such as data features, selection requirements, and interaction environment.

\textbf{Target and Context Awareness.} The MeTACAST techniques are both target- and context-aware, taking into account various data features during the selection process such as position, structure, and density distribution to realize the user's intention based on their input.
\point identifies the cluster nearest to the user's pointing location and dynamically adjusts or switches selections based on ongoing input. \brush analyzes the user's input stroke and infers their intention by identifying the primary filament branch close to the path. Finally, \paint selects targets based on how the user perceives the structure and distinguishes them from the surrounding context.
These strategies are well-suited for data selection in immersive environments due to their inherent flexibility, which enables users to focus fully on the data features without separating input location from data position.

\textbf{Density- vs.\ Region-based Control.}
Our results partially support \textbf{H1}: MeTACAST techniques are generally more accurate than the re\-gion-based \baseline, except in T5 where the \baseline method was slightly more accurate than \point and \paint---but the difference was minor. In T5, \brush was the second fastest technique (after \baseline), and some participants preferred the region-based interaction when they had limited time to see the target filaments. T5 required participants to rapidly identify features of the target filaments (position, density). 
Here, if participants were able to quickly recall the target's location, the region-based method offers a straightforward way to complete the task.
This observation also explains why some participants ranked \baseline highly in this task, despite being aware that their performance would not be good with this method. Nonetheless, for this dataset, \brush---which considers both the position and density distribution in the selection process---demonstrated clear advantages in terms of accuracy, completion time, and user preference.

\textbf{Target Shape.} Our findings strongly support the notion that different selection strategies are well-adapted to distinct target shapes. Specifically, \point and \paint demonstrated substantially faster performance for \textit{Shell} and \textit{Disk}, whereas \brush and \baseline exhibited advantages in selecting filament-like structures such as \textit{Rings} and \textit{Filaments}. It is important, however, to note that complex data environments typically involve multiple factors, making it challenging to determine the optimal selection method based on a single data feature alone. For instance, although \brush is ideal for filament-like structures, it was not consistently the fastest method in such tasks (\eg, \textit{Strings}). This is primarily due to the fact that, in the \textit{Strings} task, participants had to brush the entire string with \brush and \baseline, while \paint and \point require only minimal painting/pointing on or around the target string.

\textbf{Partial Selection.} In general, our results provide strong evidence to support \textbf{H3}: \point and \paint are \textbf{\textit{faster}} than the two brushing techniques in \textbf{\textit{selecting whole components}} such as for \textit{Disk} and \textit{Shell}, in particular when the target is well-separated from other clusters. This also limits, however, their use when multiple clusters or a subset of cluster are required to be selected (\eg, \textit{Filaments})---then \brush and region-based techniques are better. 

\textbf{Occlusion.} Three datasets had complex occlusion features: \textit{Shell} (partially surrounded half-ball), \textit{Strings} (wrapped inner string), and \textit{Filaments} (comprising many small details). Therefore, participants had to avoid mistakenly selecting interfering particles from other clusters. Our results show that for \textit{Shell} and \textit{Strings}, \paint and \point were faster than \brush and \baseline, and \paint was slightly faster than \point. The main reason is that in these two tasks, the whole cluster needs to be selected. As long as the user can recognize the partial area of the target cluster, they can accurately and quickly select the entire cluster. A reason why \point was slightly slower than \paint may be that we derive the threshold of \point from the density at the input position. Participants thus needed to be careful when dragging the controller. In contrast, the \textit{Filaments} task was particularly challenging for \paint and \point since they treated clusters as individual objects. Future work could explore how to use \paint and \point to identify specific internal structures such as the half-ball (\textit{Shell}) and inner string (\textit{Strings}) in situations where the user lacks a complete understanding of the target object.
\vspace{-1ex}

\subsection{Comparison of existing spatial selection techniques}
\label{subsec:comparison}
Target- and context-aware techniques are especially important in three situations: first, when the user is unfamiliar with the dataset's structure and features; second, when occlusion and perception distortion impede the user's ability to see, judge, or reach the target, or when they need to trace the target's border carefully; and third, when the user has a clear idea about the target and expects to use a simple and fast input to select it. We compared MeTACAST with existing spatial selection techniques' characteristics (summarized in \autoref{tab:comparison} in \autoref{appendix-comparison}) and now discuss their pros and cons under various situations, to offer design guidelines and suggest appropriate methods for different scenarios.

Previous taxonomies have categorized spatial selection techniques according to various factors such as targets (\eg, object, ROI) \cite{Poupyrev:1999:ManipulatingOI}, input strategies and degree of freedom of the input \cite{Argelaguet:2013:ASO}, and the users' level of control during the selection process \cite{Besancon:2019:TangibleBrush}. In our own work we emphasized different aspects of spatial selection, including data characteristics, selection requirements (such as the selection basis and target shape), selection strategies (such as metaphor and selection strategy), and the users' level of control (including precision demand and post-adjustment).
Note that the required level of input precision depends on the user's desired level of control and the level of performance needed for the selection. For instance, precise input (2D or 3D) could always be required (high), it could be imprecise or incomplete (low), or imprecise input could be allowed but may cause inaccurate results (medium). To narrow our focus, we only focus on spatial selection where the user specifies a region in space in which the targets are located. 

\textbf{Partial Selection and Interaction Metaphor.} 
An important question is to examine how whole or partial selection intents relate to the chosen interaction metaphor. It would be interesting to investigate, for instance, whether the lasso approach is an effective method for making partial selections. The lasso approach is indeed the most widely used approach for defining selection ranges for multiple targets or partial selection on 2D surfaces (\eg, CylinderSelection, CloudLasso \cite{Yu:2012:ESA}, SpaceCast \cite{Yu:2016:CEE}, LassoNet\cite{Chen:2020:LassoNet}, Hybrid AR selection \cite{sereno:2022:HTTinAR}, and Tangible brush \cite{Besancon:2019:TangibleBrush}). A potential approach for defining/extending the selection range in 3D is through extrusion using mobile devices, as in Hybrid AR selection \cite{sereno:2022:HTTinAR} and Tangible Brush \cite{Besancon:2019:TangibleBrush}. 3D brushing provides a direct and intuitive way for users to make partial selections in 3D, as demonstrated in techniques such as Touching the cloud \cite{Lubos:2014:TBP}, Fiducial-Based Tangible \cite{gomez:2010:FBT}, Neuron Tracing \cite{mcdonald:2020:IUV}, and \brush. Unlike the lasso or extrusion, however, the selection range for 3D brushing is not always clear. Designers thus need to consider specific data features and selection contexts when developing such techniques. In contrast to for single-object selection, Raycasting is a less commonly used approach for making partial spatial selections---defining the selection range in 3D through a ray can be challenging. Other possibilities for defining a 3D range include using tangible devices such as Embodied Axes \cite{Cordeil:2020:EmbodiedAxes}, which can achieve partial selection in all dimensions.

\textbf{Selection Strategy and Precision Needs.} Our suggestion for spatial selections is to allow users to express the selection intent in their own way, and then employ selection heuristics that take all relevant input information and data characteristics into account to provide an initial result. Manual selections \cite{Roberto:2020:HMS, Besancon:2019:TangibleBrush, Cordeil:2020:EmbodiedAxes, gomez:2010:FBT, sereno:2022:HTTinAR, Lubos:2014:TBP} always require a high input precision, thus increase the users' physical and mental workload. 
Several semi-automatic methods are available for context-aware selection on 2D surfaces (\hspace{-.001ex}\cite{Yu:2012:ESA, Yu:2016:CEE, Owada:2005:VC, Chen:2020:LassoNet, Wiebel:2012:WYSIWYP}) and 3D space (MeTACAST, \cite{malmberg:2006:3DLW, jackson:2013:LTI, mcdonald:2020:IUV}), specifically designed to consider users' intention and data characteristics. They significantly enhance selection accuracy and efficiency, while also reducing the users' physical workload.

\textbf{Partial Selection and Selection Strategy and Precision Needs.}
Following the previous point, a more focused question needs to investigate if the needs for whole or partial selection increase the demand for input precision, leading to greater mental and physical workload. In addition, it would be interesting to explore how selection strategies can mitigate these challenges. 
We thus focus on selection techniques that have low input precision needs. 
One scenario is that, for selecting single clusters, users may not need to provide a precise or complete input. The whole target cluster can be selected with a short\discretionary{/}{}{/}in\-com\-plete stroke (\paint, Volume Catcher \cite{Owada:2005:VC}, TraceCast \cite{Yu:2016:CEE}) or a single click (\point, PointCast \cite{Yu:2016:CEE}, WYSIWYP \cite{Wiebel:2012:WYSIWYP}). 
Another scenario is that, for selecting a target subset, users may not need to define a precise selection range. The intended selection can, \eg, be made through a complete stroke in LassoNet \cite{Chen:2020:LassoNet}. Yet there are two powerful 3D brushing methods do not require a selection range but still precisely select the intended partial target: \brush and Neuron Tracing \cite{mcdonald:2020:IUV}. \vspace{-1.5em}

\subsection{Limitations}
\textls[-5]{As with most work, ours too has some limitations. First, for \brush we currently use a pre-defined range and search for particles that flow into this range by following the gradient direction. 
While adjusting this range (spatial adjust) and adjusting the density threshold (density adjust) are both possible, it may cause confusion for users. A more reliable approach would be to compute these parameters automatically based on the data distribution, edge, or topological structures. 
Second, MeTACAST is currently designed for handheld or table-size visualization, and may not be suitable for room-size settings. In such environments users may face difficulties in accurately estimating the density distribution and identifying dense clusters if they are located within them.
Third, MeTACAST methods are density-based methods which perform well when variations in density distribution are visible. When users encounter difficulties, however, in distinguishing data features, such as when the density remains consistent across the entire point cloud, shape-based techniques may be more suitable choices.
Last, it is important to note that our interaction technique design requires $6$DOF input devices.}

\section{Conclusion}
\label{sec:Conclusion}
We presented a family of spatial 3D data selection methods, MeTACAST (\point, \brush, and \paint), for VR environments. 
With MeTACAST, users can explore immersive 3D spaces and select data in their preferred way such as pointing, brushing, or drawing. 
We demonstrated MeTACAST for particle data and density as the measure, yet it can work also with volumetric data and any other scalar field.
The basic idea behind our techniques is that users select visually interesting or important locations that contain key information (\eg, density distribution, geometric shape, and position) that causes the feature to be visible. As such, two aspects are essential.
First, we need to understand the data and determine what is considered key information or what should attract a user's attention. This aspect involves analyzing the data features and seeking the expertise of domain specialists to determine the data characteristics and what they are trying to find.
Second, we need to understand the users and how they behave when encountering key information. To accomplish this part, think-aloud studies are effective to collect user behavior and their intentions.

Our target- and context-aware methods can then infer user intentions from their input, which means that more input can provide more information about the goal. 3D immersive environments thus offer greater flexibility than 2D projection-based approaches, allowing users to flexibly interact with data and seamlessly integrate their actions across modalities.
By incorporating information from various modalities (\eg, head and eye movements; path, direction, and speed of input; gestures; user location) more sophisticated target- and context-aware methods can be developed for complex data exploration.

\acknowledgments{
The work was partially supported by NSFC (62272396).}

\section*{Supplemental Material Pointers}

We share our additional materials (appendix, study results/data and analysis scripts, videos) at \href{https://osf.io/dvj9n/}{\texttt{osf.io/dvj9n}}. We also make our prototypical implementation of the MeTACAST techniques available at \href{https://github.com/LixiangZhao98/MeTACAST}{\texttt{github.com\discretionary{/}{}{/}LixiangZhao98\discretionary{/}{}{/}MeTACAST}}.

\section*{Images/graphs/plots/tables/data license/copyright}
We as authors state that all of our figures, graphs, plots, and data tables in this article are and remain under our own personal copyright, with the permission to be used here. We also make them available under the \href{https://creativecommons.org/licenses/by/4.0/}{Creative Commons At\-tri\-bu\-tion 4.0 International (\ccLogo\,\ccAttribution\ \mbox{CC BY 4.0})} license and share them at \href{https://osf.io/dvj9n/}{\texttt{osf.io/dvj9n}}.

\bibliographystyle{abbrv-doi-hyperref-narrow}

\bibliography{abbreviations,Reference}

\begin{thebibliography}{10}
\renewcommand*{\sfdefault}{PTSansNarrow-TLF}

\bibitem{Ferran:2009:EPS}
F.~Argelaguet and C.~Andujar.
\newblock Efficient {3D} pointing selection in cluttered virtual environments.
\newblock {\em IEEE Comput Graph Appl}, 29(6):34--43, 2009.
  \href{https://doi.org/10.1109/MCG.2009.117}
{doi: \textsf{%
10\hspace{.1pt}\discretionary{.}{%
}{.}\hspace{.4pt}1109\discretionary{/}{%
}{/}MCG\hspace{.1pt}\discretionary{.}{%
}{.}\hspace{.4pt}2009\hspace{.1pt}\discretionary{.}{%
}{.}\hspace{.4pt}117}}


\bibitem{Argelaguet:2013:ASO}
F.~Argelaguet and C.~Andujar.
\newblock A survey of {3D} object selection techniques for virtual
  environments.
\newblock {\em Comput Graph}, 37(3):121--136, 2013.
  \href{https://doi.org/10.1016/j.cag.2012.12.003}
{doi: \textsf{%
10\hspace{.1pt}\discretionary{.}{%
}{.}\hspace{.4pt}1016\discretionary{/}{%
}{/}j\hspace{.1pt}\discretionary{.}{%
}{.}\hspace{.4pt}cag\hspace{.1pt}\discretionary{.}{%
}{.}\hspace{.4pt}2012\hspace{.1pt}\discretionary{.}{%
}{.}\hspace{.4pt}12\hspace{.1pt}\discretionary{.}{%
}{.}\hspace{.4pt}003}}


\bibitem{Baguley:2009:ES}
T.~Baguley.
\newblock Standardized or simple effect size: What should be reported?
\newblock {\em Br J Psychol}, 100(3):603--617, Aug. 2009.
  \href{https://doi.org/10.1348/000712608X377117}
{doi: \textsf{%
10\hspace{.1pt}\discretionary{.}{%
}{.}\hspace{.4pt}1348\discretionary{/}{%
}{/}000712608X377117}}


\bibitem{baloup:2019:raycursor}
M.~Baloup, T.~Pietrzak, and G.~Casiez.
\newblock Raycursor: A {3D} pointing facilitation technique based on
  raycasting.
\newblock In {\em Proc.\ CHI}, pp. 101:1--101:12. ACM, New York, 2019.
  \href{https://doi.org/10.1145/3290605.3300331}
{doi: \textsf{%
10\hspace{.1pt}\discretionary{.}{%
}{.}\hspace{.4pt}1145\discretionary{/}{%
}{/}3290605\hspace{.1pt}\discretionary{.}{%
}{.}\hspace{.4pt}3300331}}


\bibitem{Jill:1994:TLAF}
J.~Bechtold.
\newblock The {L}yman-{A}lpha forest near 34 quasi-stellar objects with
  $z>2.6$.
\newblock {\em Astrophys J Suppl}, 91:1--78, 1994.
  \href{https://doi.org/10.1086/191937}
{doi: \textsf{%
10\hspace{.1pt}\discretionary{.}{%
}{.}\hspace{.4pt}1086\discretionary{/}{%
}{/}191937}}


\bibitem{besanccon:2021:state-of-art}
L.~Besan{\c{c}}on, A.~Ynnerman, D.~F. Keefe, L.~Yu, and T.~Isenberg.
\newblock The state of the art of spatial interfaces for {3D} visualization.
\newblock {\em Computer Graphics Forum}, 40(1):293--326, 2021.
  \href{https://doi.org/10.1111/cgf.14189}
{doi: \textsf{%
10\hspace{.1pt}\discretionary{.}{%
}{.}\hspace{.4pt}1111\discretionary{/}{%
}{/}cgf\hspace{.1pt}\discretionary{.}{%
}{.}\hspace{.4pt}14189}}


\bibitem{Besancon:2019:TangibleBrush}
L.~Besançon, M.~Sereno, L.~Yu, M.~Ammi, and T.~Isenberg.
\newblock Hybrid touch/tangible spatial {3D} data selection.
\newblock {\em Comput Graph Forum}, 38(3):553--567, 2019.
  \href{https://doi.org/10.1111/cgf.13710}
{doi: \textsf{%
10\hspace{.1pt}\discretionary{.}{%
}{.}\hspace{.4pt}1111\discretionary{/}{%
}{/}cgf\hspace{.1pt}\discretionary{.}{%
}{.}\hspace{.4pt}13710}}


\bibitem{Bond:1996:HFG}
J.~Bond, L.~Kofman, and D.~Pogosyan.
\newblock How filaments of galaxies are woven into the cosmic web.
\newblock {\em Nature}, 380(6575):603--606, 1996.
  \href{https://doi.org/10.1038/380603a0}
{doi: \textsf{%
10\hspace{.1pt}\discretionary{.}{%
}{.}\hspace{.4pt}1038\discretionary{/}{%
}{/}380603a0}}


\bibitem{Valve}
R.~Brown.
\newblock Valve index.
\newblock Web site:
  \href{https://vr-compare.com/headset/valveindex}{\texttt{{vr\discretionary{}{-}{-}compare\discretionary{.}{}{.}com\discretionary{/}{}{/}headset\discretionary{/}{}{/}valveindex}}}.
\newblock Last accessed: March 2023.

\bibitem{Nicholas:2016:SPP}
N.~Brunhart-Lupo, B.~W. Bush, K.~Gruchalla, and S.~Smith.
\newblock Simulation exploration through immersive parallel planes.
\newblock In {\em Workshop on Immersive Analytics}, pp. 19--24. IEEE Comp.\
  Soc., Los Alamitos, 2016.
  \href{https://doi.org/10.1109/IMMERSIVE.2016.7932377}
{doi: \textsf{%
10\hspace{.1pt}\discretionary{.}{%
}{.}\hspace{.4pt}1109\discretionary{/}{%
}{/}IMMERSIVE\hspace{.1pt}\discretionary{.}{%
}{.}\hspace{.4pt}2016\hspace{.1pt}\discretionary{.}{%
}{.}\hspace{.4pt}7932377}}


\bibitem{Chen:2009:ANI}
W.~Chen, Z.~Ding, S.~Zhang, A.~MacKay-Brandt, S.~Correia, H.~Qu, J.~A. Crow,
  D.~F. Tate, Z.~Yan, and Q.~Peng.
\newblock A novel interface for interactive exploration of {DTI} fibers.
\newblock {\em IEEE Trans Vis Comput Graph}, 15(6):1433--1440, 2009.
  \href{https://doi.org/10.1109/TVCG.2009.112}
{doi: \textsf{%
10\hspace{.1pt}\discretionary{.}{%
}{.}\hspace{.4pt}1109\discretionary{/}{%
}{/}TVCG\hspace{.1pt}\discretionary{.}{%
}{.}\hspace{.4pt}2009\hspace{.1pt}\discretionary{.}{%
}{.}\hspace{.4pt}112}}


\bibitem{Chen:2020:LassoNet}
Z.~Chen, W.~Zeng, Z.~Yang, L.~Yu, C.-W. Fu, and H.~Qu.
\newblock {LassoNet}: Deep lasso-selection of {3D} point clouds.
\newblock {\em IEEE Trans Vis Comput Graph}, 26(1):195--204, 2020.
  \href{https://doi.org/10.1109/TVCG.2019.2934332}
{doi: \textsf{%
10\hspace{.1pt}\discretionary{.}{%
}{.}\hspace{.4pt}1109\discretionary{/}{%
}{/}TVCG\hspace{.1pt}\discretionary{.}{%
}{.}\hspace{.4pt}2019\hspace{.1pt}\discretionary{.}{%
}{.}\hspace{.4pt}2934332}}


\bibitem{Cordeil:2020:EmbodiedAxes}
M.~Cordeil, B.~Bach, A.~Cunningham, B.~Montoya, R.~T. Smith, B.~H. Thomas, and
  T.~Dwyer.
\newblock Embodied axes: Tangible, actuated interaction for {3D} augmented
  reality data spaces.
\newblock In {\em Proc.\ CHI}, pp. 486:1--486:12. ACM, New York, 2020.
  \href{https://doi.org/10.1145/3313831.3376613}
{doi: \textsf{%
10\hspace{.1pt}\discretionary{.}{%
}{.}\hspace{.4pt}1145\discretionary{/}{%
}{/}3313831\hspace{.1pt}\discretionary{.}{%
}{.}\hspace{.4pt}3376613}}


\bibitem{Cordeil:2017:DSS}
M.~Cordeil, B.~Bach, Y.~Li, E.~Wilson, and T.~Dwyer.
\newblock Design space for spatio-data coordination: Tangible interaction
  devices for immersive information visualisation.
\newblock In {\em Proc.\ PacificVis}, pp. 46--50. IEEE Comp.\ Soc., Los
  Alamitos, 2017. \href{https://doi.org/10.1109/PACIFICVIS.2017.8031578}
{doi: \textsf{%
10\hspace{.1pt}\discretionary{.}{%
}{.}\hspace{.4pt}1109\discretionary{/}{%
}{/}PACIFICVIS\hspace{.1pt}\discretionary{.}{%
}{.}\hspace{.4pt}2017\hspace{.1pt}\discretionary{.}{%
}{.}\hspace{.4pt}8031578}}


\bibitem{Cumming:2013:NS}
G.~Cumming.
\newblock The new statistics: Why and how.
\newblock {\em Psychol Sci}, 25(1):7--29, Jan. 2014.
  \href{https://doi.org/10.1177/0956797613504966}
{doi: \textsf{%
10\hspace{.1pt}\discretionary{.}{%
}{.}\hspace{.4pt}1177\discretionary{/}{%
}{/}0956797613504966}}


\bibitem{danyluk:2020:TBC}
K.~Danyluk, T.~T. Ulusoy, W.~Wei, and W.~Willett.
\newblock Touch and beyond: Comparing physical and virtual reality
  visualizations.
\newblock {\em IEEE Trans Vis Comput Graph}, 28(4):1930--1940, 2022.
  \href{https://doi.org/10.1109/TVCG.2020.3023336}
{doi: \textsf{%
10\hspace{.1pt}\discretionary{.}{%
}{.}\hspace{.4pt}1109\discretionary{/}{%
}{/}TVCG\hspace{.1pt}\discretionary{.}{%
}{.}\hspace{.4pt}2020\hspace{.1pt}\discretionary{.}{%
}{.}\hspace{.4pt}3023336}}


\bibitem{DeHaan:2005:IntenSelect}
G.~de~Haan, M.~Koutek, and F.~H. Post.
\newblock {IntenSelect}: Using dynamic object rating for assisting {3D} object
  selection.
\newblock In {\em Proc.\ EGVE}, pp. 201--209. EG Assoc., Goslar, 2005.
  \href{https://doi.org/10.2312/EGVE/IPT_EGVE2005/201-209}
{doi: \textsf{%
10\hspace{.1pt}\discretionary{.}{%
}{.}\hspace{.4pt}2312\discretionary{/}{%
}{/}EGVE\discretionary{/}{%
}{/}IPT\_EGVE2005\discretionary{/}{%
}{/}201\discretionary{%
}{-}{-}209}}


\bibitem{Dragicevic:2016:FSC}
P.~Dragicevic.
\newblock Fair statistical communication in {HCI}.
\newblock In J.~Robertson and M.~Kaptein, eds., {\em Modern Statistical Methods
  for HCI}, chap.~13, pp. 291--330. Springer, Cham, 2016.
  \href{https://doi.org/10.1007/978-3-319-26633-6_13}
{doi: \textsf{%
10\hspace{.1pt}\discretionary{.}{%
}{.}\hspace{.4pt}1007\discretionary{/}{%
}{/}978\discretionary{%
}{-}{-}3\discretionary{%
}{-}{-}319\discretionary{%
}{-}{-}26633\discretionary{%
}{-}{-}6\_13}}


\bibitem{Dragicevic:2014:RAH}
P.~Dragicevic, F.~Chevalier, and S.~Huot.
\newblock Running an {HCI} experiment in multiple parallel universes.
\newblock In {\em CHI Extended Abstracts}, pp. 607--618. ACM, New York, 2014.
  \href{https://doi.org/10.1145/2559206.2578881}
{doi: \textsf{%
10\hspace{.1pt}\discretionary{.}{%
}{.}\hspace{.4pt}1145\discretionary{/}{%
}{/}2559206\hspace{.1pt}\discretionary{.}{%
}{.}\hspace{.4pt}2578881}}


\bibitem{Ferdosi:2011:DEM}
B.~Ferdosi, H.~Buddelmeijer, S.~Trager, M.~Wilkinson, and J.~Roerdink.
\newblock Comparison of density estimation methods for astronomical.
\newblock {\em Astron Astrophys}, 531:A114:1--A114:16, 2011.
  \href{https://doi.org/10.1051/0004-6361/201116878}
{doi: \textsf{%
10\hspace{.1pt}\discretionary{.}{%
}{.}\hspace{.4pt}1051\discretionary{/}{%
}{/}0004\discretionary{%
}{-}{-}6361\discretionary{/}{%
}{/}201116878}}


\bibitem{Franzluebbers:2022:VRP}
A.~Franzluebbers, C.~Li, A.~Paterson, and K.~Johnsen.
\newblock Virtual reality point cloud annotation.
\newblock In {\em Proc.\ SUI}, pp. 14:1--14:11. ACM, New York, 2022.
  \href{https://doi.org/10.1145/3565970.3567696}
{doi: \textsf{%
10\hspace{.1pt}\discretionary{.}{%
}{.}\hspace{.4pt}1145\discretionary{/}{%
}{/}3565970\hspace{.1pt}\discretionary{.}{%
}{.}\hspace{.4pt}3567696}}


\bibitem{gomez:2010:FBT}
S.~R. Gomez, R.~Jianu, and D.~H. Laidlaw.
\newblock A fiducial-based tangible user interface for white matter
  tractography.
\newblock In {\em Proc.\ ISVC}, vol.~2, pp. 373--381. Springer, Berlin, 2010.
  \href{https://doi.org/10.1007/978-3-642-17274-8_37}
{doi: \textsf{%
10\hspace{.1pt}\discretionary{.}{%
}{.}\hspace{.4pt}1007\discretionary{/}{%
}{/}978\discretionary{%
}{-}{-}3\discretionary{%
}{-}{-}642\discretionary{%
}{-}{-}17274\discretionary{%
}{-}{-}8\_37}}


\bibitem{Grossman:2006:DES}
T.~Grossman and R.~Balakrishnan.
\newblock The design and evaluation of selection techniques for {3D} volumetric
  displays.
\newblock In {\em Proc.\ UIST}, pp. 3--12. ACM, New York, 2006.
  \href{https://doi.org/10.1145/1166253.1166257}
{doi: \textsf{%
10\hspace{.1pt}\discretionary{.}{%
}{.}\hspace{.4pt}1145\discretionary{/}{%
}{/}1166253\hspace{.1pt}\discretionary{.}{%
}{.}\hspace{.4pt}1166257}}


\bibitem{Martin:2001:morse}
M.~Guest.
\newblock Morse theory in the 1990's.
\newblock arXiv preprint math/0104155, 2001.
  \href{https://doi.org/10.48550/arXiv.math/0104155}
{doi: \textsf{%
10\hspace{.1pt}\discretionary{.}{%
}{.}\hspace{.4pt}48550\discretionary{/}{%
}{/}arXiv\hspace{.1pt}\discretionary{.}{%
}{.}\hspace{.4pt}math\discretionary{/}{%
}{/}0104155}}


\bibitem{TLXScale}
S.~Hart.
\newblock Nasa-task load index ({NASA-TLX}); 20 years later.
\newblock {\em Proc Hum Factors Ergon Soc Annu Meet}, 50(9):904--908, 2006.
  \href{https://doi.org/10.1177/154193120605000909}
{doi: \textsf{%
10\hspace{.1pt}\discretionary{.}{%
}{.}\hspace{.4pt}1177\discretionary{/}{%
}{/}154193120605000909}}


\bibitem{Bernd:2009:MVT}
B.~Hentschel, M.~Wolter, and T.~Kuhlen.
\newblock Virtual reality-based multi-view visualization of time-dependent
  simulation data.
\newblock In {\em Proc.\ VR}, pp. 253--254. IEEE Comp.\ Soc., Los Alamitos,
  2009. \href{https://doi.org/10.1109/VR.2009.4811041}
{doi: \textsf{%
10\hspace{.1pt}\discretionary{.}{%
}{.}\hspace{.4pt}1109\discretionary{/}{%
}{/}VR\hspace{.1pt}\discretionary{.}{%
}{.}\hspace{.4pt}2009\hspace{.1pt}\discretionary{.}{%
}{.}\hspace{.4pt}4811041}}


\bibitem{Hong:2021:DET}
J.~Hong, F.~Argelaguet, A.~Trubuil, and T.~Isenberg.
\newblock Design and evaluation of three selection techniques for tightly
  packed {3D} objects in cell lineage specification in botany.
\newblock In {\em Proc.\ GI}, pp. 213--223. CHCCS, Mississauga, 2021.
  \href{https://doi.org/10.20380/GI2021.33}
{doi: \textsf{%
10\hspace{.1pt}\discretionary{.}{%
}{.}\hspace{.4pt}20380\discretionary{/}{%
}{/}GI2021\hspace{.1pt}\discretionary{.}{%
}{.}\hspace{.4pt}33}}


\bibitem{Christophe:2018:FCS}
C.~Hurter, N.~H. Riche, S.~M. Drucker, M.~Cordeil, R.~Alligier, and
  R.~Vuillemot.
\newblock {FiberClay}: Sculpting three dimensional trajectories to reveal
  structural insights.
\newblock {\em IEEE Trans Vis Comput Graph}, 25(1):704--714, 2019.
  \href{https://doi.org/10.1109/TVCG.2018.2865191}
{doi: \textsf{%
10\hspace{.1pt}\discretionary{.}{%
}{.}\hspace{.4pt}1109\discretionary{/}{%
}{/}TVCG\hspace{.1pt}\discretionary{.}{%
}{.}\hspace{.4pt}2018\hspace{.1pt}\discretionary{.}{%
}{.}\hspace{.4pt}2865191}}


\bibitem{jackson:2013:LTI}
B.~Jackson, T.~Y. Lau, D.~Schroeder, K.~C. Toussaint, and D.~F. Keefe.
\newblock A lightweight tangible {3D} interface for interactive visualization
  of thin fiber structures.
\newblock {\em IEEE Trans Vis Comput Graph}, 19(12):2802--2809, 2013.
  \href{https://doi.org/10.1109/TVCG.2013.121}
{doi: \textsf{%
10\hspace{.1pt}\discretionary{.}{%
}{.}\hspace{.4pt}1109\discretionary{/}{%
}{/}TVCG\hspace{.1pt}\discretionary{.}{%
}{.}\hspace{.4pt}2013\hspace{.1pt}\discretionary{.}{%
}{.}\hspace{.4pt}121}}


\bibitem{Keefe:2008:DDIT}
D.~F. Keefe, R.~C. Zeleznik, and D.~H. Laidlaw.
\newblock Tech-note: Dynamic dragging for input of {3D} trajectories.
\newblock {\em Proc.\ 3DUI}, pp. 51--54, 2008.
  \href{https://doi.org/10.1109/3DUI.2008.4476591}
{doi: \textsf{%
10\hspace{.1pt}\discretionary{.}{%
}{.}\hspace{.4pt}1109\discretionary{/}{%
}{/}3DUI\hspace{.1pt}\discretionary{.}{%
}{.}\hspace{.4pt}2008\hspace{.1pt}\discretionary{.}{%
}{.}\hspace{.4pt}4476591}}


\bibitem{konig:2009:APD}
W.~A. K{\"o}nig, J.~Gerken, S.~Dierdorf, and H.~Reiterer.
\newblock Adaptive pointing---{D}esign and evaluation of a precision enhancing
  technique for absolute pointing devices.
\newblock In {\em Proc.\ INTERACT}, pp. 658--671. Springer, Berlin, 2009.
  \href{https://doi.org/10.1007/978-3-642-03655-2_73}
{doi: \textsf{%
10\hspace{.1pt}\discretionary{.}{%
}{.}\hspace{.4pt}1007\discretionary{/}{%
}{/}978\discretionary{%
}{-}{-}3\discretionary{%
}{-}{-}642\discretionary{%
}{-}{-}03655\discretionary{%
}{-}{-}2\_73}}


\bibitem{Kopper:2011:RapidAA}
R.~Kopper, F.~Bacim, and D.~A. Bowman.
\newblock Rapid and accurate {3D} selection by progressive refinement.
\newblock In {\em Proc.\ 3DUI}, pp. 67--74. IEEE Comp.\ Soc., Los Alamitos,
  2011. \href{https://doi.org/10.1109/3DUI.2011.5759219}
{doi: \textsf{%
10\hspace{.1pt}\discretionary{.}{%
}{.}\hspace{.4pt}1109\discretionary{/}{%
}{/}3DUI\hspace{.1pt}\discretionary{.}{%
}{.}\hspace{.4pt}2011\hspace{.1pt}\discretionary{.}{%
}{.}\hspace{.4pt}5759219}}


\bibitem{Kraus:2020:TII}
M.~Kraus, N.~Weiler, D.~Oelke, J.~Kehrer, D.~A. Keim, and J.~Fuchs.
\newblock The impact of immersion on cluster identification tasks.
\newblock {\em IEEE Trans Vis Comput Graph}, 26(1):525--535, 2020.
  \href{https://doi.org/10.1109/TVCG.2019.2934395}
{doi: \textsf{%
10\hspace{.1pt}\discretionary{.}{%
}{.}\hspace{.4pt}1109\discretionary{/}{%
}{/}TVCG\hspace{.1pt}\discretionary{.}{%
}{.}\hspace{.4pt}2019\hspace{.1pt}\discretionary{.}{%
}{.}\hspace{.4pt}2934395}}


\bibitem{Lee:2003:TICV}
S.~Lee, J.~Seo, G.~J. Kim, and C.-M. Park.
\newblock Evaluation of pointing techniques for ray casting selection in
  virtual environments.
\newblock In {\em Proc.\ SPIE}, vol. 4756, pp. 38--44. SPIE, Bellingham, 2003.
  \href{https://doi.org/10.1117/12.497665}
{doi: \textsf{%
10\hspace{.1pt}\discretionary{.}{%
}{.}\hspace{.4pt}1117\discretionary{/}{%
}{/}12\hspace{.1pt}\discretionary{.}{%
}{.}\hspace{.4pt}497665}}


\bibitem{Lorensen:1987:MCH}
W.~E. Lorensen and H.~E. Cline.
\newblock Marching {C}ubes: A high resolution {3D} surface construction
  algorithm.
\newblock {\em ACM SIGGRAPH Comput Graph}, 21(4):163--169, 1987.
  \href{https://doi.org/10.1145/37402.37422}
{doi: \textsf{%
10\hspace{.1pt}\discretionary{.}{%
}{.}\hspace{.4pt}1145\discretionary{/}{%
}{/}37402\hspace{.1pt}\discretionary{.}{%
}{.}\hspace{.4pt}37422}}


\bibitem{Lubos:2014:TBP}
P.~Lubos, R.~Beimler, M.~Lammers, and F.~Steinicke.
\newblock Touching the cloud: Bimanual annotation of immersive point clouds.
\newblock In {\em Proc.\ 3DUI}, pp. 191--192. IEEE Comp.\ Soc., Los Alamitos,
  2014. \href{https://doi.org/10.1109/3DUI.2014.6798885}
{doi: \textsf{%
10\hspace{.1pt}\discretionary{.}{%
}{.}\hspace{.4pt}1109\discretionary{/}{%
}{/}3DUI\hspace{.1pt}\discretionary{.}{%
}{.}\hspace{.4pt}2014\hspace{.1pt}\discretionary{.}{%
}{.}\hspace{.4pt}6798885}}


\bibitem{Lucas:2005:DE3}
J.~F. Lucas.
\newblock Design and evaluation of {3D} multiple object selection techniques.
\newblock Master's thesis, Virginia Polytechnic Institute and State University,
  USA, 2005.
\newblock \href{http://hdl.handle.net/10919/31769}{hdl: \textsf{10919/31769}}.

\bibitem{Lucas:2005:DEM}
J.~F. Lucas, D.~A. Bowman, J.~Chen, and C.~A. Wingrave.
\newblock Design and evaluation of {3D} multiple object selection techniques.
\newblock Report, Virginia Polytechnic Institute and State University, USA,
  2005.
\newblock \href{https://www.researchgate.net/publication/228764561}{url:
  \texttt{researchgate\discretionary{}{.}{.}net\discretionary{/}{}{/}publication\discretionary{/}{}{/}228764561}}.

\bibitem{Jonathan:2001:VOS}
J.~I. Maletic, J.~Leigh, A.~Marcus, and G.~Dunlap.
\newblock Visualizing object-oriented software in virtual reality.
\newblock In {\em Proc.\ IWPC}, pp. 26--35. IEEE Comp.\ Soc., Los Alamitos,
  2001. \href{https://doi.org/10.1109/WPC.2001.921711}
{doi: \textsf{%
10\hspace{.1pt}\discretionary{.}{%
}{.}\hspace{.4pt}1109\discretionary{/}{%
}{/}WPC\hspace{.1pt}\discretionary{.}{%
}{.}\hspace{.4pt}2001\hspace{.1pt}\discretionary{.}{%
}{.}\hspace{.4pt}921711}}


\bibitem{malmberg:2006:3DLW}
F.~Malmberg, E.~Vidholm, and I.~Nystr{\"o}m.
\newblock A {3D} live-wire segmentation method for volume images using haptic
  interaction.
\newblock In {\em Proc.\ DGCI}, pp. 663--673. Springer, Berlin, 2006.
  \href{https://doi.org/10.1007/11907350_56}
{doi: \textsf{%
10\hspace{.1pt}\discretionary{.}{%
}{.}\hspace{.4pt}1007\discretionary{/}{%
}{/}11907350\_56}}


\bibitem{Maslych:2023:TIA}
M.~Maslych, Y.~Hmaiti, R.~Ghamandi, P.~Leber, R.~K. Kattoju, J.~Belga, and
  J.~J. LaViola.
\newblock Toward intuitive acquisition of occluded {VR} objects through an
  interactive disocclusion mini-map.
\newblock In {\em Proc.\ VR}, pp. 460--470. IEEE Comp.\ Soc., Los Alamitos,
  2023. \href{https://doi.org/10.1109/VR55154.2023.00061}
{doi: \textsf{%
10\hspace{.1pt}\discretionary{.}{%
}{.}\hspace{.4pt}1109\discretionary{/}{%
}{/}VR55154\hspace{.1pt}\discretionary{.}{%
}{.}\hspace{.4pt}2023\hspace{.1pt}\discretionary{.}{%
}{.}\hspace{.4pt}00061}}


\bibitem{mcdonald:2020:IUV}
T.~McDonald, W.~Usher, N.~Morrical, A.~Gyulassy, S.~Petruzza, F.~Federer,
  A.~Angelucci, and V.~Pascucci.
\newblock Improving the usability of virtual reality neuron tracing with
  topological elements.
\newblock {\em IEEE Trans Vis Comput Graph}, 27(2):744--754, 2021.
  \href{https://doi.org/10.1109/TVCG.2020.3030363}
{doi: \textsf{%
10\hspace{.1pt}\discretionary{.}{%
}{.}\hspace{.4pt}1109\discretionary{/}{%
}{/}TVCG\hspace{.1pt}\discretionary{.}{%
}{.}\hspace{.4pt}2020\hspace{.1pt}\discretionary{.}{%
}{.}\hspace{.4pt}3030363}}


\bibitem{Roberto:2020:HMS}
R.~A. Montano-Murillo, C.~Nguyen, R.~H. Kazi, S.~Subramanian, S.~DiVerdi, and
  D.~Martinez-Plasencia.
\newblock Slicing-{V}olume: Hybrid {3D}\discretionary{/}{}{/}{2D} multi-target
  selection technique for dense virtual environments.
\newblock In {\em Proc.\ VR}, pp. 53--62. IEEE Comp.\ Soc., Los Alamitos, 2020.
  \href{https://doi.org/10.1109/VR46266.2020.00023}
{doi: \textsf{%
10\hspace{.1pt}\discretionary{.}{%
}{.}\hspace{.4pt}1109\discretionary{/}{%
}{/}VR46266\hspace{.1pt}\discretionary{.}{%
}{.}\hspace{.4pt}2020\hspace{.1pt}\discretionary{.}{%
}{.}\hspace{.4pt}00023}}


\bibitem{Owada:2005:VC}
S.~Owada, F.~Nielsen, and T.~Igarashi.
\newblock Volume catcher.
\newblock In {\em Proc.\ I3D}, pp. 111--116. ACM, New York, 2005.
  \href{https://doi.org/10.1145/1053427.1053445}
{doi: \textsf{%
10\hspace{.1pt}\discretionary{.}{%
}{.}\hspace{.4pt}1145\discretionary{/}{%
}{/}1053427\hspace{.1pt}\discretionary{.}{%
}{.}\hspace{.4pt}1053445}}


\bibitem{Pierce:1997:IPI}
J.~S. Pierce, A.~S. Forsberg, M.~J. Conway, S.~Hong, R.~C. Zeleznik, and M.~R.
  Mine.
\newblock Image plane interaction techniques in {3D} immersive environments.
\newblock In {\em Proc.\ I3D}, pp. 39--43. ACM, New York, 1997.
  \href{https://doi.org/10.1145/253284.253303}
{doi: \textsf{%
10\hspace{.1pt}\discretionary{.}{%
}{.}\hspace{.4pt}1145\discretionary{/}{%
}{/}253284\hspace{.1pt}\discretionary{.}{%
}{.}\hspace{.4pt}253303}}


\bibitem{Pivovar:2022:VRS}
J.~Pivovar, J.~DeGuzman, and E.~S. Rosenberg.
\newblock Virtual reality on a {SWIM}: Scalable world in miniature.
\newblock In {\em Proc.\ VR: Abstracts and Workshops}, pp. 912--913. IEEE
  Comp.\ Soc., Los Alamitos, 2022.
  \href{https://doi.org/10.1109/VRW55335.2022.00307}
{doi: \textsf{%
10\hspace{.1pt}\discretionary{.}{%
}{.}\hspace{.4pt}1109\discretionary{/}{%
}{/}VRW55335\hspace{.1pt}\discretionary{.}{%
}{.}\hspace{.4pt}2022\hspace{.1pt}\discretionary{.}{%
}{.}\hspace{.4pt}00307}}


\bibitem{Poupyrev:1999:ManipulatingOI}
I.~Poupyrev and T.~Ichikawa.
\newblock Manipulating objects in virtual worlds: Categorization and empirical
  evaluation of interaction techniques.
\newblock {\em J Vis Lang Comput}, 10(1):19--35, 1999.
  \href{https://doi.org/10.1006/jvlc.1998.0112}
{doi: \textsf{%
10\hspace{.1pt}\discretionary{.}{%
}{.}\hspace{.4pt}1006\discretionary{/}{%
}{/}jvlc\hspace{.1pt}\discretionary{.}{%
}{.}\hspace{.4pt}1998\hspace{.1pt}\discretionary{.}{%
}{.}\hspace{.4pt}0112}}


\bibitem{sereno:2022:HTTinAR}
M.~Sereno, S.~Gosset, L.~Besan{\c{c}}on, and T.~Isenberg.
\newblock Hybrid touch/tangible spatial selection in augmented reality.
\newblock {\em Comput Graph Forum}, 41(3):403--415, 2022.
  \href{https://doi.org/10.1111/cgf.14550}
{doi: \textsf{%
10\hspace{.1pt}\discretionary{.}{%
}{.}\hspace{.4pt}1111\discretionary{/}{%
}{/}cgf\hspace{.1pt}\discretionary{.}{%
}{.}\hspace{.4pt}14550}}


\bibitem{Shneiderman:1996:ETD}
B.~Shneiderman.
\newblock The eyes have it: A task by data type taxonomy for information
  visualizations.
\newblock In {\em Proc.\ VL}, pp. 336--343. IEEE Comp.\ Soc., Los Alamitos,
  1996. \href{https://doi.org/10.1109/VL.1996.545307}
{doi: \textsf{%
10\hspace{.1pt}\discretionary{.}{%
}{.}\hspace{.4pt}1109\discretionary{/}{%
}{/}VL\hspace{.1pt}\discretionary{.}{%
}{.}\hspace{.4pt}1996\hspace{.1pt}\discretionary{.}{%
}{.}\hspace{.4pt}545307}}


\bibitem{springel:2008:aquarius}
V.~Springel, J.~Wang, M.~Vogelsberger, A.~Ludlow, A.~Jenkins, A.~Helmi, J.~F.
  Navarro, C.~S. Frenk, and S.~D. White.
\newblock Aquarius project: The subhaloes of galactic haloes.
\newblock {\em Mon Not R Astron Soc}, 391(4):1685--1711, 2008.
  \href{https://doi.org/10.1111/j.1365-2966.2008.14066.x}
{doi: \textsf{%
10\hspace{.1pt}\discretionary{.}{%
}{.}\hspace{.4pt}1111\discretionary{/}{%
}{/}j\hspace{.1pt}\discretionary{.}{%
}{.}\hspace{.4pt}1365\discretionary{%
}{-}{-}2966\hspace{.1pt}\discretionary{.}{%
}{.}\hspace{.4pt}2008\hspace{.1pt}\discretionary{.}{%
}{.}\hspace{.4pt}14066\hspace{.1pt}\discretionary{.}{%
}{.}\hspace{.4pt}x}}


\bibitem{Springel:2005:SimulationsOT}
V.~Springel, S.~D.~M. White, A.~Jenkins, C.~S. Frenk, N.~Yoshida, L.~Gao, J.~F.
  Navarro, R.~J. Thacker, D.~J. Croton, J.~C. Helly, J.~A. Peacock, S.~Cole,
  P.~A. Thomas, H.~M.~P. Couchman, A.~E. Evrard, J.~M. Colberg, and F.~R.
  Pearce.
\newblock Simulations of the formation, evolution and clustering of galaxies
  and quasars.
\newblock {\em Nature}, 435:629--636, 2005.
  \href{https://doi.org/10.1038/nature03597}
{doi: \textsf{%
10\hspace{.1pt}\discretionary{.}{%
}{.}\hspace{.4pt}1038\discretionary{/}{%
}{/}nature03597}}


\bibitem{Stenholt:2012:SMOS}
R.~Stenholt.
\newblock Efficient selection of multiple objects on a large scale.
\newblock In {\em Proc.\ VRST}, pp. 105--112. ACM, New York, 2012.
  \href{https://doi.org/10.1145/2407336.2407357}
{doi: \textsf{%
10\hspace{.1pt}\discretionary{.}{%
}{.}\hspace{.4pt}1145\discretionary{/}{%
}{/}2407336\hspace{.1pt}\discretionary{.}{%
}{.}\hspace{.4pt}2407357}}


\bibitem{Stoakley:1995:WIM}
R.~Stoakley, M.~J. Conway, and R.~Pausch.
\newblock Virtual reality on a {WIM}: Interactive worlds in miniature.
\newblock In {\em Proc.\ CHI}, pp. 265--272. ACM, New York, 1995.
  \href{https://doi.org/10.1145/223904.223938}
{doi: \textsf{%
10\hspace{.1pt}\discretionary{.}{%
}{.}\hspace{.4pt}1145\discretionary{/}{%
}{/}223904\hspace{.1pt}\discretionary{.}{%
}{.}\hspace{.4pt}223938}}


\bibitem{Tietjen:2008:METKT}
C.~Tietjen, K.~M{\"u}hler, F.~Ritter, O.~Konrad, M.~Hindennach, and B.~Preim.
\newblock {METK} -- {T}he medical exploration toolkit.
\newblock In {\em Bildverarbeitung f{\"u}r die Medizin}, pp. 407--411.
  Springer, Berlin, 2008. \href{https://doi.org/10.1007/978-3-540-78640-5_82}
{doi: \textsf{%
10\hspace{.1pt}\discretionary{.}{%
}{.}\hspace{.4pt}1007\discretionary{/}{%
}{/}978\discretionary{%
}{-}{-}3\discretionary{%
}{-}{-}540\discretionary{%
}{-}{-}78640\discretionary{%
}{-}{-}5\_82}}


\bibitem{Teylingen:1997:VDV}
R.~van Teylingen, W.~Ribarsky, and C.~van~der Mast.
\newblock Virtual data visualizer.
\newblock {\em IEEE Trans Vis Comput Graph}, 3(1):65--74, 1997.
  \href{https://doi.org/10.1109/2945.582350}
{doi: \textsf{%
10\hspace{.1pt}\discretionary{.}{%
}{.}\hspace{.4pt}1109\discretionary{/}{%
}{/}2945\hspace{.1pt}\discretionary{.}{%
}{.}\hspace{.4pt}582350}}


\bibitem{VandenBos:2009:PMAPA}
G.~R. VandenBos, ed.
\newblock {\em Publication Manual of the American Psychological Association}.
\newblock APA, Washington, DC, 6\textsuperscript{th} ed., 2009.
\newblock url:
  \href{http://www.apastyle.org/manual/}{\texttt{apastyle\discretionary{}{.}{.}org\discretionary{/}{}{/}manual}}.

\bibitem{Wei:2023:PGT}
Y.~Wei, R.~Shi, D.~Yu, Y.~Wang, Y.~Li, L.~Yu, and H.-N. Liang.
\newblock Predicting gaze-based target selection in augmented reality headsets
  based on eye and head endpoint distributions.
\newblock In {\em Proc.\ CHI}, pp. 283:1--283:14. ACM, New York, 2023.
  \href{https://doi.org/10.1145/3544548.3581042}
{doi: \textsf{%
10\hspace{.1pt}\discretionary{.}{%
}{.}\hspace{.4pt}1145\discretionary{/}{%
}{/}3544548\hspace{.1pt}\discretionary{.}{%
}{.}\hspace{.4pt}3581042}}


\bibitem{Wiebel:2012:WYSIWYP}
A.~Wiebel, F.~M. Vos, D.~Foerster, and H.-C. Hege.
\newblock {WYSIWYP}: What you see is what you pick.
\newblock {\em IEEE Trans Vis Comput Graph}, 18(12):2236--2244, 2012.
  \href{https://doi.org/10.1109/TVCG.2012.292}
{doi: \textsf{%
10\hspace{.1pt}\discretionary{.}{%
}{.}\hspace{.4pt}1109\discretionary{/}{%
}{/}TVCG\hspace{.1pt}\discretionary{.}{%
}{.}\hspace{.4pt}2012\hspace{.1pt}\discretionary{.}{%
}{.}\hspace{.4pt}292}}


\bibitem{Wills:1996:S5W}
G.~J. Wills.
\newblock Selection: 524,288 ways to say ``this is interesting''.
\newblock In {\em Proc.\ InfoVis}, pp. 54--60. IEEE Comp.\ Soc., Los Alamitos,
  1996. \href{https://doi.org/10.1109/INFVIS.1996.559216}
{doi: \textsf{%
10\hspace{.1pt}\discretionary{.}{%
}{.}\hspace{.4pt}1109\discretionary{/}{%
}{/}INFVIS\hspace{.1pt}\discretionary{.}{%
}{.}\hspace{.4pt}1996\hspace{.1pt}\discretionary{.}{%
}{.}\hspace{.4pt}559216}}


\bibitem{Wingrave:2005:ERS}
C.~Wingrave, R.~Tintner, B.~Walker, D.~Bowman, and L.~Hodges.
\newblock Exploring individual differences in raybased selection: Strategies
  and traits.
\newblock In {\em Proc.\ VR}, pp. 163--170. IEEE Comp.\ Soc., Los Alamitos,
  2005. \href{https://doi.org/10.1109/VR.2005.1492770}
{doi: \textsf{%
10\hspace{.1pt}\discretionary{.}{%
}{.}\hspace{.4pt}1109\discretionary{/}{%
}{/}VR\hspace{.1pt}\discretionary{.}{%
}{.}\hspace{.4pt}2005\hspace{.1pt}\discretionary{.}{%
}{.}\hspace{.4pt}1492770}}


\bibitem{Wyvill:1986:DSS}
G.~Wyvill, C.~McPheeters, and B.~Wyvill.
\newblock Data structure for \emph{soft} objects.
\newblock {\em Vis Comput}, 2(4):227--234, 1986.
  \href{https://doi.org/10.1007/BF01900346}
{doi: \textsf{%
10\hspace{.1pt}\discretionary{.}{%
}{.}\hspace{.4pt}1007\discretionary{/}{%
}{/}BF01900346}}


\bibitem{xu:2012:lazy}
P.~Xu, H.~Fu, O.~K.-C. Au, and C.-L. Tai.
\newblock Lazy selection: A scribble-based tool for smart shape elements
  selection.
\newblock {\em ACM Trans Graph}, 31(6):142:1--142:9, 2012.
  \href{https://doi.org/10.1145/2366145.2366161}
{doi: \textsf{%
10\hspace{.1pt}\discretionary{.}{%
}{.}\hspace{.4pt}1145\discretionary{/}{%
}{/}2366145\hspace{.1pt}\discretionary{.}{%
}{.}\hspace{.4pt}2366161}}


\bibitem{Yu:2012:ESA}
L.~Yu, K.~Efstathiou, P.~Isenberg, and T.~Isenberg.
\newblock Efficient structure-aware selection techniques for {3D} point cloud
  visualizations with {2DOF} input.
\newblock {\em IEEE Trans Vis Comput Graph}, 18(12):2245--2254, 2012.
  \href{https://doi.org/10.1109/TVCG.2012.217}
{doi: \textsf{%
10\hspace{.1pt}\discretionary{.}{%
}{.}\hspace{.4pt}1109\discretionary{/}{%
}{/}TVCG\hspace{.1pt}\discretionary{.}{%
}{.}\hspace{.4pt}2012\hspace{.1pt}\discretionary{.}{%
}{.}\hspace{.4pt}217}}


\bibitem{Yu:2016:CEE}
L.~Yu, K.~Efstathiou, P.~Isenberg, and T.~Isenberg.
\newblock {CAST}: Effective and efficient user interaction for context-aware
  selection in {3D} particle clouds.
\newblock {\em IEEE Trans Vis Comput Graph}, 22(1):886--895, 2016.
  \href{https://doi.org/10.1109/TVCG.2015.2467202}
{doi: \textsf{%
10\hspace{.1pt}\discretionary{.}{%
}{.}\hspace{.4pt}1109\discretionary{/}{%
}{/}TVCG\hspace{.1pt}\discretionary{.}{%
}{.}\hspace{.4pt}2015\hspace{.1pt}\discretionary{.}{%
}{.}\hspace{.4pt}2467202}}


\bibitem{Zhao:2022:LWIM}
L.~Zhao, N.~Cao, S.~He, H.-N. Liang, and L.~Yu.
\newblock {L-WiM}: Collaborative exploration in immersive environments.
\newblock In {\em Proc.\ ISMAR-Adjunct}, pp. 118--123. IEEE Comp.\ Soc., Los
  Alamitos, 2022. \href{https://doi.org/10.1109/ISMAR-Adjunct57072.2022.00031}
{doi: \textsf{%
10\hspace{.1pt}\discretionary{.}{%
}{.}\hspace{.4pt}1109\discretionary{/}{%
}{/}ISMAR\discretionary{%
}{-}{-}Adjunct57072\hspace{.1pt}\discretionary{.}{%
}{.}\hspace{.4pt}2022\hspace{.1pt}\discretionary{.}{%
}{.}\hspace{.4pt}00031}}


\end{thebibliography}

\clearpage

\begin{strip}
\noindent\begin{minipage}{\textwidth}
\makeatletter
\centering%
\sffamily\bfseries\fontsize{15}{16.5}\selectfont
\vgtc@title\\[.5em]
\large Appendix\\[.75em]
\makeatother
\normalfont\rmfamily\normalsize\noindent\raggedright In this appendix we provide additional tables, plots, and charts that show data beyond the material that we could include in the main paper due to space limitations or because it was not essential for explaining our approach. For access to the source code, application, datasets, and analysis scripts used in this work, please refer to \href{https://osf.io/dvj9n/}{\texttt{osf.io/dvj9n}} and to \href{https://github.com/LixiangZhao98/MeTACAST}{\texttt{github.com/LixiangZhao98/MeTACAST}}.
\end{minipage}
\end{strip}

\appendix

\section{Additional Illustrations from the Elicitation Study}
\label{appendix-think-aloud}

\begin{figure}[ht]
  \centering
    \captionsetup[subfigure]{labelformat=empty}
      \includegraphics[width=1\columnwidth]{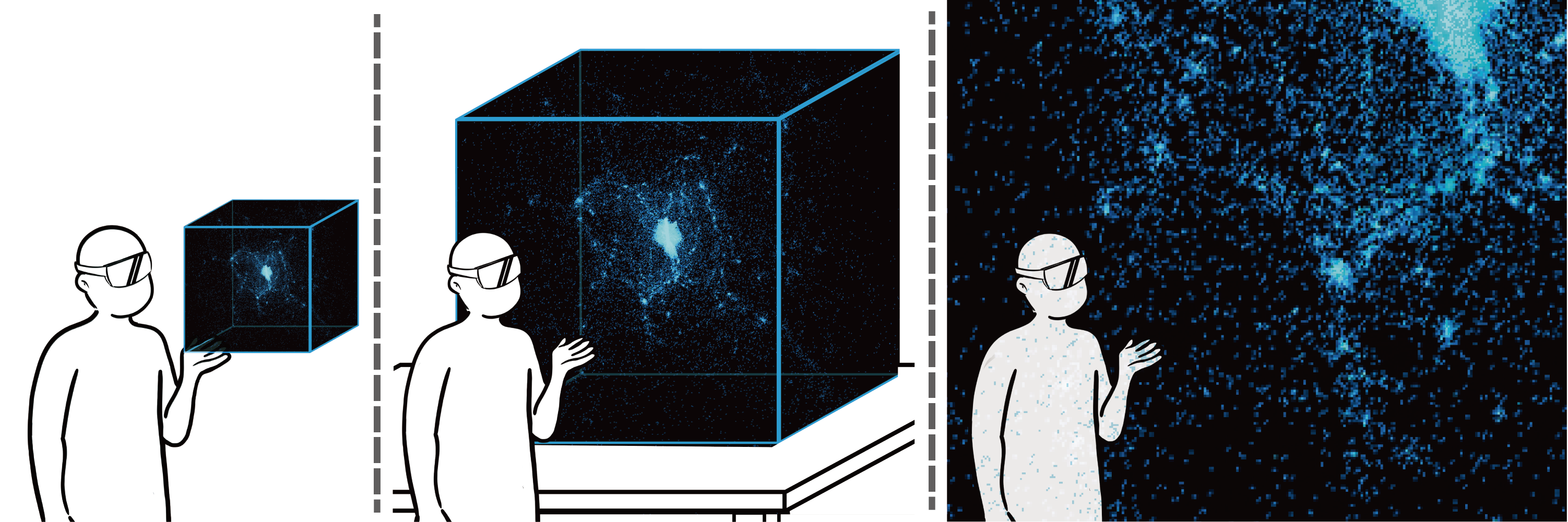}
 \caption{Three visualization sizes: (a) hand, (b) table, and (c) room size.}
 \label{fig:visualizationsize}
\end{figure}

\autoref{fig:visualizationsize} shows a visual representation of the visual presentation size concepts that we used in the study in \autoref{sec:Think-aloud Study}, while \autoref{fig:ThinkAloudData} shows the four datasets we used in this experiment.

\newlength{\imagewidth}
\setlength{\imagewidth}{0.47\columnwidth}
\begin{figure}[h]
    \centering
    \captionsetup[subfigure]{labelformat=empty} 
    \subfloat[\label{fig:cluster_pilot} ]
        {\begin{overpic}[width=\imagewidth]{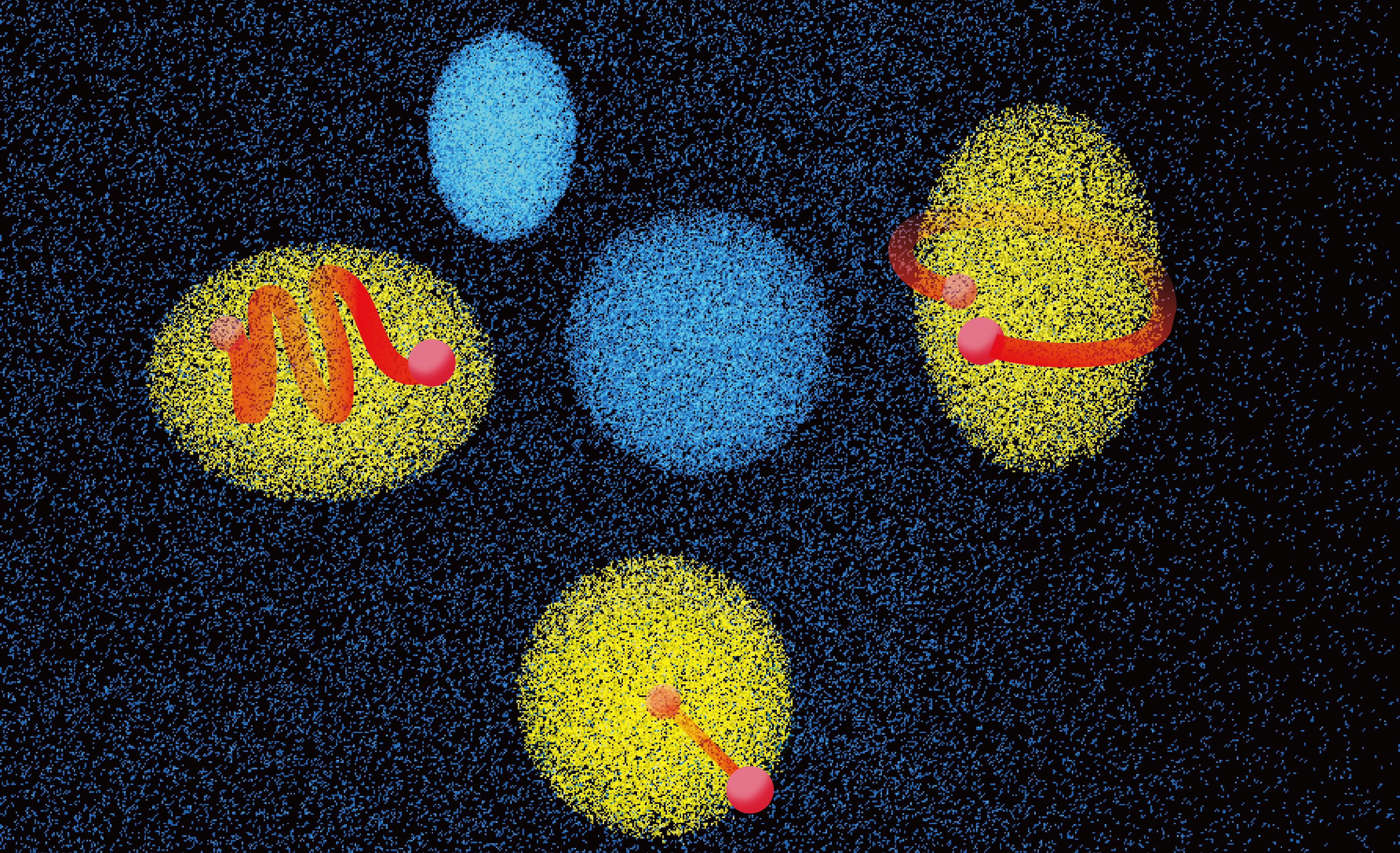}
            \put(5,5){\textcolor{white}{(a)}}
        \end{overpic}} \hspace{1 pt}
    \subfloat[\label{fig:N-body_pilot} ]
        {\begin{overpic}[width=\imagewidth]{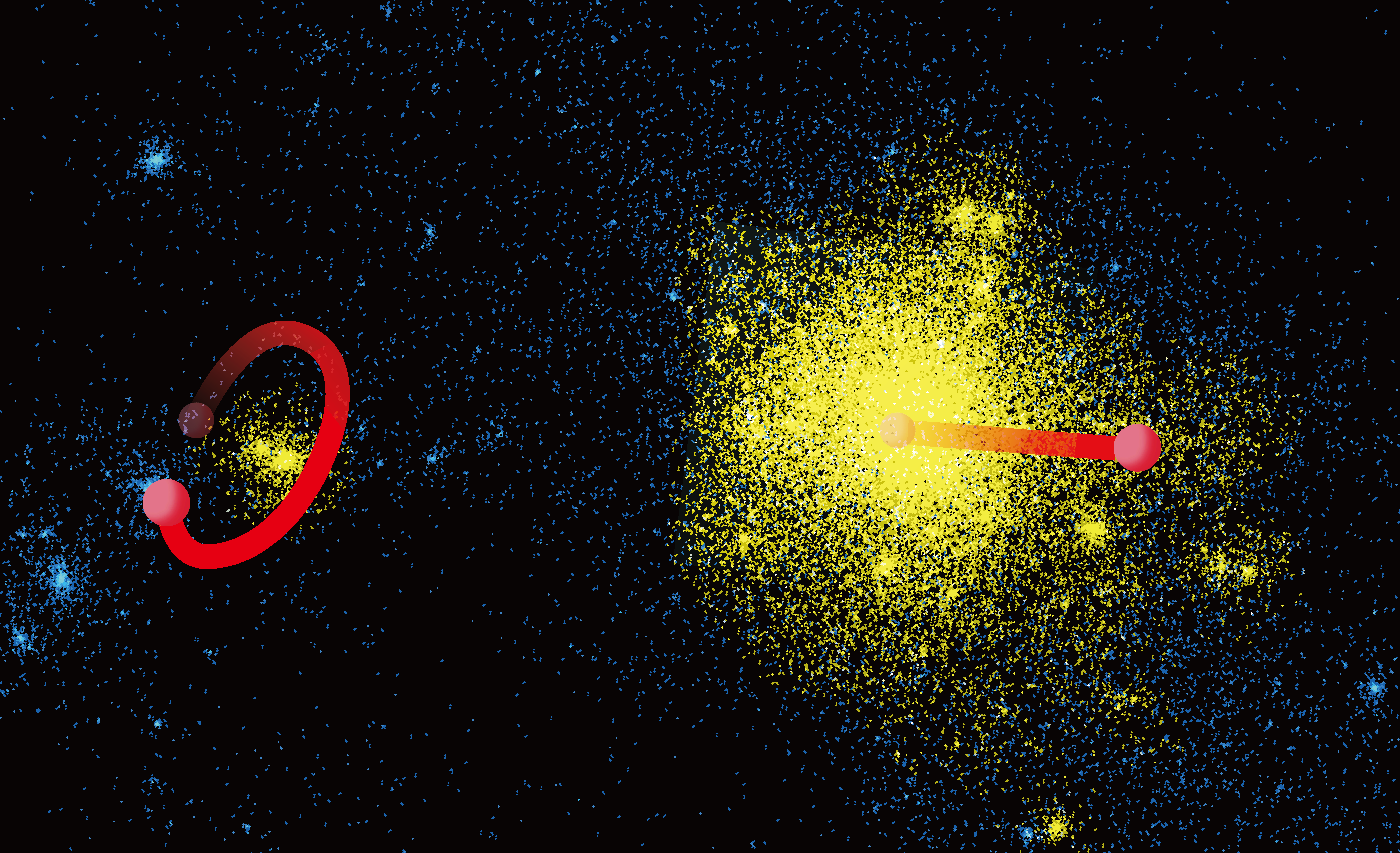}
              \put(5,5){\textcolor{white}{(b)}}
          \end{overpic}}\\[-3.5ex]
    \subfloat[\label{fig:filament_pilot} ]
        {\begin{overpic}[width=\imagewidth]{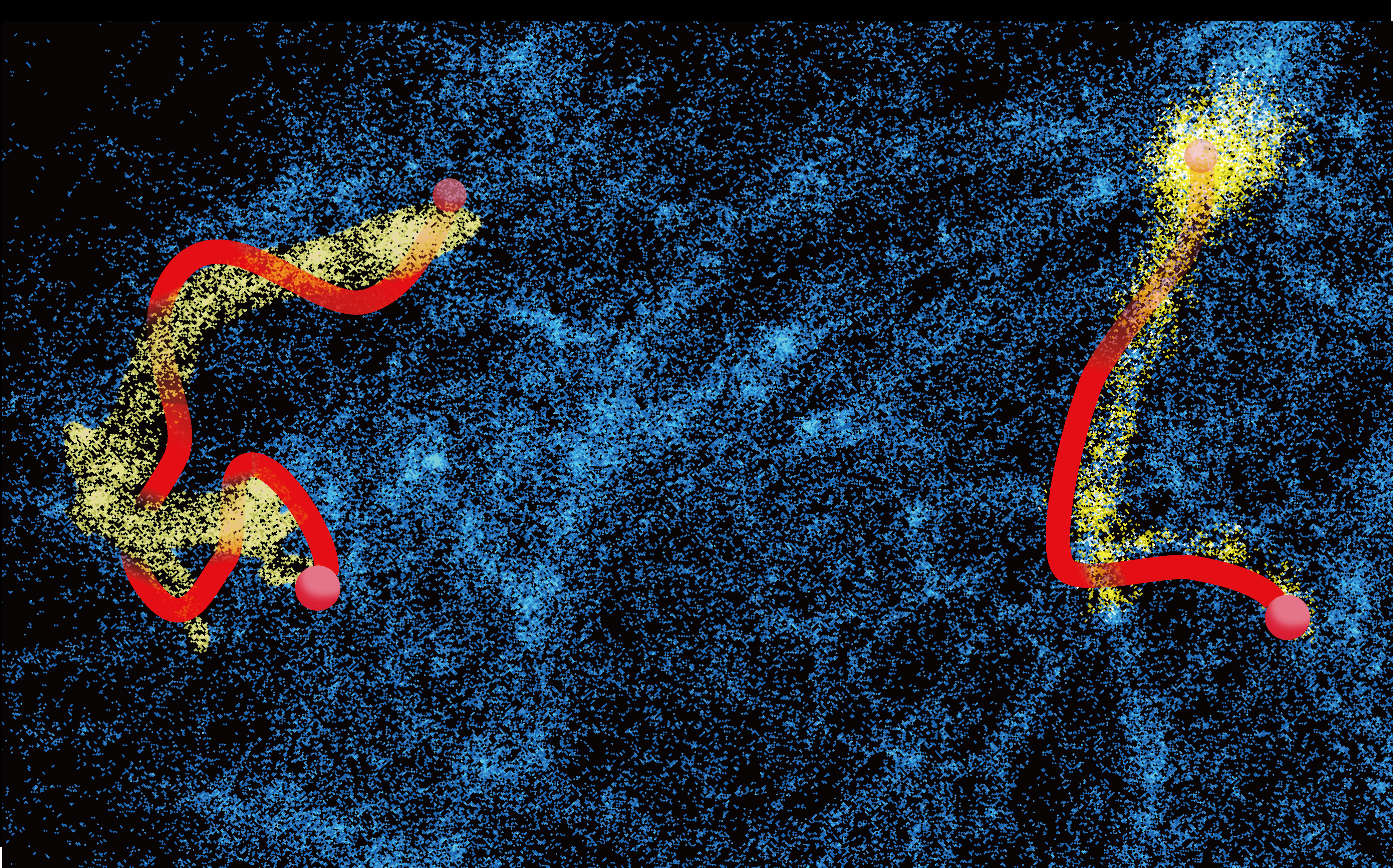}
              \put(5,5){\textcolor{white}{(c)}}
        \end{overpic}} \hspace{1 pt}
    \subfloat[\label{fig:shell_pilot} ]
        {\begin{overpic}[width=\imagewidth,trim={0 0 45px 0},clip]{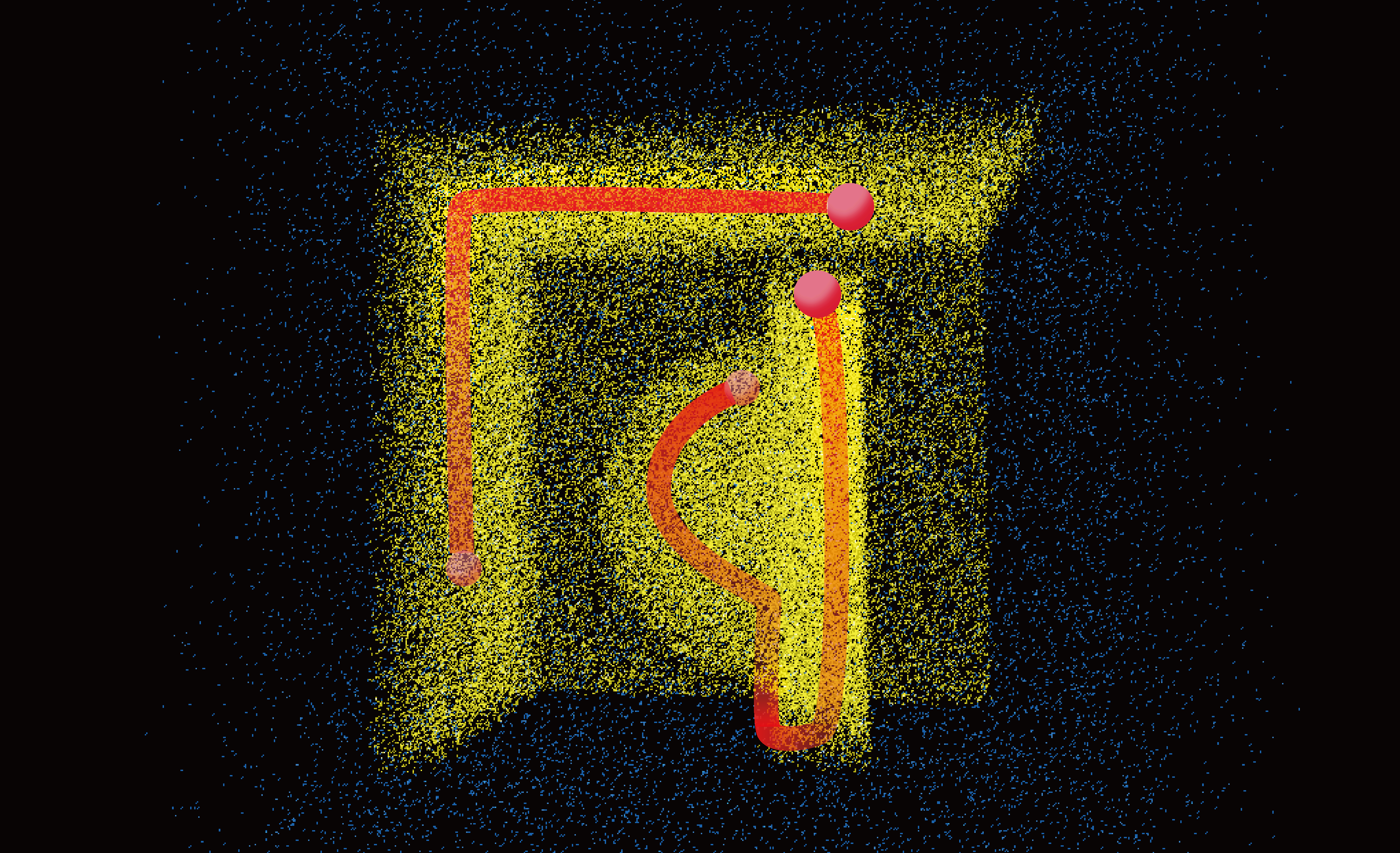}
              \put(5,5){\textcolor{white}{(d)}}
         \end{overpic}}
				\vspace{-4ex}
 \caption{User input trajectories in selecting four point cloud datasets employed in the think-aloud study: (a) \textit{Clusters}; (b) \textit{N-body Simulation}; (c) \textit{Filament}; (d) \textit{Complex geometries}.} 
 \label{fig:ThinkAloudData}
\end{figure}

\section{Density estimation}
\label{appendix-densityestimation}
\setlength{\lineskiplimitbackup}{\lineskiplimit}
\setlength{\lineskiplimit}{-\maxdimen}
We used the same density estimation method as Yu et al. \cite{Yu:2012:ESA,Yu:2016:CEE}. For efficiency reasons we implemented it on the GPU. We define a box $B$ covering the point cloud data and divide the space into a 100\,\texttimes\,100\,\texttimes\,100 grid. 
For each direction $k$ = $x$,$y$,$z$, we define the smoothing length
\setlength{\lineskiplimit}{\lineskiplimitbackup}
 \begin{equation}
     \ell_{k}=2({P}_{k}^{(80)}-{P}_{k}^{(20)}) / \log N
 \end{equation}
where $N$ is the particle count in the box $B$ and ${P}_{j}^{(q)}$ is coordinate $k$’s $q$-th percentile value. 
For the $i$\textsuperscript{th} grid-node at position $\mathbf{r}^{(i)}$, we compute the density  $\rho(\mathbf{r}^{(i)})$ using the modified Breiman kernel density estimation with a finite-support adaptive Epanechnikov kernel \cite{Ferdosi:2011:DEM}, given by
a
\setlength{\lineskiplimit}{\lineskiplimitbackup}
 \begin{equation}
     \rho(\mathbf{r}^{(i)})=\frac{15}{8 \pi N} \sum_{j} \frac{1}{\ell_{x}^{(j)} \ell_{y}^{(j)} \ell_{z}^{(j)}} E(\|\mathbf{r}^{(j;i)}\|),
 \end{equation}
with 
\begin{equation}
    \mathbf{r}_{k}^{(j ; i)}=(\mathbf{r}_{k}^{(j)}-\mathbf{r}_{k}^{(i)}) / \ell_{k}^{(j)}
\end{equation}
where $\mathbf{r}^{(j)}$ is the position of the $j$\textsuperscript{th} particle and $\ell_{k}^{(j)}$ are the smoothing lengths of the $j$\textsuperscript{th} particle along the $k$\textsuperscript{th} direction ($k=x,y,z$) generated from the pilot density calculation \cite{Yu:2012:ESA,Yu:2016:CEE}. The Epanechnikov kernel $E(x)$ is given by
 \begin{equation}
 E(x)=\left\{\begin{array}{ll}
 1-x^{2}, & |x| < 1, \\
 0, & |x| \ge 1. 
 \end{array}\right.
 \end{equation}

The density field of the data is pre-computed offline and the selection geometry construction with Marching Cubes is performed on GPU. 

\section{System Performance}
\label{appendix-system performance}

We have tested the performance of the MeTACAST with Unity on Intel Core\texttrademark{} i9, GeForce RTX3090. \autoref{tab:system performance} shows the performance of the MeTACAST, including point cloud dataset sizes, the performance of density estimation, selection algorithm and geometry construction.

\begin{table}[htbp]
    \caption{MeTACAST performance. Times are in seconds.}
    \label{tab:system performance} 
    \centering
    \scriptsize%
	\centering%
     \begin{tabular*}{\hsize}{l@{\extracolsep{\fill}}c c c c c}
    \toprule
        Dataset & \thead{Particle \\ Size}& \thead{ Density \\Estimation (offline)}& \thead{Selection \\Algorithm} &\thead{ Marching \\ Cubes}  \\ \midrule
       \autoref{fig:teaser}(left)& 76k & 0.38 &0.06 &0.005  \\ 
       \autoref{fig:teaser}(mid)& 442k &3.47 &1.69  &0.008   \\ 
       \autoref{fig:teaser}(right)& 146k & 0.80 &0.17  &0.009 \\ 
        \bottomrule
    \end{tabular*}
\end{table}

\section{Additional Result Data from the Study}
\label{appendix:B:}
\label{appendix-overallresults}
\label{appendix-userpreference}

\begin{figure}[t]
    \centering
    \captionsetup[subfigure]{labelformat=empty}
      \includegraphics[width=1\columnwidth]{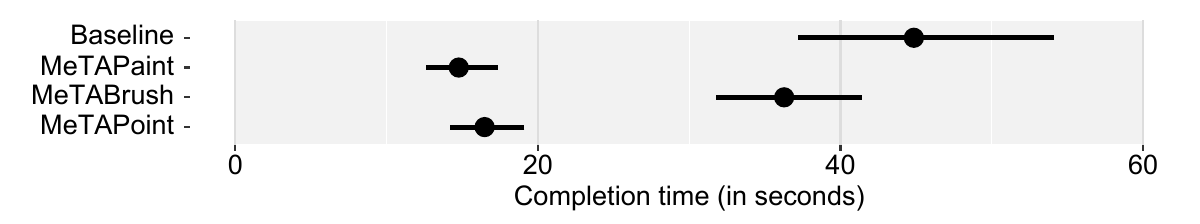}
 \caption{The overall (T1 to T4) geometric means of completion time for each selection technique. Error bars: 95\% CIs. }
 \label{fig:overall time}
\end{figure}

\begin{figure}[t]
    \centering
    \captionsetup[subfigure]{labelformat=empty}
      \includegraphics[width=1\columnwidth]{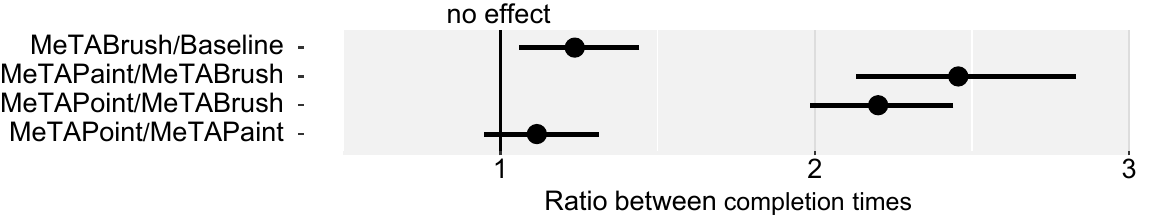}
 \caption{The overall (T1 to T4) pairwise ratio of completion time. Error bars show 95\% confidence intervals (CIs).}
 \label{fig:overall time ratio}
\end{figure}

\begin{figure}[t]
    \centering
    \captionsetup[subfigure]{labelformat=empty}
      \includegraphics[width=1\columnwidth]{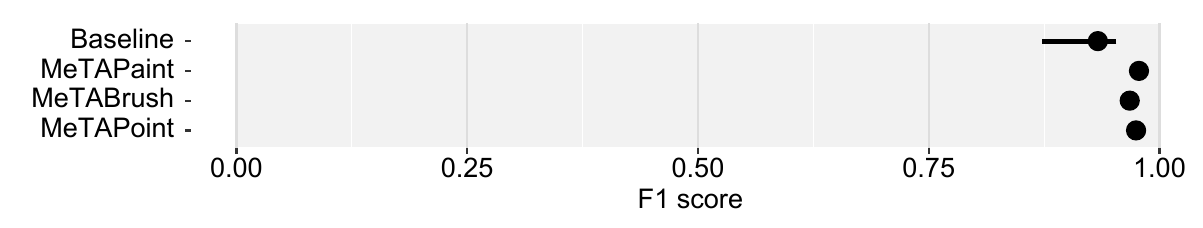}
 \caption{The overall (T1 to T4) F1 score. Error bars: 95\% CIs.}
 \label{fig:Overall F1 score}
\end{figure}

\begin{figure}[t]
    \centering
    \captionsetup[subfigure]{labelformat=empty}
      \includegraphics[width=1\columnwidth]{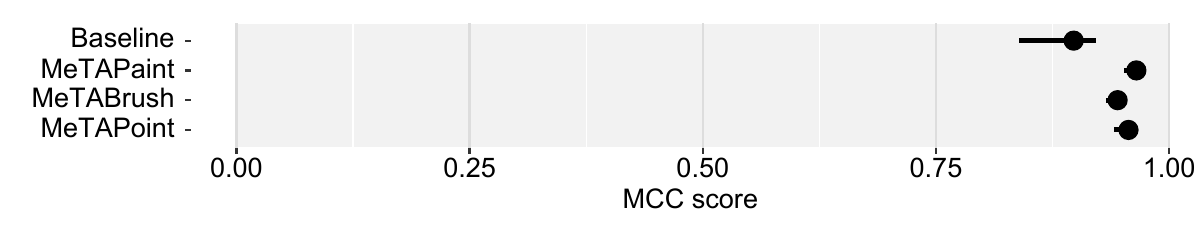}
 \caption{The overall (T1 to T4) MCC score. Error bars: 95\% CIs.}
 \label{fig:Overall MCC score}
\end{figure}

\begin{figure}[t]
    \centering
    \captionsetup[subfigure]{labelformat=empty}
      \includegraphics[width=1\columnwidth]{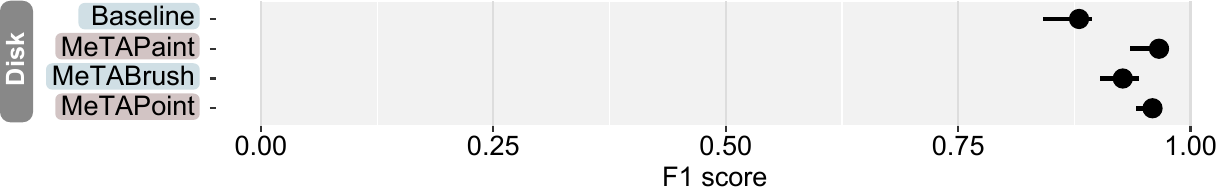}
 \caption{The F1 score for T1 (\textit{Disk}). Error bars: 95\% CIs.}
 \label{fig:F1 score for T1}
\end{figure}

\begin{figure}[t]
    \centering

      \includegraphics[width=1\columnwidth]{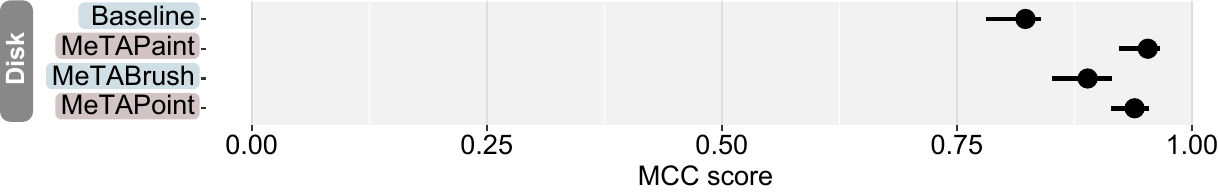}
 \caption{The MCC score for T1 (\textit{Disk}). Error bars: 95\% CIs.}
 \label{fig:MCC score for T1}
\end{figure}

\begin{figure}[t]
    \centering
    \captionsetup[subfigure]{labelformat=empty}
      \includegraphics[width=1\columnwidth]{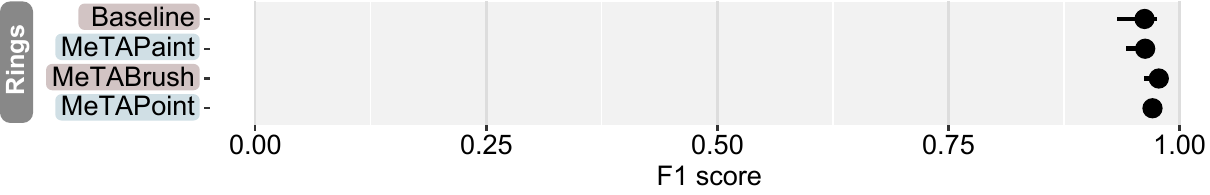}
 \caption{The F1 score for T2 (\textit{Rings}). Error bars: 95\% CIs.}
 \label{fig:F1 score for T2}
\end{figure}

\begin{figure}[t]
    \centering
    \captionsetup[subfigure]{labelformat=empty}
      \includegraphics[width=1\columnwidth]{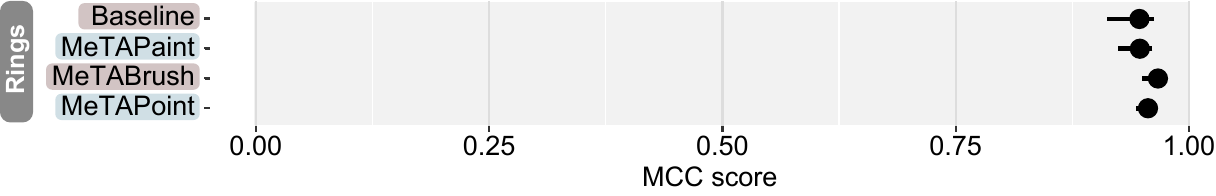}
 \caption{The MCC score for T2 (\textit{Rings}). Error bars: 95\% CIs.}
 \label{fig:MCC score for T2}
\end{figure}

\begin{figure}[t]
    \centering
    \captionsetup[subfigure]{labelformat=empty}
      \includegraphics[width=1\columnwidth]{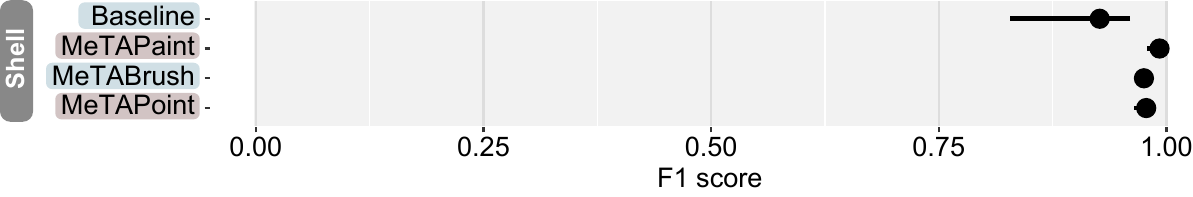}
 \caption{The F1 score for T3 (\textit{Shell}). Error bars: 95\% CIs.}
 \label{fig:F1 score for T3}
\end{figure}

\begin{figure}[t]
    \centering
    \captionsetup[subfigure]{labelformat=empty}
      \includegraphics[width=1\columnwidth]{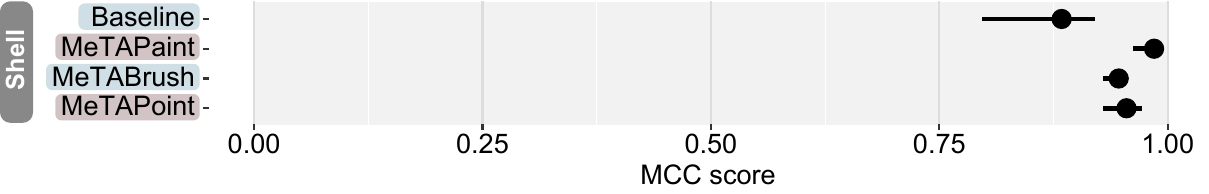}
 \caption{The MCC score for T3 (\textit{Shell}). Error bars: 95\% CIs.}
 \label{fig:MCC score for T3}
\end{figure}

\begin{figure}[t]
    \centering
    \captionsetup[subfigure]{labelformat=empty}
      \includegraphics[width=1\columnwidth]{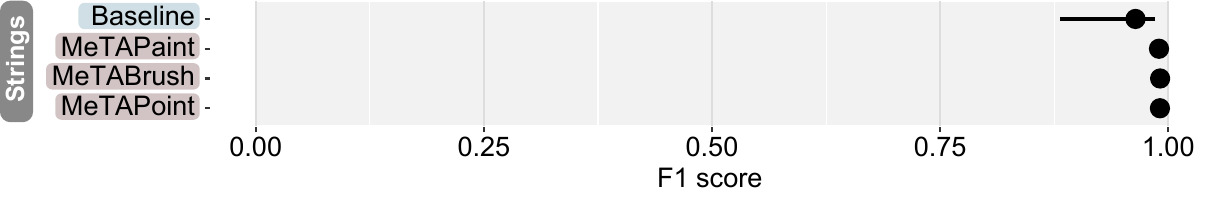}
 \caption{The F1 score for T4 (\textit{Strings}). Error bars: 95\% CIs.}
 \label{fig:F1 score for T4}
\end{figure}

\begin{figure}[t]
    \centering
    \captionsetup[subfigure]{labelformat=empty}
      \includegraphics[width=1\columnwidth]{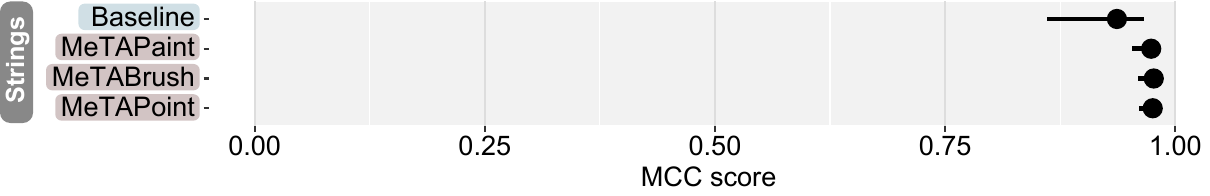}
 \caption{The MCC score for T4 (\textit{Strings}). Error bars: 95\% CIs.}
 \label{fig:MCC score for T4}
\end{figure}

\begin{figure}[t]
    \centering
    \captionsetup[subfigure]{labelformat=empty}
      \includegraphics[width=1\columnwidth]{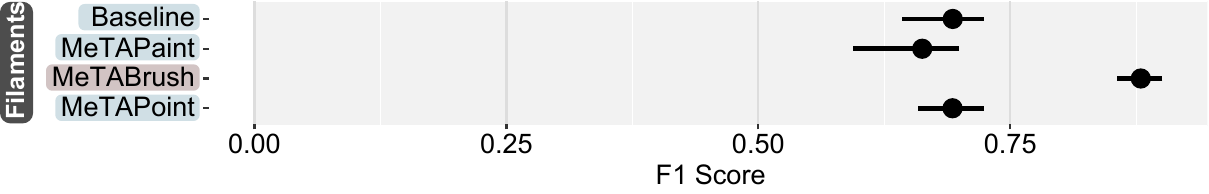}
 \caption{The F1 score for T5 (\textit{Filaments}). Error bars: 95\% CIs.}
 \label{fig:F1 score for T5}
\end{figure}

\begin{figure}[t]
    \centering
    \captionsetup[subfigure]{labelformat=empty}
      \includegraphics[width=1\columnwidth]{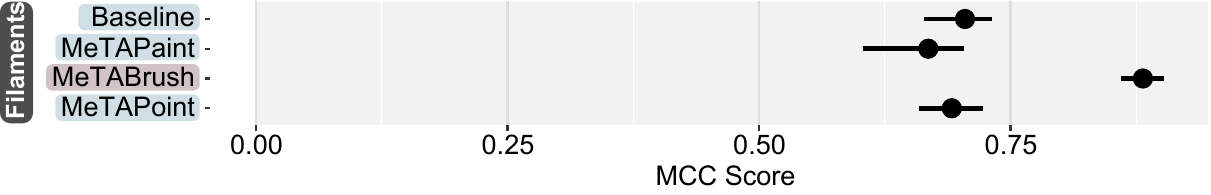}
 \caption{The MCC score for T5 (\textit{Filaments}). Error bars: 95\% CIs.}
 \label{fig:MCC score for T5}
\end{figure}

\begin{figure}[t]
    \centering
        \includegraphics[width=1\columnwidth]{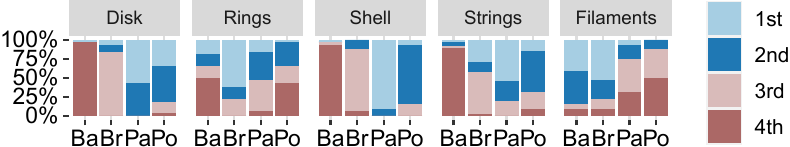}
    \caption{User preference for techniques with each dataset.}
   \label{fig:TechniqueRank}
\end{figure}

The following tables and graphs show additional results from the experiment we described in \autoref{sec:User study}. \autoref{tab:perdata} shows all mean task completion times, accuracy scores, and their corresponding 95\% confidence intervals for T1 to T5. \autoref{tab:overall table} shows the overall mean task completion times, accuracy scores, and their corresponding 95\% confidence intervals. \autoref{fig:overall time} shows the overall geometric means of completion time for each technique and \autoref{fig:overall time ratio} shows the pairwise ratio of completion times for T1--T4. \autoref{fig:Overall F1 score} and \ref{fig:Overall MCC score} show the overall accuracy scores for T1--T4. \autoref{fig:F1 score for T1}--\ref{fig:MCC score for T5} show the accuracy results per dataset. \autoref{fig:TechniqueRank} shows the rank across all datasets (\textit{Disk}, \textit{Rings}, \textit{Shell}, \textit{Strings}, \textit{Filaments}): \paint (Pa), \point (Po), \brush (Br), \baseline (Ba).

\begin{table}[h]
    \caption{The mean task completion times, accuracy scores, and their corresponding 95\% confidence intervals for T1 to T5.}
    \label{tab:perdata}
    \scriptsize%
    \centering%
    \begin{tabular*}{\hsize}{@{\extracolsep{\fill}}rl c@{\hspace{2pt}}c c@{\hspace{2pt}}c c@{\hspace{2pt}}c}
    \toprule
        \rule{0pt}{8pt}
        & Technique & Time & 95\%\,CI & F1 & 95\%\,CI & MCC & 95\%\,CI \\
        \midrule
         \multirow{4}{*}{\STAB{\rotatebox[origin=c]{90}{\textit{Disk}}}} & \point & 11s & [9,13] & .95  & [.94,.96] & .93 & [.91,.95] \\ 
        & \brush & 42s & [35,50] & .92  & [.90,.94] & .88 & [.85,.91] \\ 
        & \paint & 10s & [8,11] & .96  & [.93,.97] & .95 & [.92,.96] \\ 
        & \baseline & 37s & [30,46] & .88  & [.84,.89] & .82 & [.78,.83] \\ 
        \midrule
        \multirow{4}{*}{\STAB{\rotatebox[origin=c]{90}{\textit{Rings}}}} & \point & 38s & [32,45] & .97  & [.96,.97] & .95 & [.94,.96] \\ 
        & \brush & 17s & [14,21] & .97  & [.96,.98] & .96 & [.94,.97] \\ 
        & \paint & 40s & [32,50] & .96  & [.94,.97] & .94 & [.92,.96] \\ 
        & \baseline & 27s & [21,33] & .96  & [.93,.97] & .94 & [.91,.96] \\ 
        \midrule
        \multirow{4}{*}{\STAB{\rotatebox[origin=c]{90}{\textit{Shell}}}} & \point & 11s & [9,14] & .97  & [.96,.98] & .95 & [.92,.97] \\ 
        & \brush & 58s & [49,67] & .97  & [.96,.97] & .94 & [.92,.95] \\ 
        & \paint & 8s & [6,11] & .99  & [.97,.99] & .98 & [.96,.99] \\ 
        & \baseline & 62s & [51,76] & .92  & [.82,.96] & .88 & [.79,.92] \\ 
        \midrule
        \multirow{4}{*}{\STAB{\rotatebox[origin=c]{90}{\textit{Strings}}}} & \point & 14s & [11,17] & .99  & [.98,.99] & .97 & [96,.98] \\ 
        & \brush & 40s & [34,46] & .99  & [.98,.99] & .97 & [.95,.98] \\ 
        & \paint & 12s & [10,16] & .98  & [.97,.99] & .97 & [.95,.98] \\ 
        & \baseline & 63s & [51,77] & .96  & [.88,.98] & .93 & [.86,.96] \\ 
          \midrule
        \multirow{4}{*}{\STAB{\rotatebox[origin=c]{90}{\textit{Filaments}}}} & \point & 43s & [35,54] & .69  & [.65,.72] & .69 & [.65,.72] \\ 
        & \brush & 33s & [28,38] & .87  & [.85,.90] & .88 & [.85,.90] \\ 
        & \paint & 39s & [32,47] & .66  & [.59,.69] & .66 & [.60,.70] \\ 
        & \baseline & 29s & [23,36] & .69  & [.64,.72] & .70 & [.66,.73] \\ \bottomrule
    \end{tabular*}
\end{table}

\begin{table}[t]
    \caption{The overall (T1 to T4) mean task completion times, accuracy scores, and their corresponding 95\% confidence intervals.}
    \label{tab:overall table} 
    \centering
    \scriptsize%
	\centering%
     \begin{tabular*}{\hsize}{@{\extracolsep{\fill}}c | c c | c c | c c}
    \toprule
        Technique & Time & CI & FI & CI & MCC & CI \\ \midrule
        \point & 16s & [14,19] & .97 & [.96,.98] & .95 & [.94,.96] \\ 
        \brush & 36s & [31,41] & .96 & [.96,.97] & .94 & [.93,.95] \\ 
        \paint & 14s & [12,17] & .97 & [.96,.98] & .96 & [.95,.97] \\ 
        \baseline & 44s & [37,54] & .93 & [.87,.95] & .89 & [.83,.92] \\ \bottomrule
    \end{tabular*}
\end{table}

\clearpage
\section{Comparison of task performance between two user groups}
\label{appendix-comparison-seperate}
\begin{figure}[t]
    \centering
    \captionsetup[subfigure]{labelformat=empty}
      \includegraphics[width=1\columnwidth]{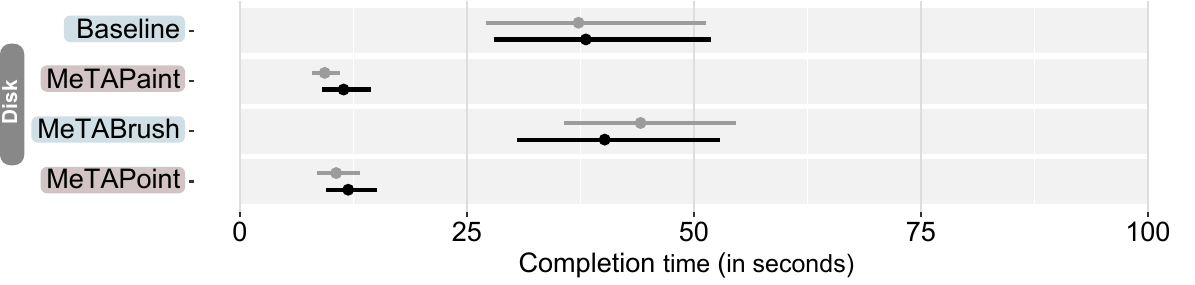}
 \caption{The geometric mean completion times in seconds for T1 (\textit{Disk}). VR users (black), non-experts (gray). Error bars: 95\% CIs.}
 \label{fig:TwoGroups_Time_T1}
\end{figure}

\begin{figure}[t]
    \centering
    \captionsetup[subfigure]{labelformat=empty}
      \includegraphics[width=1\columnwidth]{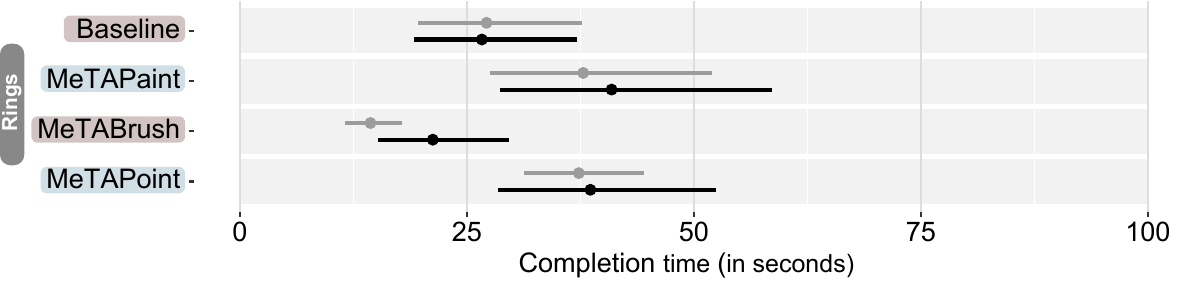}
 \caption{The geometric mean completion times in seconds for T2 (\textit{Rings}). VR users (black), non-experts (gray). Error bars: 95\% CIs.}
 \label{fig:TwoGroups_Time_T2}
\end{figure}

\begin{figure}[t]
    \centering
    \captionsetup[subfigure]{labelformat=empty}
      \includegraphics[width=1\columnwidth]{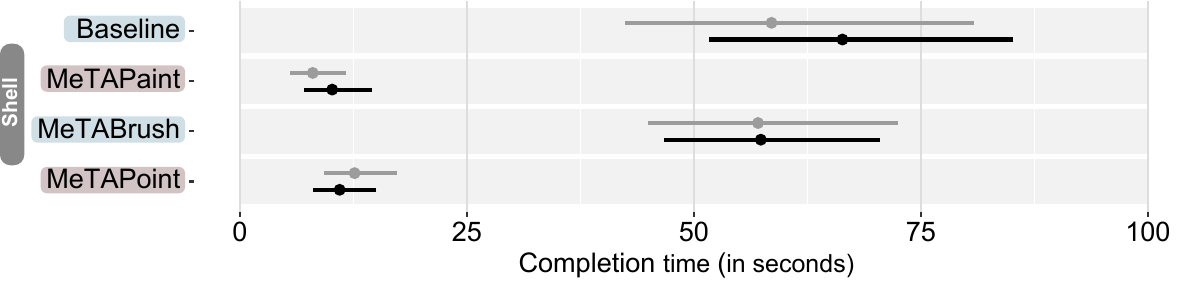}
 \caption{The geometric mean completion times in seconds for T3 (\textit{Shell}). VR users (black), non-experts (gray). Error bars: 95\% CIs.}
 \label{fig:TwoGroups_Time_T3}
\end{figure}

\begin{figure}[t]
    \centering
    \captionsetup[subfigure]{labelformat=empty}
      \includegraphics[width=1\columnwidth]{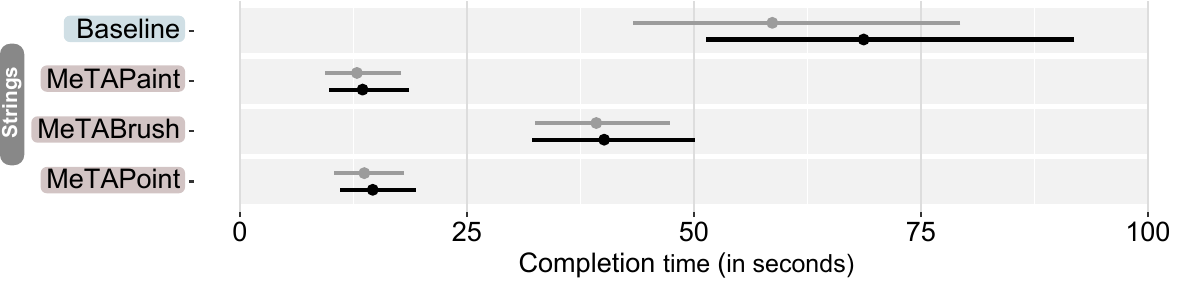}
 \caption{The geometric mean completion times in seconds for T4 (\textit{Strings}). VR users (black), non-experts (gray). Error bars: 95\% CIs.}
 \label{fig:TwoGroups_Time_T4}
\end{figure}

\begin{figure}[t]
    \centering
    \captionsetup[subfigure]{labelformat=empty}
      \includegraphics[width=1\columnwidth]{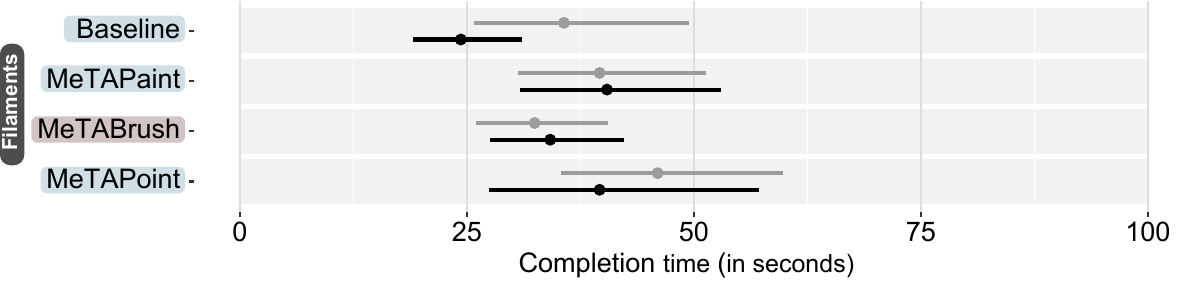}
 \caption{The geometric mean completion times in seconds for T5 (\textit{Filaments}). VR users (black), non-experts (gray). Error bars: 95\% CIs.}
 \label{fig:TwoGroups_Time_T5}
\end{figure}

\begin{figure}[t]
    \centering
    \captionsetup[subfigure]{labelformat=empty}
      \includegraphics[width=1\columnwidth]{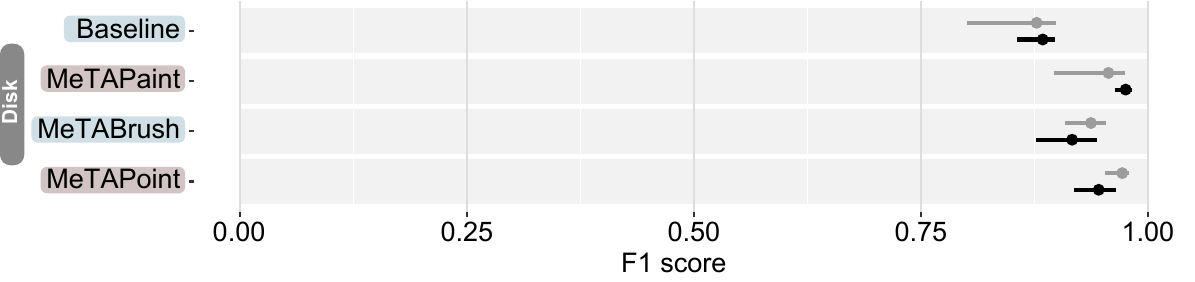}
 \caption{The F1 score for T1 (\textit{Disk}). VR users (black), non-experts (gray). Error bars: 95\% CIs.}
 \label{fig:TwoGroups_F1_T1}
\end{figure}

\begin{figure}[t]
    \centering
    \captionsetup[subfigure]{labelformat=empty}
      \includegraphics[width=1\columnwidth]{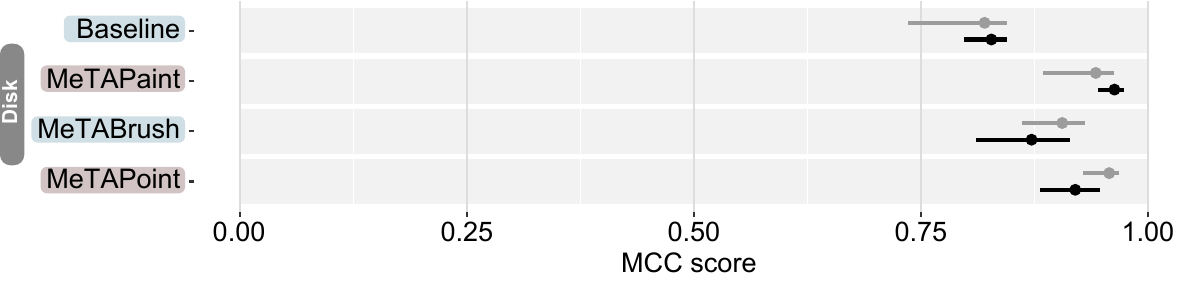}
 \caption{The MCC score for T1 (\textit{Disk}). VR users (black), non-experts (gray). Error bars: 95\% CIs.}
 \label{fig:TwoGroups_MCC_T1}
\end{figure}

\begin{figure}[t]
    \centering
    \captionsetup[subfigure]{labelformat=empty}
      \includegraphics[width=1\columnwidth]{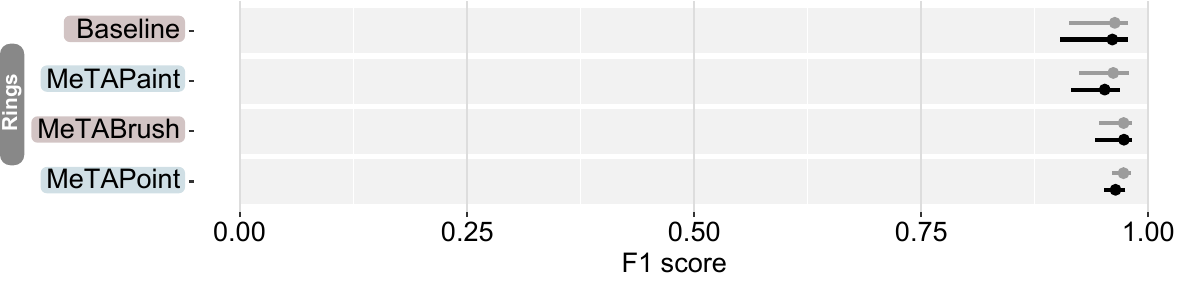}
 \caption{The F1 score for T2 (\textit{Rings}). VR users (black), non-experts (gray). Error bars: 95\% CIs.}
 \label{fig:TwoGroups_F1_T2}
\end{figure}

\begin{figure}[t]
    \centering
    \captionsetup[subfigure]{labelformat=empty}
      \includegraphics[width=1\columnwidth]{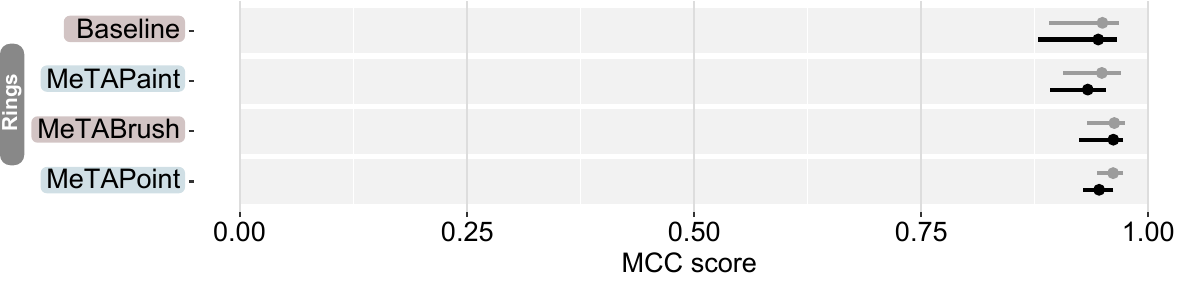}
 \caption{The MCC score for T2 (\textit{Rings}). VR users (black), non-experts (gray). Error bars: 95\% CIs.}
 \label{fig:TwoGroups_MCC_T2}
\end{figure}

\begin{figure}[t]
    \centering
    \captionsetup[subfigure]{labelformat=empty}
      \includegraphics[width=1\columnwidth]{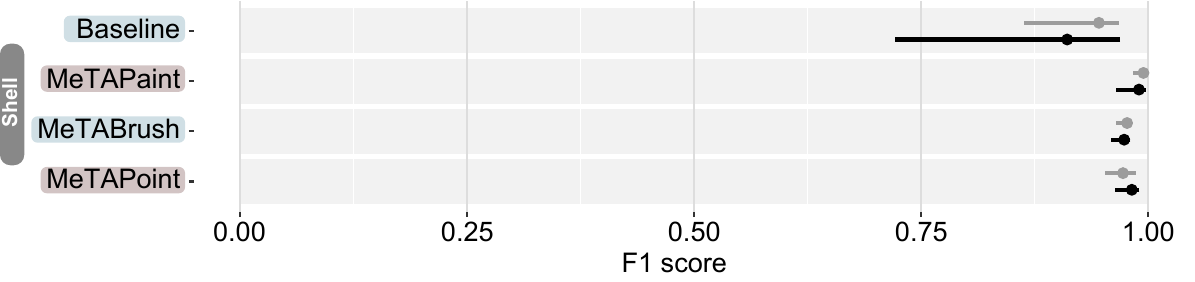}
 \caption{The F1 score for T3 (\textit{Shell}). VR users (black), non-experts (gray). Error bars: 95\% CIs.}
 \label{fig:TwoGroups_F1_T3}
\end{figure}

\begin{figure}[t]
    \centering
    \captionsetup[subfigure]{labelformat=empty}
      \includegraphics[width=1\columnwidth]{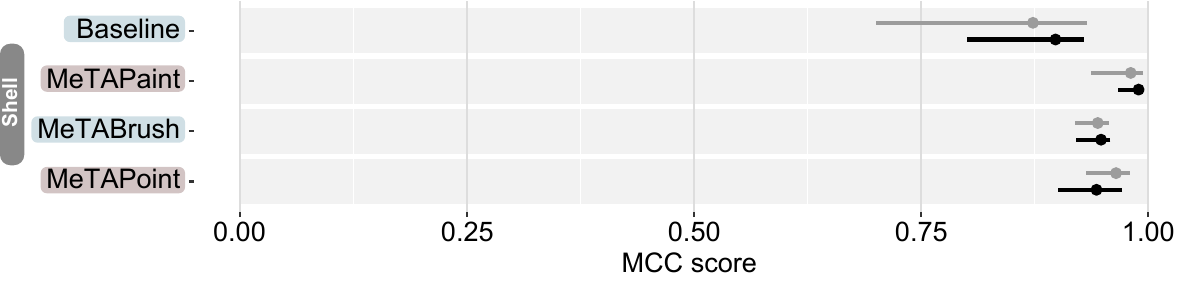}
 \caption{The MCC score for T3 (\textit{Shell}). VR users (black), non-experts (gray). Error bars: 95\% CIs.}
 \label{fig:TwoGroups_MCC_T3}
\end{figure}

\begin{figure}[t]
    \centering
    \captionsetup[subfigure]{labelformat=empty}
      \includegraphics[width=1\columnwidth]{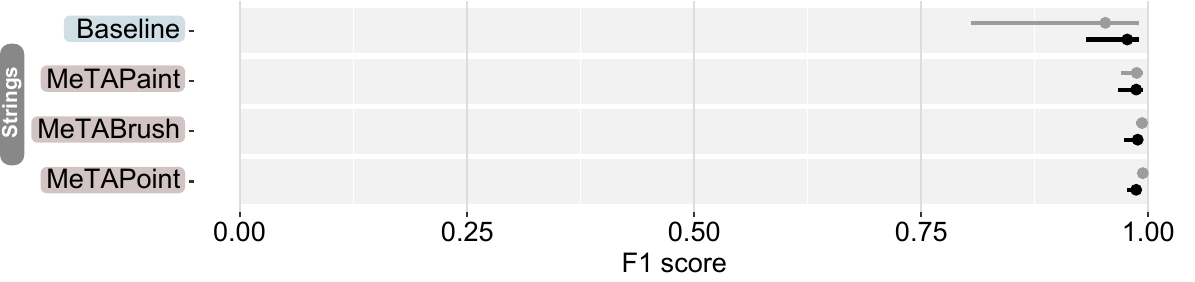}
 \caption{The F1 score for T4 (\textit{Strings}). VR users (black), non-experts (gray). Error bars: 95\% CIs.}
 \label{fig:TwoGroups_F1_T4}
\end{figure}

\begin{figure}[t]
    \centering
    \captionsetup[subfigure]{labelformat=empty}
      \includegraphics[width=1\columnwidth]{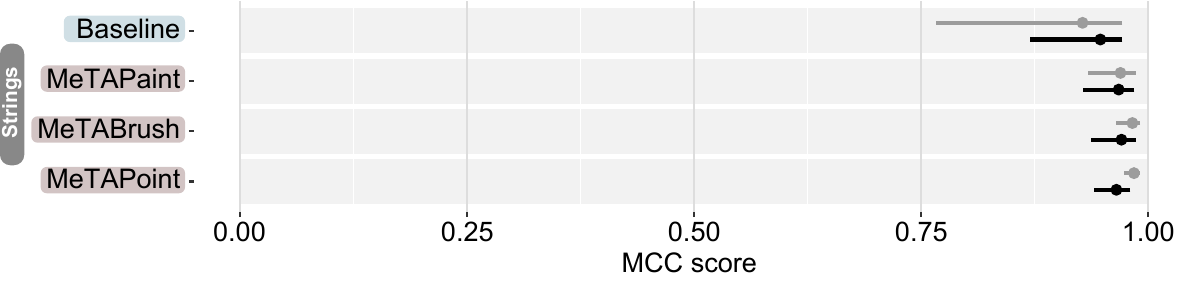}
 \caption{The MCC score for T4 (\textit{Strings}). VR users (black), non-experts (gray). Error bars: 95\% CIs.}
 \label{fig:TwoGroups_MCC_T4}
\end{figure}

\begin{figure}[t]
    \centering
    \captionsetup[subfigure]{labelformat=empty}
      \includegraphics[width=1\columnwidth]{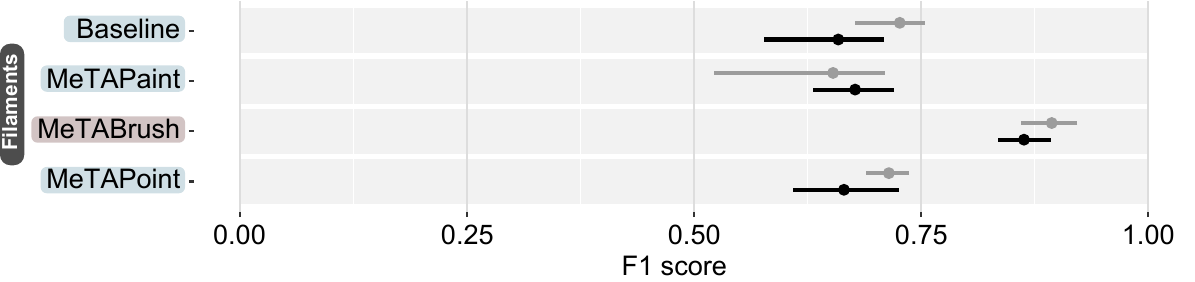}
 \caption{The F1 score for T5 (\textit{Filaments}). VR users (black), non-experts (gray). Error bars: 95\% CIs.}
 \label{fig:TwoGroups_F1_T5}
\end{figure}

\begin{figure}[t]
    \centering
    \captionsetup[subfigure]{labelformat=empty}
      \includegraphics[width=1\columnwidth]{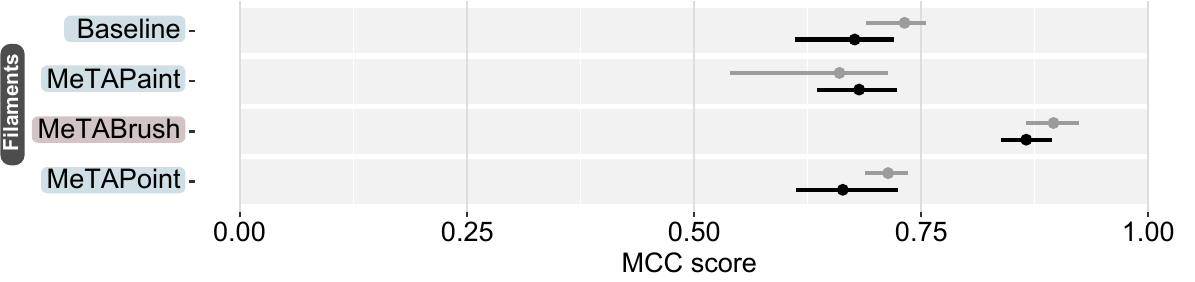}
 \caption{The MCC score for T5 (\textit{Filaments}). VR users (black), non-experts (gray). Error bars: 95\% CIs.}
 \label{fig:TwoGroups_MCC_T5}
\end{figure}

We further conducted a separate analysis of task performance, taking into consideration the participants' level of VR fluency. We divided the participants into two groups: $15$ VR users (weekly experience) and 17 non-experts (yearly experience and novices). Overall, our comparison revealed that both groups exhibited comparable task performance with MeTACAST methods. However, with Baseline method, both groups were equally fast in the \textit{Disk} and \textit{Rings} tasks. In the \textit{Shell} and \textit{Strings} tasks, which have more complex occlusion features as indicated in ~\autoref{fig:datafeature_selectiontechnique},  VR users were slower than non-experts. On the other hand, VR users were faster in the most challenging task, \textit{Filaments}. We assume that VR users were slower in the complex tasks due to the higher demands of task completion, but they had better direction sense in VR environment and were able achieve better performance in the task which requires a good sense of data location and structure. This finding also indicates that task performance with MeTACAST is independent of the influence of VR experience---even novice users are able to achieve similar performance to VR users, thanks to the intuitive and effective design of the selection techniques. The following graphs show task performance results for the two groups (black: VR users, gray: non-experts). \autoref{fig:TwoGroups_Time_T1}--\ref{fig:TwoGroups_Time_T5} show the geometric means of completion times for each technique per dataset, and \autoref{fig:TwoGroups_F1_T1}--\ref{fig:TwoGroups_MCC_T5} show the accuracy results per dataset, comparing the two user groups.

\newlength{\tablesep}
\setlength{\tablesep}{6.5pt}
\begin{table*}[t]
	\centering
	\footnotesize
	\caption{An overview of spatial selection techniques for 3D data.}
	\begin{tabu}{l@{\hspace{\tablesep}}l@{\hspace{\tablesep}}l@{\hspace{\tablesep}}l@{\hspace{\tablesep}}l@{\hspace{\tablesep}}l@{\hspace{\tablesep}}l@{\hspace{\tablesep}}l}
	  \toprule
		technique 								            & data type 			        & {\makecell[l]{selection \\ basis}}  	& whole\,/\,partial 			& selection strategy 		    & {\makecell[l]{precision \\ needs}}  			& metaphor 		& shape adjustment\\
	  \midrule
		Cylinder Selection \cite{Lucas:2005:DE3,Lucas:2005:DEM} 			& ROI 				            & region 			    & whole\,/\,partial 				& semi-autom., 2D           & high          & lasso     	& no control 	\\
		Cloudlasso \cite{Yu:2012:ESA} 		                & point cloud\,/\,scalar        & value 			    & whole\,/\,partial 				& semi-autom., 2D		    & medium        & lasso		    & threshold adjust \\
		SpaceCAST \cite{Yu:2016:CEE}		                & point cloud\,/\,scalar        & value			        & whole\,/\,partial					& semi-autom., 2D		    & medium        & lasso			& threshold adjust	\\
		Volume Catcher \cite{Owada:2005:VC}	                & scalar 				        & value 			    & whole						        & semi-autom., 2D		    & low           & lasso		    & no control	\\
		TraceCAST \cite{Yu:2016:CEE}	                    & point cloud\,/\,scalar 	    & value			        & whole				                & semi-autom., 2D		    & low           & lasso		    & threshold adjust	\\
		PointCAST \cite{Yu:2016:CEE}		                & point cloud\,/\,scalar 	    & value			        & whole				                & semi-autom., 2D		    & low           & raycasting	& threshold adjust	\\
		LassoNet \cite{Chen:2020:LassoNet}					& point cloud 				    & deep learning			& whole\,/\,partial				    & semi-autom., 2D		    & low           & lasso		    & no control	\\
		WYSIWYP \cite{Wiebel:2012:WYSIWYP}					& scalar  		                & value	                & none						        & semi-autom., 2D		    & low           & raycasting	& no need to control	\\
        Slicing Volume \cite{Roberto:2020:HMS}	            & point cloud                   & region                & partial                           & manual                    & high          & raycasting    & region adjust\\
        Hybrid AR selection \cite{sereno:2022:HTTinAR}      & point cloud\,/\,scalar        & region                & partial                           & manual                    & high          & lasso\,+\,extrusion   & no control \\
        TangibleBrush \cite{Besancon:2019:TangibleBrush}    & point cloud\,/\,scalar        & region                & partial                           & manual                    & high          & lasso\,+\,extrusion   & no control \\
        Touching the cloud \cite{Lubos:2014:TBP}            & point cloud                   & region                & partial                           & manual                    & high          & brush         & region adjust \\         
        Embodied Axes \cite{Cordeil:2020:EmbodiedAxes}      & ROI                           & region			    & whole\,/\,partial                 & manual                    & high          & tangible (box) & region adjust\\
        Fiducial-Based Tangible \cite{gomez:2010:FBT}       & tensor                        & region                & partial                           & manual                    & high          & brush    & no control \\
        Live-Wire \cite{malmberg:2006:3DLW}                 & scalar                        & value                 & partial                           & semi-autom., 3D           & medium        & seed points            & no control \\
        Lightweight Tangible \cite{jackson:2013:LTI}        & vector                        & value                 & whole                 & semi-autom., 3D           & medium        & tangible (orien.) & no control \\
        Neuron Tracing \cite{mcdonald:2020:IUV}   & scalar                        & value                 & partial                           & semi-autom., 3D           & low           & brush                 & no control \\
        \point (this paper)                                           & point cloud\,/\,scalar 	    & value			        & whole	                            & semi-autom., 3D           & low           & point\,+\,drag    & threshold adjust	\\
        \brush (this paper)                                           & point cloud\,/\,scalar 	    & value                  & whole\,/\,partial                 & semi-autom., 3D          & low           & brush         & threshold adjust	\\
        \paint (this paper)                                    & point cloud\,/\,scalar 	    & value			        & whole	                            & semi-autom., 3D           & low           & brush         & threshold adjust	\\
		\bottomrule%
	\end{tabu}\vspace{-1ex}
	\label{tab:comparison}
\end{table*}

\section{Comparison of existing spatial selection techniques}
\label{appendix-comparison}
\autoref{tab:comparison} shows an overview of spatial selection techniques for 3D data focusing on data characteristics, selection requirements (such as the selection basis and target shape), selection strategies (such as metaphor and selection strategy), and users’ level of control (including precision demand and post-adjustment), as described in \autoref{subsec:comparison}.

\section*{Images/graphs/plots/tables/data license/copyright}
We as authors state that all of our figures, graphs, plots, and data tables in this appendix are and remain under our own personal copyright, with the permission to be used here. We also make them available under the \href{https://creativecommons.org/licenses/by/4.0/}{Creative Commons At\-tri\-bu\-tion 4.0 International (\ccLogo\,\ccAttribution\ \mbox{CC BY 4.0})} license and share them at \href{https://osf.io/dvj9n/}{\texttt{osf.io/dvj9n}}.

\end{document}